\documentclass[11pt,oneside]{book}%
\usepackage[dvips]{graphics}%

\usepackage[centertags]{amsmath}
\usepackage{amsfonts}
\usepackage{amssymb}
\usepackage{amsthm}
\usepackage{newlfont}

\begin{document}%

\setlength{\unitlength}{1mm}
\baselineskip .85cm
\large \baselineskip .85cm
\begin{titlepage}
\title{\Huge\vspace{-2cm} {\bf CLASSICAL SOLUTIONS IN EINSTEIN'S GRAVITY AND STUDY OF SOME COLLAPSING MODELS} \vspace*{5cm} \\
\Large {\bf THESIS SUBMITTED FOR THE DEGREE OF DOCTOR OF
PHILOSOPHY (SCIENCE)
   OF  JADAVPUR UNIVERSITY}\\\vspace{1cm}
    }

\author{ {\bf BY} \vspace*{2cm} \\
\LARGE{\bf U~J~J~A~L~~ D~E~B~N~A~T~H}  \vspace*{2cm}\\
\large {\bf DEPARTMENT OF MATHEMATICS} \\
\large {\bf JADAVPUR UNIVERSITY}\\
\small  {\bf CALCUTTA - 700 032} \\
\large {\bf INDIA}\\\\ {\bf 2004}}
\date{}
\maketitle
\end{titlepage}
\pagenumbering{roman}
\newpage
\vspace*{2.5cm}
\begin{center}
 {\large {\bf\Large DECLARATION} }
\end{center}
\vspace*{1.5cm} This is to certify that the thesis entitled
``{\it CLASSICAL SOLUTIONS IN EINSTEIN'S GRAVITY AND STUDY OF SOME
COLLAPSING MODELS}" submitted by Ujjal Debnath who got his name
registered on 18.01.2002 for the award of Ph.D.(Science) degree of
Jadavpur University, is absolutely based upon his own work under
the supervision of Prof. Subenoy Chakraborty (D. Sc.) and Prof.
Narayan Chandra Chakraborty, Department of Mathematics, Jadavpur
University, Calcutta 700032, and that neither this thesis nor any
part of its has been submitted for any degree / diploma or any
other academic award anywhere before.\\

\vspace*{3cm}
\noindent(Subenoy Chakraborty) \hspace{2in}(Narayan Chandra Chakraborty)\\
Professor,                     \hspace{3in} Professor,\\
Department of Mathematics      \hspace{1.7in} Department of Mathematics \\
Jadavpur University            \hspace{2.3in} Jadavpur University \\
Calcutta 700032                \hspace{2.6in} Calcutta 700032\\
India                          \hspace{3.4in} India \\


\newpage
\vspace*{1cm}
\begin{center}
 {\huge Acknowledgements}
\end{center}
\vspace*{1cm}

I express my deep gratitude to my supervisors Prof. Subenoy
Chakraborty and Prof. Narayan Chandra Chakraborty for their
guidance and support throughout my work. It was an enlighting
experience for me to work under them. I further thank them for
stimulating discussions and careful analysis of the problems. I
am also grateful to the members of Relativity and Cosmology
Research Centre (J.U.) for their valuable suggestions. I would
like to thank the teachers and staff at the  Department of
Mathematics, Jadavpur University, for offering their valuable
resources, advice and help for completion of this work. I would
like to thank Prof. John D. Barrow of Cambridge University and
Prof. Asit Banerjee of Jadavpur University for their help and
affection. I also thanks to  IUCAA (Pune, India) for  worm
hospitality and facilities for some works.

\vspace*{2cm}
\noindent Ujjal Debnath\\
Department of Mathematics\\
Jadavpur University \\
Calcutta - 700032\\
India

\baselineskip .7cm
\tableofcontents

\newpage

\pagenumbering{arabic}

\addcontentsline{toc}{part}{Part A~: Classical Solutions in
Einstein's Gravity}
 \baselineskip
.81cm \markright{ }

\vspace*{8cm}
\begin{center}
{\Huge{{\bf Part A} \\ \vspace{1cm} Classical Solutions in
Einstein's Gravity }}
\end{center}

\large \baselineskip .85cm

\chapter{General Introduction}
\setcounter{page}{2} \markright{\it CHAPTER~\ref{chap1}. General
Introduction}
\label{chap1}%

\section{Cosmological Implications of the Theory of varying Speed of Light}
The {\it constants} of nature are not constant at all. During the
early phase of the universe, the constants of nature were varying
significantly, although now-a-days this variation is quite
insignificant. Assuming their constancy at all times requires
massive extrapolation, with no observational basis. Could the
universe come into being riding the back of wildly varying
constants? Several constants of nature have been stripped off
their status in theories proposed in the past. Physicists have
long entertained the possibility of a varying gravitational
constant $G$ (e.g., Brans-Dicke theories [Brans and Dicke, 1961]),
a varying electron charge $e$ [Bekenstein et al, 1982, 2002] and
more generally varying coupling constants. Indeed with the advent
of string theory, the variation of these {\it constants} seems to
be fashionable. In sharp contrast, the constancy of the speed of
light has remain scared and the term {\it heresy} is occasionally
used in relation to {\it Varying Speed of Light Theories} (i.e.,
VSL Theories) [Barrow, 1999]. The reason is clear: the constancy
of $c$, unlike the constancy of $G$ or $e$, is the pillar of
special relativity and thus of modern physics. Varying $c$
theories are expected to cause much more structural damage to
physics formalism than other varying constant
theories.\\

The first {\it varying constant} was the speed of light as
suggested by Kelvin and Tait [1874], which was thirty years before
Einstein's proposal of special relativity. Thereby a varying $c$
did not shok anyone, as indeed $c$, unlike $G$ played no special
role in the formalism of physics. This is to be contrasted with
the state of affairs after 1905, when Eddington commented ``{\it A
variation in $c$ is self-contradictory}'' [Eddington, 1946]. This
astonishing statement does a disservice to the experimental
testability and scientific respectability of the theory of
relativity. The Eddington's statement has now been transmuted
into ``asking whether $c$ has varied over cosmic history is like
asking whether the numbers of liters to the gallon has varied ''
[Duff, 2002]. The implication is that the constancy of the speed
of light is a logical necessity, a definition that could not have
been otherwise. For centuries the constancy of the speed of light
played no role in physics and presumably physics did not start
being logically consistent in 1905. Furthermore, the postulate of
the constancy of $c$ in special relativity was prompted by
experiments (including those leading to Maxwell's theory) rather
than issues of consistency. Seen from an angle, even in a world
where all seems to vary and nothing is constant, it is always
possible to define units such that $c$ remains a constant. So
varying $c$ is tautological and is tied to the definition of a
system of units.\\

In classical electromagnetism the speed of light is only constant
in vacuum and it `varies' in dielectric media. The changing $c$
breaks Lorentz invariance and conservation of energy. The
geometry of Universe is not affected by a changing $c$. One can
allowed a changing $c$ to do the job normally done by {\it
superluminal expansion}. For changing $c$, the gravitational laws
should be modified. The basic assumption is that a variable $c$
does not induce corrections to curvature in the cosmological
frame and therefore, Einstein's equations, relating curvature to
stress energy are still valid. The reason behind this postulate
is that $c$ changes in the Local Lorentzian frames associated
with cosmological expansion. The effect is a special relativistic
effect and not a gravitational effect. Therefore curvature should
not be related with the variation of $c$.\\

The good news for experimentalists is that once the theoretical
choice is made, the different theories typically lead to very
different predictions. For instance, dilaton theories violate the
weak equivalence principle but maintain the Lorentz invariance
whereas VSL theories maintain weak equivalence principle
[Magueijo et al, 2002; Moffat, 2001] but break Lorentz invariance.
These differences have clear observational implications. For
instance the ``STEP'' satellite could soon rule out the dilaton
theories capable of explaining the results given by Webb et al
[http://einstein.stanford.edu/ STEP/]. Some VSL critics fail to
notice that there are theories for which a varying $c$ is
actually a dimensionless statement. This does not include the
results of Davies et al [2002], where a dimensionless
statement on varying $c$ is achieved only because a constant $G$ was assumed.\\

Even after the proposal of special relativity in 1905 many
varying speed of light theories were considered, most notably by
Einstein himself [1911]. VSL was then rediscovered and forgotten
on several occasions. For instance, in the 1930's VSL was used as
an alternative explanation for the cosmological redshift. This
theory conflicts with fine structure observations [Stewart, 1931;
Bue, 1932; Wold, 1935]. None of these efforts relates to recent
VSL theories, which are firmly entrenched in the success of the
hot big bang theory of the universe. According to the concept,
first modern VSL theory was Moffat's ground breaking researches
[Moffat, 1993], where spontaneous symmetry breaking of Lorentz
symmetry leads to VSL and an elegant solution to the horizon
problem.\\

The basic dynamical postulate is that Einstein's field equations
are valid even when $\dot{c}\ne 0$, with minimal coupling (i.e.,
with $c$ replaced by a field in the relevant equations) in this
particular frame:
\begin{equation}
G_{\mu\nu}-\Lambda g_{\mu\nu}=\frac{8\pi G}{c^{4}}~T_{\mu\nu}
\end{equation}

This is inspired by the statement of Maxwell's equations in
dielectric media. The first strong assumption is that the field
$c$ does not contribute to the stress-energy tensor. More
importantly, this postulate can only be true in one frame. Hence
the action remains as shown by Barrow et al [1998, 1999]

\begin{equation}
S=\int dx^{4}\left[\sqrt{-g}\left(\frac{c^{4}(R+2\Lambda)}{16\pi
G} +{\cal L_{M}}\right)+{\cal L_{c}} \right]
\end{equation}

The dynamical variables are the metric $g_{\mu\nu}$, any matter
field variables contained in the matter Lagrangian ${\cal L_{M}}$
and the field $c$ itself. The Riemann tensor and Ricci scalar are
to be computed from $g_{\mu\nu}$ at constant $c$ in the usual way.
This can be true in one frame. Therefore Einstein's equations do
not acquire new terms in the preferred frame. Minimal coupling at
the level of Einstein's equations is at the heart of the model's
ability to solve the cosmological problems. It requires of any
action principle formulation that the contribution ${\cal L_{c}}$
must not contain the metric explicitly and so does not contribute
to
the energy-momentum tensor.\\

Like inflation [Guth, 1981; Albretch et al, 1982; Linde, 1982,
83], modern VSL theories were motivated by the {\it cosmological
problems}$-$ the flatness, horizon, entropy, homogeneity, isotropy
and cosmological constant problems of big bang cosmology
([Albretch et al, 1999; Magueijo et al, 2000] for a review and
[Avelino et al, 2003] for a dimensionless description). At its
most basic VSL was inspired by the horizon problem. It does not
take much to see that a larger speed of light in the early
universe could `open up the horizons' [Moffat, 1993; Albretch et
al, 1999]. More mathematically, the comoving horizon is given by
$r_{h}=\frac{c}{\dot{a}}$ which indicates that in the past
$r_{h}$ must have decreased so as to causally connect the large
region at present. Thus

\begin{equation}
\frac{\ddot{a}}{\dot{a}}-\frac{\dot{c}}{c}>0
\end{equation}

that is, either one can get accelerated expansion (inflation) or
a decreasing speed of light or a combination of both. This
argument is far from general: a contraction period ($\dot{a}<0$,
as in the bouncing universe) or a static start for the universe
($\dot{a}=0$) are contradictory examples. However the horizon
problem is just a warm up for the other problems.\\

The VSL cosmological model first proposed by Albretch and
Magueijo [1999] making use of a preferred frame, thereby
violating the principle of relativity. In this respect a few
remarks are in order. Physicists don't like preferred frames but
they often ignore the very obvious fact that they have a great
candidate for a preferred frame: the cosmological frame. Modern
physicists invariably choose to formulate their laws without
reference to this preferred frame. The VSL theory proposed by
Albretch et al [1999], makes a radically different choice in this
respect: it ties the formulation of the physical laws to the
cosmological frame. Equation (1.1) can only be true in one frame;
although the dynamics of Einstein's gravity is preserved to a
large extent, it is no longer a geometrical theory. However, if
$c$ does not vary by much, the effects of the preferred frame are
negligible, of the order
$\left(\frac{\dot{c}}{c}\right)^{n}\frac{v}{c}$, where $n$ is the
rank of the corresponding tensor in general relativity. Inertial
forces in these VSL theories may turn out to be a strong
experimental probe and are currently under
investigation.\\

Since Friedmann equations are still valid even when $\dot{c}\ne
0$. Thus the Friedmann equations are
\begin{equation}
\frac{\dot{a}^{2}}{a^{2}}=\frac{8\pi G}{3}~\rho-\frac{k
c^{2}}{a^{2}}
\end{equation}
and
\begin{equation}
\frac{\ddot{a}}{a}=-\frac{4\pi
G}{3}\left(\rho+\frac{3p}{c^{2}}\right)
\end{equation}

where $\rho c^{2}$ and $p$ are the energy density and pressure
respectively, $k=0, \pm 1$ is the spatial curvature and a dot
denotes a derivative with respect to cosmological time. If the
universe is flat ($k=0$) and radiation dominated
$(p=\frac{1}{3}\rho c^{2})$, we have as usual $a\propto t^{1/2}$.
As expected from minimal coupling, the Friedmann equations remain
valid, with $c$ being replaced by a variable in much the same way
that Maxwell equations in media may be obtain by simply replacing
the
dielectric constant of the vacuum by that of the medium.\\

Combining the two Friedmann equations (1.4) and (1.5) now leads
to:

\begin{equation}
\dot{\rho}+3\frac{\dot{a}}{a}\left(\rho+\frac{p}{c^{2}}\right)=\frac{3kc\dot{c}}{4\pi
G a^{2}}
\end{equation}

i.e., there is a source term in the energy conservation equation.
This turns out to be a general feature of this VSL theory. It is
to be noted that in general relativity stress-energy conservation
results directly from Einstein's equations, as an integrability
condition via Bianchi identities. By trying this theory to a
preferred frame, violations of Bianchi identities must occur
[Besset et al, 2000] and furthermore the link between them and
energy conservation is broken. Hence one expect violations of
energy conservation and these are proportional to gradients of
$c$. In the brane world realization of VSL [Youm, 2001], these
violations of energy conservation are nothing but matter sticking
to or falling off the brane.\\

In the context of preset $c(t)$, two scenarios were considered:
phase transitions and Machian. In phase transitions, the speed of
light varies abruptly at a critical temperature as in the models
of Moffat [1993] and Albretch et al [1999]. This could be related
to the spontaneous breaking of local Lorentz symmetry [Moffat,
1993]. Later Barrow considered scenarios in which the speed of
light varies like a power of the expansion factor i.e., $c\propto
a^{n}$, the so-called Machian scenarios. Taken at face value such
scenarios are inconsistent with experiment (Barrow et al [2000]
for an example of late time constraint on $n$). Such variations
must therefore be confined to the very early universe and so the
$c$-function considered by Barrow [1999] and Barrow et al [1999]
should really be understood as
\begin{equation}
c=c_{0}\left\{1+\left(\frac{a}{a_{0}}\right)^{n} \right\}
\end{equation}

where $a_{0}$ is the scale at which VSL switches off. The problem
with putting in `by hand' a function $c(t)$ is that the
predictive power of the theory is severely reduced [Moffat, 2001].\\

Now one may consider more quantitative regarding conditions for a
solution to the cosmological problems. Now starting with the
horizon problem, one can consider the phase transition scenario
[Moffat, 1993; Albretch et al, 1999] for the time $t_{c}$ when the
speed of light changes from $c^{-}$ to $c^{+}$. The past light
cone intersects $t=t_{c}$ at a sphere with comoving radius
$r=c^{+}(\eta_{0}-\eta_{c})$ where $\eta_{0}$ and $\eta_{c}$ are
the conformal times now and at $t_{c}$ respectively. The horizon
size at $t_{c}$, on the other hand, has comoving radius
$r_{h}=c^{-}\eta_{c}$. If $\frac{c^{-}}{c^{+}}\gg
\frac{\eta_{0}}{\eta_{c}}$ then $r\ll r_{h}$, meaning that the
whole observable universe today has in fact always been in causal
contact. This requires

\begin{equation}
log_{_{10}}\frac{c^{-}}{c^{+}}\gg
32-\frac{1}{2}~log_{_{10}}z_{eq}+\frac{1}{2}~log_{_{10}}\frac{T^{+}_{c}}{T^{+}_{p}}
\end{equation}

where $z_{eq}$ is the redshift at matter radiation equality and
$T^{+}_{c}$ and $T^{+}_{p}$ are the universe and Planck
temperature after the phase transition. If $T^{+}_{c}\approx
T^{+}_{p}$ this implies light traveling more than 30 orders of
magnitude faster before the phase transition. It is tempting, for
symmetry reasons, simply to postulate that $c^{-}=\infty$ but
this is not strictly necessary.\\

Considering now the flatness problem, let $\rho_{c}$ be the
critical density of the universe:

\begin{equation}
\rho_{c}=\frac{3}{8\pi G}\left(\frac{\dot{a}}{a}\right)^{2}
\end{equation}

i.e., the mass density corresponding to the flat model ($k=0$)
for a given value of $\frac{\dot{a}}{a}$. Now defining deviations
from flatness in terms of $\epsilon=\Omega-1$ with
$\Omega=\frac{\rho}{\rho_{c}}$, the equations (1.4), (1.5) and
(1.6) are simplified to

\begin{equation}
\dot{\epsilon}=\epsilon(1+\epsilon)(2+3\gamma)\frac{\dot{a}}{a}+2\epsilon~\frac{\dot{c}}{c}
\end{equation}

where $\gamma-1=\frac{p}{\rho c^{2}}$ is the equation of state
($\gamma=1,~4/3$ for matter/radiation). In the standard big bang
theory $\epsilon$ grows like $a^{2}$ in the radiation era and
like `$a$' in the matter era, leading to a total growth by 32
orders of magnitude since the Planck epoch. The observational
fact that $\epsilon$ can be at most of order 1 nowadays requires
either that $\epsilon=0$ strictly or that an amazing fine tuning
must have existed in the initial conditions ($\epsilon<10^{-32}$
at $t=t_{p}$). This is the flatness puzzle.\\

As equation (1.10) shows, a decreasing speed of light
($\frac{\dot{c}}{c}<0$) would drive $\epsilon$ to 0, achieving
the required tuning. If the speed of light changes in a sharp
phase transition with $\mid\frac{\dot{c}}{c}\mid\gg
\frac{\dot{a}}{a}$, then a decrease in $c$ by more than 32 orders
or magnitude would suitably flatten the universe. But this should
be obvious even before doing any numeries, from inspection of the
{\it non-conservation} equation (1.6). Indeed if $\rho$ is above
its critical value (as in the case for a closed universe with
$k=1$) then equation (1.6) tells that energy is destroyed. If
$\rho<\rho_{c}$ (as for an open model, for which $k=-1$) then
energy is produced. Either way the energy density is pushed
towards the critical value $\rho_{c}$. In contrast to the big
bang model, during a period with $\frac{\dot{c}}{c}<0$ only the
flat, critical universe is stable. This is the VSL solution to
the flatness problem.\\

VSL cosmology has had further success in resolving other problems
of big bang cosmology that are usually tackled by inflation. It
solves the entropy, isotropy and homogeneity problems [Moffat,
1993; Albretch et al, 1999]. Also it solves at least one version
of the cosmological constant problem [Albretch et al, 1999]
([Moffat, 2001] for a discussion of the quantum version of the
Lambda problem). Further VSL cosmology has a quasi-flat and
quasi-Lambda attractor [Barrow et al, 1999] (i.e., an attractor
with non-vanishing, but also non-dominating Lambda or curvature).\\

These cosmological implications of VSL have led to much further
work. The stability of the various solutions was demonstrated by
Szydowski et al [2002] and Biesiada et al [2000] using dynamical
systems methods [Cimento et al, 2001; Gopakumar et al, 2001]. The
role of Lorentz symmetry breaking in the ability of these models
to solve the cosmological problems was discussed by Avelino and
Martins [1999]. Also combinations of varying $c$ and varying $G$
have been studied by Barrow et al [Barrow et al, 1999]. Barrow
[2002] has cast doubts on the ability of this model to solve the
isotropy problem if the universe starts very anisotropic. It was
also found that if $c$ falls fast enough to solve the flatness
and horizon problems then the quantum wavelengths of massive
particles states and the radii of primordial black holes will
grow to exceed the scale of the particle horizon. However this
statement depends crucially on how
the Planck's constant $h$ scales with $c$ in VSL theories.\\

The relation between VSL and the second law of thermodynamics was
investigated recently by Barrow [2002], Youm [2002], Coule [1999]
and Cimento et al [2001]. In particular Cimento et al [2001] found
that if the second law of thermodynamics is to be retained in open
universes, $c$ can only decrease, whereas in flat and close
models it must stay constant. A similar discussion in the context
of black holes can be found in Davies et al [2002]. It is not
surprising if the second law of thermodynamics were violated in
VSL models with hard breaking of Lorentz invariance. After all
the first law is violated in these models. However a deviation
from first principles remains elusive.\\

It is nevertheless interesting that several observational puzzles
can be solved with VSL. With supplementary the redshift
dependence in fine structure constant $\alpha$ could be seen as a
result of a variation of $c$. Another puzzle was the observation
of rare very high energy cosmic rays, in conflict with standard
kinematic calculations based on special relativity which predict
a cut-off well below the observed energies. This could present
the first experimental mishap of special relativity and evidence
for some VSL theories. Finally, the recently observed accelerating
universe may be well explained with the variation of $c$.\\

\section{Quintessence}
For the last few decades, it is generally believed that after a
small period of inflationary era, when the universe was
accelerated, the universe has been evolving in a big bang
scenario (deceleration of the universe) and most of the present
day observations are in accord with the {\it Standard Cosmological
Model} (SCM). The SCM can give a satisfactory explanation to
other observational properties of the present universe (e.g.,
primordial nucleosynthesis, extragalactic sources redshift,
cosmic microwave radiation). But difficulty has started when data
from high redshift type Ia Supernovae ($z\sim 1$) observations
[Perlmutter et al, 1998, 1999; Riess et al, 1998; Garnavich et
al, 1998] suggest an accelerated universe at the present epoch. So
there must be some matter field which is either neglected or
unknown responsible for this accelerated universe [Bachall et al,
1999]. This type of matter field is called {\it quintessence
matter}, shortly {\it Q-matter} and the problem is called {\it
quintessence problem}. According to Caldwell et al [1998] and
Sahni et al [2000] this Q-matter can behave like a cosmological
constant by combining positive energy density and negative
pressure related by the equation of state $p=-\rho$ [Turner, 1997;
Ostriker et al, 1995].\\

However this does not constitute the only possibility: among all
the other candidates, the missing energy can be associated to a
dynamical time-dependent and spatially homogeneous or
inhomogeneous scalar field $\phi$ evolving slowly down its
potential $V(\phi)$. The resulting cosmological models are known
as {\it quintessence models} [Caldwell et al, 1998; Turner et al,
1984, 1997]. In these models, the scalar field can be seen as a
perfect fluid with a negative pressure given by $p=(\gamma-1)\rho$
($0<\gamma<1$). In what follows one shall focus on equations of
state with a constant $\gamma$ and disregard the case of time
varying equations of state invoked in some quintessence models,
called {\it tracker models}, to solve the {\it cosmic coincidence
problem} [Zlatev et al, 1999; Cooray et al, 1999].\\

Going on with the analysis, very interesting question is whether
it is possible to construct a successful common scheme for the
two cosmological mechanisms involving rolling scalar fields i.e.,
quintessence and inflation. This perspective has the appealing
feature of providing a unified view of the past and recent
history of the universe, but can also remove some weak points of
the two mechanisms when considered separately. Indeed, inflation
could provide the initial conditions for quintessence without any
need to fix them by hand and quintessence could hope to give some
more hints in constraining the inflation potential on
observational grounds.\\

A homogeneous and isotropic universe is characterized by the
Friedmann-Robertson-Walker (FRW) line element

\begin{equation}
ds^{2}=dt^{2}-a^{2}(t)\left(\frac{dr^{2}}{1-kr^{2}}+r^{2}d\theta^{2}+r^{2}sin^{2}\theta
d\phi^{2} \right)
\end{equation}

where $a$ is the scale factor and $k$ is the spatial curvature
index (i.e., $k=0,~\pm 1$). Assuming the universe is filled with
perfect normal matter plus quintessence fluid corresponding to
some scalar field governed by Klein-Gordon equation, the over all
stress-energy tensor of the cosmic fluid without the dissipative
pressure has the form

\begin{equation}
T_{\mu\nu}=(\rho+p)u_{\mu}u_{\nu}+pg_{\mu\nu},~~~~u_{\mu}u^{\mu}=-1
\end{equation}

where $\rho=\rho_{_{m}}+\rho_{_{\psi}}$ and
$p=p_{_{m}}+p_{_{\psi}}$. Here $\rho_{_{m}}$ and $p_{_{m}}$ are
the energy density and pressure of the matter whose equation of
state is $p_{_{m}}=(\gamma_{m}-1)\rho_{_{m}}$ with adiabatic
index in the interval $1\le\gamma_{m}\le 2$. Likewise
$\rho_{_{\psi}}$ and $p_{_{\psi}}$, the energy density and
pressure of the minimally coupled self-interacting Q-matter field
$\psi$ i.e.,

\begin{equation}
\rho_{_{\psi}}=\frac{1}{2}\dot{\psi}^{2}+V(\psi),~~~p_{_{\psi}}=\frac{1}{2}\dot{\psi}^{2}-V(\psi)
\end{equation}

are related by an equation of state similar to that of the
matter, viz., $p_{_{\psi}}=(\gamma_{\psi}-1)\rho_{_{\psi}}$, so
that its adiabatic index is given by
\begin{equation}
\gamma_{\psi}=\frac{\dot{\psi}^{2}}{\frac{1}{2}\dot{\psi}^{2}+V(\psi)}
\end{equation}

where for non-negative potential $V(\psi)$ one has
$0\le\gamma_{\psi}\le 2$. The scalar field can be properly
interpreted as Q-matter provided $\gamma_{\psi}<1$ [Zlatev et al,
1999; Steinhardt et al, 1999] . As usual an overdot means
derivative with respect to cosmic time. In general
$\gamma_{\psi}$ varies as the universe expands and the same is
true for $\gamma_{m}$, since the massive and massless components
of the matter fluid redshift at different
rates.\\

The Friedmann equation together with the energy conservation of
the normal matter fluid and quintessence (Klein-Gordon equation)
are

\begin{equation}
H^{2}+\frac{k}{a^{2}}=\frac{1}{3}(\rho_{_{m}}+\rho_{_{\psi}})
\end{equation}
\vspace{-5mm}
\begin{equation}
\dot{\rho}_{_{m}}+3H\gamma_{m}\rho_{_{m}}=0
\end{equation}
\vspace{-5mm}
\begin{equation}
\ddot{\psi}+3H\dot{\psi}+\frac{dV(\phi)}{d\phi}=0
\end{equation}

where $H\equiv \frac{\dot{a}}{a}$ denotes the Hubble factor.
Introducing $\Omega_{m}\equiv \frac{\rho_{_{m}}}{\rho_{_{c}}},~
\Omega_{\psi}\equiv
\frac{\rho_{_{\psi}}}{\rho_{_{c}}},~\Omega_{k}=-\frac{k}{(aH)^{2}}$
and $\Omega=\Omega_{m}+\Omega_{\psi}$ with $\rho_{_{c}}\equiv
3H^{2}$ as the critical density, the set of equations
(1.15)-(1.17) can be recast as [Ellis et al, 1997]
\begin{equation}
\Omega_{m}+\Omega_{\phi}+\Omega_{k}=1
\end{equation}
\vspace{-5mm}
\begin{equation}
\dot{\Omega}=\Omega(\Omega-1)(3\gamma-2)H
\end{equation}
\vspace{-5mm}
\begin{equation}
\dot{\Omega}_{\psi}=[2+(3\gamma-2)\Omega-3\gamma_{\psi}]\Omega_{\psi}H
\end{equation}

where $\gamma$ is the average adiabatic index given by
\begin{equation}
\gamma~\Omega=\gamma_{m}\Omega_{m}+\gamma_{\psi}\Omega_{\psi}~.
\end{equation}

The combined measurements of the Cosmic Microwave Background
(CMB) temperature fluctuations and the distribution of galaxies
on large scales seem to imply that the universe may be flat or
nearly flat [Turner, 1997; Ostriker et al, 1995]. Hence the
interesting solution at late times of equation (1.19) is
$\Omega=1$ (i.e., $k=0$) and discard the solution $\Omega=0$ as
incompatible with observation. The solution $\Omega=1$ is
asymptotically stable for expanding universes ($H>0$) provided
that the condition
$\frac{\partial\dot{\Omega}}{\partial\Omega}<0$ holds in a
neighbourhood of $\Omega=1$ and this implies $\gamma<2/3$. Hence
the matter stress violates the strong energy condition
$\rho+3p\le 0$ and so the deceleration parameter
\begin{equation}
q=-\frac{a\ddot{a}}{\dot{a}^{2}}<0
\end{equation}

(because, $\frac{\ddot{a}}{a}=-\frac{1}{6}(\rho+3p)>0$). Hence as
a consequence the universe accelerates its expansion.\\

From theoretical point of view a lot of works [Caldwell et al,
1998; Ostriker et al, 1995; Peebles, 1984; Wang et al, 2000;
Perlmutter et al, 1999; Dodelson et al, 2000; Faraoni, 2000] have
been done to solve the quintessence problem and possible
candidates for Q-matter are cosmological constant (or more
generally a variable cosmological term), a scalar field [Peebles
et al, 1988, 2002; Ratra et al, 1988; Ott, 2001; Hwang et al,
2001; Ferreira et al, 1998] with a potential giving rise to a
negative pressure at the present epoch, a dissipative fluid with
an effective negative stress [Cimento et al, 2000] and more exotic
matter like a frustrated network of non-abelian cosmic strings or
domain wall [Bucher et al, 1999; Battye et al, 1999].\\

Unfortunately, most of the fields (Q-matter field, tracker field)
work only for a spatially flat ($k=0$) FRW model. Recently,
Cimento et al [2000] showed that a combination of dissipative
effects such as a bulk viscous stress and a quintessence scalar
field gives an accelerated expansion for an open universe
($k=-1$) as well. This model also provides a solution for the
`coincidence problem' as the ratio of the density parameters
corresponding to the normal matter and the quintessence field
asymptotically approaches a constant value. Bertolami and Martins
[2000] obtained an accelerated expansion for the universe in a
modified Brans-Dicke (BD) theory by introducing a potential which
is a function of the Brans-Dicke scalar field itself. Banerjee
and Pavon [2001] have shown that it is possible to have an
accelerated universe with BD
theory in Friedmann model without any matter.\\\\

\large \baselineskip .85cm
\chapter{Brans-Dicke  Cosmology  in  an  Anisotropic  Model when  Velocity  of  Light  Varies}
\label{chap2}\markright{\it CHAPTER~\ref{chap2}. Brans-Dicke
Cosmology  in  an  Anisotropic  Model when  Velocity  of Light
Varies}

\section{Prelude}
In  recent  years  theories  with  varying  speed  of  light
(VSL)  has  been  attracted considerable  attention  [Albretch et
al, 1999; Barrow et al, 1998, 1999, 2000; Magueijo, 2000, 2001,
2003] due  to  its  ability  to  solve  the  so-called
cosmological puzzles -  the horizon, flatness  and  Lambda
problems  of big-bang cosmology. All  previous  attempts to
overcome  these difficulties  invoke  the  basic inflationary
form, where  the observable  universe  experiences  a  period of
`superluminal' expansion. Here,  one has  to  modify  the matter
content  of the  universe  such  that  ordinary  Einstein
gravity  becomes repulsive  and  halts  the  exponential
expansion. One  can however resolve to theories  with  varying
speed  of light  as the  alternative  method  to solve  the
above  mentioned cosmological puzzles. Here, instead  of
changing  the matter content  of  the  universe, one has  to
change  the  speed  of light  in  the  early  universe. The
universe  is assumed  to be  radiation  dominated  in the  early
stage  and  the  matter content is assumed to be the  same  as
in  the  standard big-bang  (SBB)  model. So  the  geometry  and
expansion  factor of  the  universe  go  in  accordance  with
the  SBB  model.\\

So  far  some  works  have  been  done  with  VSL  by  Magueijo
and  co-workers [Albretch et al, 1999; Barrow et al, 1998, 1999,
2000; Magueijo, 2000, 2001] and others [Jacobson et al, 2000;
Landau et al, 2000; Besset et al, 2000; Kiritsis, 1999; Harko et
al, 1999]  specially  in  isotropic  cosmological models (which
also include BD-theory). Very  recently, Magueijo [2001] has
investigated, the possibility  of  black  holes formation by
studying  spherically  symmetric  solutions with  VSL.\\

In  this  chapter, we study  Brans-Dicke  Cosmology  in
anisotropic  Kantowski-Sachs  space-time  model, considering
variation  of  the  speed of  light. We address the flatness
problem considering  both  the  cases  namely, (i) when  the
Brans-Dicke scale  field $\phi$ is  constant (ii) when $\phi$
varies, specially  for  radiation  dominated  era perturbatively
and non-perturbatively. For radiation dominated era we give the
exact  solution to  the  flatness  problem. We study solutions
to  the lambda, quasi-lambda and quasi-flatness problems  and
examine their asymptotic behaviour.

\section{The  Basic  Equations}
For  anisotropic  Kantowski-Sachs (KS) model  with  metric  ansatz

$$
ds^{2}=-c^{2}dt^{2}+a^{2}(t)dr^{2}+b^{2}(t)d\Omega_{2}^{2}~,
$$

the  Brans-Dicke (BD) field  equations  with  varying  speed  of
light are

\begin{equation}
\frac{\ddot{a}}{a}+2\frac{\ddot{b}}{b}=-\frac{8\pi}{(3+2\omega)\phi}\left[
(2+\omega)\rho+3(1+\omega)\frac{p}{c^{2}}\right]-\omega\left(\frac{\dot{\phi}}{\phi}
\right)^{2}-\frac{\ddot{\phi}}{\phi}
\end{equation}

\begin{equation}
\left(\frac{\dot{b}}{b}
\right)^{2}+2\frac{\dot{a}}{a}\frac{\dot{b}}{b}=\frac{8\pi\rho}{\phi}-\frac{c^{2}}{b^{2}}-\left(\frac{\dot{a}}{a}+2\frac{\dot{b}}{b}
\right)\frac{\dot{\phi}}{\phi}+\frac{\omega}{2}\left(\frac{\dot{\phi}}{\phi}
\right)^{2}
\end{equation}

and  the  wave  equation  is

\begin{equation}
\ddot{\phi}+\left(\frac{\dot{a}}{a}+2\frac{\dot{b}}{b}
\right)\dot{\phi}=\frac{8\pi}{3+2\omega}\left(\rho-\frac{3p}{c^{2}}
\right)
\end{equation}

Here  the  velocity  of  light  $c$  is  an  arbitrary  function
of  time, $\omega$ is  the  BD  coupling  parameter  and  the BD
scalar  field  is $\phi=\frac{1}{G}$. From  the  above  field
equations  we have  the  {\it non-conservation}  equation
\begin{equation}
\dot{\rho}+\left(\frac{\dot{a}}{a}+2\frac{\dot{b}}{b}
\right)\left(\rho+\frac{p}{c^{2}} \right)=\frac{c\dot{c}}{4\pi
b^{2}}\phi
\end{equation}

Now  $p=\frac{1}{3}\rho c^{2}$ is  the  equation  of  state  in
the radiation  era for  which  the  general  solution  $\phi$ is
(from equation (2.3))
\begin{equation}
\phi=\phi_{0}+\alpha\int{\frac{dt}{a b^{2}}}.
\end{equation}

The  above  field  equations  in  BD-theory  have  been
formulated  in  Jordan  frame. To  switch  over  to  Einstein
frame  we  shall  have  to  make  the  following  transformations:
\begin{eqnarray*}
d\hat{t}=\sqrt{G\phi}~dt,~~\hat{a}=\sqrt{G\phi}~a,~~\hat{b}=\sqrt{G\phi}~b,
~~\sigma=\left(\omega+\frac{3}{2}\right)^{1/2}\text{ln}(G\phi),
\end{eqnarray*}
\vspace{-5mm}
\begin{equation}
~~\hat{\rho}=(G\phi)^{-2}\rho,~~\hat{p}=(G\phi)^{-2}p~,
\end{equation}

and  the  above  field  equations  become ($c$ is  treated  as
constant)

\begin{equation}
\frac{\hat{a}''}{\hat{a}}+2\frac{\hat{b}''}{\hat{b}}=-4\pi
G\left(\hat{\rho}+\frac{3\hat{p}}{c^{2}}\right)-\sigma'^{2}~,
\end{equation}

\begin{equation}
\left(\frac{\hat{b}'}{\hat{b}}\right)^{2}+2\frac{\hat{a}'}{\hat{a}}\frac{\hat{b}'}{\hat{b}}
+\frac{c^{2}}{\hat{b}^{2}}=8\pi
G\hat{\rho}+\frac{\hat{\sigma}'^{2}}{2}~,
\end{equation}
and
\begin{equation}
\sigma''+\left(\frac{\hat{a}'}{\hat{a}}+2\frac{\hat{b}'}{\hat{b}}\right)\sigma'=\frac{8\pi
G}{\sqrt{6+4\omega}}\left(\hat{\rho}-\frac{3\hat{p}}{c^{2}}\right)
\end{equation}

with  the prime `` $'$ '' and `` . '' represent the
differentiation with respect to $\hat{t}$ and $t$ respectively.\\

One  can  interpret  these  field  equations  as  standard  KS
equations  with  constant $G$  and  a  scalar  field  $\sigma$  is
added to  the  normal  matter . The  scalar  field  behaves like
a ``stiff''  perfect  fluid  with  equation  of  state

\begin{equation}
\hat{p}_{\sigma}=\hat{\rho}_{\sigma}=\frac{\sigma'^{2}}{16\pi G}~.
\end{equation}

If  the  velocity  of  light  is  constant, then  in  Einstein
frame  total  stress-energy  tensor  is  conserved  but  there
is  an  exchange  of  energy  between  the  scalar  field  and
normal  matter  according  to  the  following  equation
\begin{equation}
\hat{\rho}'+\left(\frac{\hat{a}'}{\hat{a}}+2\frac{\hat{b}'}{\hat{b}}\right)
\left(\hat{\rho}+\frac{\hat{p}}{c^{2}}\right)=-\left[\hat{\rho}'_{\sigma}+
\left(\frac{\hat{a}'}{\hat{a}}+2\frac{\hat{b}'}{\hat{b}}\right)\left(
\hat{\rho}_{\sigma}+\frac{\hat{p}_{\sigma}}{c^{2}}\right)\right]=
-\frac{\sigma'}{\sqrt{6+4\omega}}\left(\hat{\rho}-\frac{3\hat{p}}{c^{2}}\right)
\end{equation}

On  the  other  hand, if  the  velocity  of  light  varies then
we  have two separate  ``non-conservation'' equations

\begin{equation}
\hat{\rho}'+\left(\frac{\hat{a}'}{\hat{a}}+2\frac{\hat{b}'}{\hat{b}}\right)
\left(\hat{\rho}+\frac{\hat{p}}{c^{2}}\right)=-\frac{\sigma'}{\sqrt{6+4\omega}}
\left(\hat{\rho}-\frac{3\hat{p}}{c^{2}}\right)+\frac{c c'}{4\pi
G\hat{b}^{2}}~,
\end{equation}
and
\begin{equation}
\hat{\rho}'_{\sigma}+
\left(\frac{\hat{a}'}{\hat{a}}+2\frac{\hat{b}'}{\hat{b}}\right)\left(
\hat{\rho}_{\sigma}+\frac{\hat{p}_{\sigma}}{c^{2}}\right)=\frac{\sigma'}
{\sqrt{6+4\omega}}\left(\hat{\rho}-\frac{3\hat{p}}{c^{2}}\right)
\end{equation}

Thus  standard  KS  model  field  equations  with  constant  $G$
for  VSL  are  also  the  field  equations  for  BD  theory  in
Einstein  frame  but  here  one  must  add  to  normal  matter
the  scalar  field  energy  and  pressure. Therefore, the total
energy  density  and  pressure  are  given  by
$\hat{\rho}_{t}=\hat{\rho}+\hat{\rho}_{\sigma}$ and
$\hat{p}_{t}=\hat{p}+\hat{p}_{\sigma}$ respectively.\\

\section{Perturbative  Solutions  to  the  Flatness  Problem}

For  a  possible  solution  to  the  flatness  problem  in  VSL
theory, let us  first  study  solutions  when  there  are  small
deviations  from  flatness. The  critical  energy  density
$\rho_{c}$ in a  BD  universe  is  given  by  the  equation

\begin{equation}
\left(\frac{\dot{b}}{b}
\right)^{2}+2\frac{\dot{a}}{a}\frac{\dot{b}}{b}=\frac{8\pi\rho_{c}}{\phi}-\left(\frac{\dot{a}}{a}+2\frac{\dot{b}}{b}
\right)\frac{\dot{\phi}}{\phi}+\frac{\omega}{2}\left(\frac{\dot{\phi}}{\phi}
\right)^{2}
\end{equation}

In  Einstein  frame  the  critical  energy  density  for  normal
matter  is

\begin{equation}
\hat{\rho}_{c}=\frac{1}{8\pi G
}\left[\left(\frac{\hat{b}'}{\hat{b}}\right)^{2}+2\frac{\hat{a}'}{\hat{a}}\frac{\hat{b}'}{\hat{b}}
-\frac{\hat{\sigma}'^{2}}{2}\right]=\frac{\rho_{c}}{G^{2}\phi^{2}}
\end{equation}

Hence, if we define  the  total  critical  energy  as
$$
\hat{p}_{\alpha}=\hat{\rho}_{c}+\hat{\rho}_{\sigma}~,
$$

then  from  (2.15)  we  have
\begin{equation}
\hat{\rho}_{\alpha}=\frac{1}{8\pi
G}\left[\left(\frac{\hat{b}}{\hat{b}}
\right)^{2}+2\frac{\hat{a}'}{\hat{a}}\frac{\hat{b}'}{\hat{b}}\right]
\end{equation}

Let  us  define  a  relative  flatness  parameter  as
\begin{equation}
\varepsilon_{t}=\frac{\hat{\rho}_{t}-\hat{\rho}_{\alpha}}{\hat{\rho}_{\alpha}}
\end{equation}

whose  evolution  equation  is given by
\begin{equation}
\hat{\varepsilon}'_{t}=\hat{\varepsilon}_{t}(1+\hat{\varepsilon}_{t})\left(
\gamma\frac{\hat{a}'}{\hat{a}}+2(\gamma-1)\frac{\hat{b}'}{\hat{b}}\right)
+2\frac{c'}{c}~\hat{\varepsilon}_{t}
\end{equation}

Here $\gamma-1=p/\rho c^{2}=\hat{p}/\hat{\rho}c^{2}$ , is  the
equation of state, with $\gamma$ constant. Since
\begin{equation}
\hat{\varepsilon}'_{t}=\frac{\varepsilon}{1+(2\omega+3)\dot{\phi}^{2}/32\pi\phi\rho_{c}}
\end{equation}

in  Jordan  frame,  the  deviation  from  flatness  is  given  by

\begin{equation}
\delta=\frac{\rho-\rho_{c}}{\rho_{c}+(2\omega+3)\dot{\phi}^{2}/32\pi\phi\rho_{c}}<\varepsilon
\end{equation}

where $\varepsilon=(\rho-\rho_{c})/\rho_{c}$  measures  natural
deviations from flatness. Now $\delta$, the  adaptation  parameter
satisfies  the differential  equation

\begin{equation}
\dot{\delta}=\delta(1+\delta)\left(\gamma\frac{\dot{a}}{a}+2(\gamma-1)\frac{\dot{b}}{b}
+\frac{1}{2}(3\gamma-2)\frac{\dot{\phi}}{\phi}\right)+2\frac{\dot{c}}{c}\delta
\end{equation}

If $\delta$ is  assumed  to  be  very  small  compare  to  unity
i.e., $\delta\ll 1$, then  neglecting  the  square  term in
(2.21), one can integrate the above to  get
\begin{equation}
\delta=\delta_{0}a^{\gamma}b^{2(\gamma-1)}\phi^{(3\gamma-2)/2}c^{2}
\end{equation}

where $\delta_{0}$ is the integration  constant. Hence we have

\begin{equation}
\frac{1}{\varepsilon}=\frac{\varepsilon_{0}}{a^{\gamma}b^{2(\gamma-1)}
\phi^{(3\gamma-2)/2}c^{2}}-\frac{(2\omega+3)\dot{\phi}^{2}}{32\pi\rho_{c}\varepsilon}
\end{equation}

In  BD  theory $G\propto\phi^{-1}$ , so  to  solve  the flatness
problem $\phi$ should decrease  in  the  early  universe and the
first term on  the right  hand  side  must  dominate  over the
second one.

In  particular, in  radiation  era  ($\gamma=4/3$) the  above
equation (2.23)  simplifies  to

\begin{equation}
\frac{1}{\varepsilon}=\frac{\varepsilon_{0}}{a^{4/3}b^{2/3}\phi
c^{2}}-\frac{(2\omega+3)\alpha^{2}}{32\pi\phi
a^{2}b^{4}\rho_{c}\varepsilon}
\end{equation}

where  we  have  used  equation  (2.5)  to  obtain  $\phi$ . If
$\alpha$ is small  and
$c=c_{0}\left[a^{n}b^{2(n-1)}\right]^{n/(3n-2)}$ then [Barrow et
al, 1999] it is still possible to  solve the  flatness  problem
for $n<-1-\sqrt{7/3}$ . Furthermore, since $\rho_{c}\propto
1/(ab^{2})^{4/3}$ at late times, so the second term is always very
small compare to the  first and hence may be neglected.\\

\section{Non-Perturbative  Solutions  with  $\phi$ = Constant:~ Matter
and Radiation  Dominated  Era}

When $\phi=\phi_{0}$ (constant) we have $\delta=\varepsilon$ in
(2.20) and the differential equation for $\delta$, (2.21) can be
written as
\begin{equation}
\dot{\varepsilon}=\varepsilon(1+\varepsilon)\left(\gamma\frac{\dot{a}}{a}
+2(\gamma-1)\frac{\dot{b}}{b}\right)+2\frac{\dot{c}}{c}~\varepsilon
\end{equation}

Now, if  we  assume a  power-law  form  in  the scale  factors
of  the  velocity  of  light  then  the  above equation  becomes

\begin{equation}
\dot{\varepsilon}=\varepsilon(1+\varepsilon)\left(\gamma\frac{\dot{a}}{a}
+2(\gamma-1)\frac{\dot{b}}{b}\right)+2\varepsilon\left(\frac{n}{3n-2}\right)
\left(n\frac{\dot{a}}{a}+2(n-1)\frac{\dot{b}}{b}\right)
\end{equation}

In  order  to  get  an  exact  analytic  solution  to  the above
equation, we  assume  $\varepsilon\ll 1$, so  we can neglect
$\varepsilon^{2}$-term. We then have  the  solution
\begin{equation}
\varepsilon=\varepsilon_{0}a^{\gamma+2n^{2}/(3n-2)}b^{2(\gamma-1)+4n(n-1)/(3n-2)}
\end{equation}

with $\varepsilon_{0}$ as the  integration  constant.

Furthermore, without  assuming  any  restriction  on  the
parameter $\varepsilon$, if  we  assume  the  two  arbitrary
constants $n$ and  $\gamma$ to  be  equal  then  also  we have an
exact  analytic solution  for $\varepsilon$,
\begin{equation}
\varepsilon=-\frac{1}{\frac{3\gamma-2}{5\gamma-2}+
A\left[a^{\gamma}b^{2(\gamma-1)}\right]^{-(5\gamma-2)/(3\gamma-2)}}
\end{equation}

where $A$  is  an  integration  constant.\\

\begin{figure}
\includegraphics{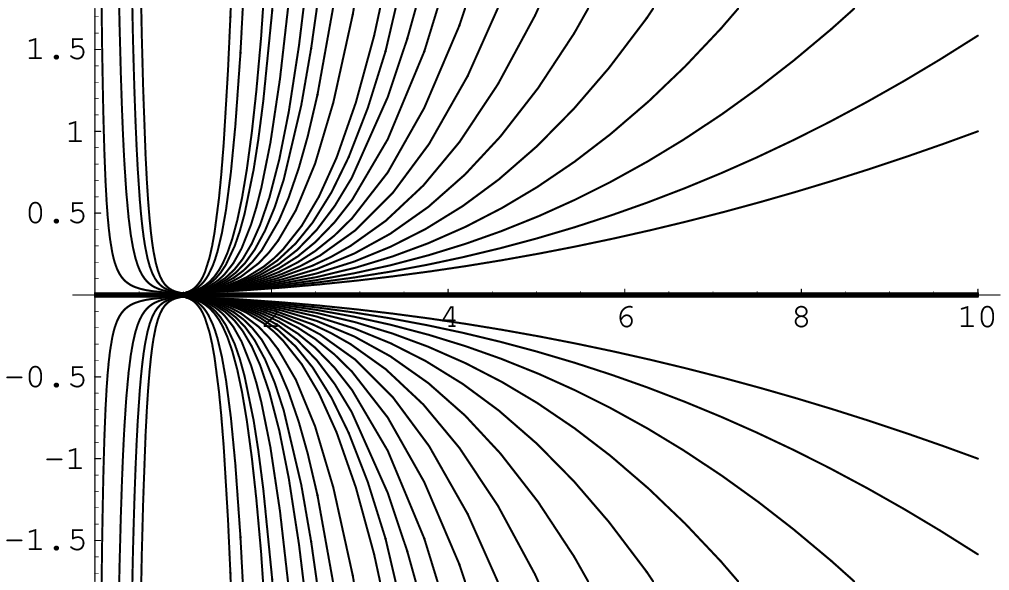}\\

Fig 2.1 : Here the flatness parameter $\varepsilon$ in (2.27) has
been plotted against $t$ for the radiation era ($\gamma=4/3$)
choosing different values for the constant $n$. The scale factors
here are
assumed to be in simple expanding form.\\\\

\includegraphics{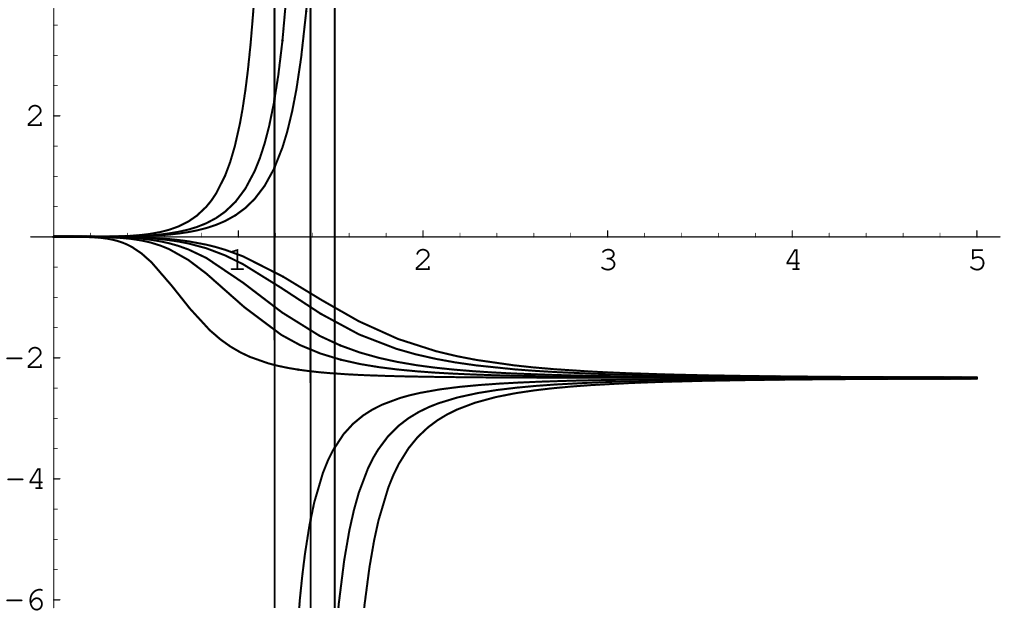}\\

Fig 2.2 : This figure shows the variation of the flatness
parameter from (2.28), where there is no restriction on
$\varepsilon$. This figure is also drawn for the radiation era
and assuming as before the simple expanding form for the scale
factors. The variation of $\varepsilon$ for different choice of
the arbitrary constant $A$ is shown in the figure.
\end{figure}

\section{Exact  Solutions  to  the  Flatness  Problem:~ Radiation
Dominated  Era} To  obtain  the  exact  solutions  to  the
flatness  problem we assume that the  variation  of  light is
governed mathematically  by  the relation [Barrow, 1999]
\begin{equation}
c=c_{0}\left[\left\{a^{n}b^{2(n-1)}\right\}^{1/(3n-2)}\sqrt{\phi G
}\right]^{n}
\end{equation}

However, in  the  Einstein  frame  the  above  relation reduces
to  $c=c_{0}\left[\hat{a}^{n}\hat{b}^{2(n-1)}\right]^{1/(3n-2)}$
and equation (2.12) becomes
\begin{equation}
\hat{\rho}'+\left(\frac{\hat{a}'}{\hat{a}}+2\frac{\hat{b}'}{\hat{b}}\right)
\left(\hat{\rho}+\frac{\hat{p}}{c^{2}}\right)=\frac{cc'}{4\pi
G\hat{b}^{2}}
\end{equation}

The  equation  of  state $\hat{p}=(\gamma-1)\hat{\rho}c^{2}$
simplifies the above equation further to

\begin{equation}
\hat{\rho}'+\gamma\hat{\rho}\left(\frac{\hat{a}'}{\hat{a}}+2\frac{\hat{b}'}{\hat{b}}\right)
=\frac{cc'}{4\pi G\hat{b}^{2}}
\end{equation}

An  exact  integral  can  be  obtained (assuming $n=\gamma$) as

\begin{equation}
\hat{\rho}'(\hat{a}\hat{b}^{2})^{\gamma}=B+\frac{\gamma
c_{0}^{2}}{4\pi
G(5\gamma-2)}\left[\hat{a}^{\gamma}\hat{b}^{2(\gamma-1)}\right]^{(5\gamma-2)/(3\gamma-2)}
\end{equation}

Hence  for  the  radiation  dominated  era, we  have  in Jordan
frame
\begin{equation}
\rho(ab^{2})^{4/3}=B+\frac{c_{0}^{2}}{14\pi G}(a^{2}b)^{14/9}(\phi
G)^{7/3}
\end{equation}

where  we  have  used
$\hat{\rho}(\hat{a}\hat{b}^{2})^{4/3}=\rho(ab^{2})^{4/3}$ .

Moreover, in the Jordan  frame  we  have  equation  (2.29)  for
the velocity of  light  and  we  can  write
\begin{equation}
\frac{\dot{c}}{c}=\frac{n}{3n-2}\left[n\frac{\dot{a}}{a}+2(n-1)\frac{\dot{b}}{b}\right]+
\frac{n}{2}\frac{\dot{\phi}}{\phi}
\end{equation}

Thus, if  $\phi$ is a decreasing function  then
$|\dot{c}/c|<|[n\dot{a}/a+2(n-1)\dot{b}/b]/(3n-2)|$ and we can
therefore conclude  that a varying speed of light in  the  early
universe (with weaker gravity) helps to solve  the flatness
problem.

Alternatively, for  general  equation  of  state
$p=(\gamma-1)\rho c^{2}$ , if  we make  the  transformation
\begin{equation}
x=\phi a^{\gamma}b^{2(\gamma-1)}
\end{equation}

and  assume  the  variation  of  the  velocity  of  light  as
$$
c=c_{0}x^{n/2}
$$

then  the  equation  of  continuity  (2.4)  becomes

\begin{equation}
\dot{\rho}+\gamma\rho\left(\frac{\dot{a}}{a}+2\frac{\dot{b}}{b}\right)=
\frac{nc_{0}^{2}\phi x^{n-1}\dot{x}}{8\pi b^{2}}
\end{equation}

Integrating this  equation  gives
\begin{equation}
\rho
a^{\gamma}b^{2\gamma}=B+\frac{nc_{0}^{2}\phi^{n+1}\left[a^{\gamma}
b^{2(\gamma-1)}\right]^{n+1}}{8\pi(n+1)}
\end{equation}

(provided $n\ne -1$) which gives  the  evolution  for  general
$\gamma$.\\

\section{Solutions  to  the  Quasi-Flatness  Problem}

\subsection{When BD Scalar  Field  $\phi$ is  Constant}

In  this  section  we  shall  investigate  whether  it  is
natural  to  have  evolution  which  asymptotes  to  a  state of
expansion  with  a  non-critical  density . Let  us  start with
the  continuity  equation  for  constant  $\phi$  and  assume as
before  the  equation  of  state to be
$$
\frac{p}{\rho c^{2}}=\gamma-1~,
$$

and  the  velocity  of  light to be
$$
c=c_{0}\left[a^{n}b^{2(n-1)}\right]^{n/(3n-2)}
$$

then  the  integral  gives
\begin{equation}
\rho(ab^{2})^{\gamma}=B+\frac{\gamma
c_{0}^{2}\phi}{4\pi(5\gamma-2)}\left[a^{\gamma}b^{2(\gamma-1)}\right]^{(5\gamma-2)/(3\gamma-2)}
\end{equation}

(It  is  to  be  noted  that  in  order  to  obtain  the
integral  one  has  to  assume $n=\gamma$).

Now  substituting  this  value  of  $\rho$ in  equation (2.2) we
have

\begin{eqnarray*}
\left(\frac{\dot{b}}{b}\right)^{2}+2\frac{\dot{a}}{a}\frac{\dot{b}}{b}
=\frac{8\pi}{\phi}-\frac{B}{(ab^{2})^{\gamma}} +\frac{\gamma
c_{0}^{2}\phi}{4\pi(5\gamma-2)}+\frac{2\left[a^{\gamma}b^{2(\gamma-1)}
\right]^{(5\gamma-2)/(3\gamma-2)}}{(ab^{2})^{\gamma}}
\end{eqnarray*}
\vspace{-9mm}

\begin{equation}
-\frac{1}{b^{2}}c_{0}^{2}\left[a^{\gamma}b^{2(\gamma-1)}
\right]^{2\gamma/(3\gamma-2)}
\end{equation}

In  particular, for  radiation  era  the  above  equation
simplifies  to

\begin{equation}
\left(\frac{\dot{b}}{b}\right)^{2}+2\frac{\dot{a}}{a}\frac{\dot{b}}{b}
=\frac{8\pi B}{\phi
(ab^{2})^{4/3}}-\frac{3}{7}c_{0}^{2}\left(a^{8}b^{-5}\right)^{2/9}
\end{equation}

If $\Omega$ is  the  density  parameter, then  it  can be  defined
as
\begin{equation}
\frac{\Omega}{\Omega-1}=\frac{8\pi G\rho}{c^{2}/b^{2}}
\end{equation}

We  note  that  the  ratio  between  the  two  terms  on  the
right  hand  side  is  almost  a  constant  for  a  quasi-flat
open  universe. However, for $\gamma=4/3$, we  get

\begin{equation}
\frac{\Omega}{\Omega-1}=\frac{4}{7}+\frac{8\pi G
B}{c_{0}^{2}(a^{2}b)^{14/9}}
\end{equation}

For  expanding  universe  $a, b$  grows  with  time. So
asymptotically  the  second  term  on  the  R.H.S  will  be
negligible  and  we  have
$$
\Omega=-\frac{4}{3}
$$

which  leads  to  an  open  universe  with  finite  $\Omega$
value today.\\

\subsection{When  BD  Scalar  Field  Varies}

For  the  radiation era ($\gamma=4/3$), let  us  define
\begin{equation}
y=\phi (ab^{2})^{2/3},~~~z=\phi(a^{2}b)^{2/3},~~~c=c_{0}z^{m}
\end{equation}

So  from  the  field  equation  (2.2)  we  get  (using  equation
(2.5))

\begin{equation}
2\frac{y'}{y}\frac{z'}{z}-\frac{z'^{2}}{z^{2}}=\frac{32\pi\rho}{3y}(ab^{2})^{4/3}-
\frac{4c^{2}}{3}\frac{z}{y}+\left(1+\frac{2\omega}{3}\right)\frac{\alpha^{2}}{y^{2}}
\end{equation}

Now, from  equation  (2.38) the expression of $\rho$ takes  the
form

\begin{equation}
\rho(ab^{2})^{4/3}=\frac{Bmc_{0}^{2}z^{2m+1}}{4\pi(2m+1)}
\end{equation}

In addition  the  critical  density $\rho_{c}$ is  obtained  from
the differential  equation

\begin{equation}
2\frac{y'}{y}\frac{z'}{z}-\frac{z'^{2}}{z^{2}}=\frac{32\pi\rho_{c}}{3y}(ab^{2})^{4/3}
+\left(1+\frac{2\omega}{3}\right)\frac{\alpha^{2}}{y^{2}}
\end{equation}

Hence  the  expression  for  the  density  parameter  is

$$
\Omega=\frac{\rho}{\rho_{c}}=\frac{2\frac{y'}{y}\frac{z'}{z}+
\frac{4c^{2}}{3}\frac{z}{y}-\left(1+\frac{2\omega}{3}\right)\frac{\alpha^{2}}{y^{2}}}
{2\frac{y'}{y}\frac{z'}{z}-\frac{z'^{2}}{z^{2}}-\left(1+\frac{2\omega}{3}\right)\frac{\alpha^{2}}{y^{2}}}
$$

or  equivalently,
\begin{equation}
\frac{\Omega}{\Omega-1}=\frac{\rho}{\rho-\rho_{c}}=\frac{8\pi
B}{c_{0}^{2}z^{2m+1}}+\frac{2m}{2m+1}
\end{equation}

Thus, if $2m+1<0$  then  as $t\rightarrow\infty,
z\rightarrow\infty$ so  we  have $\Omega\rightarrow 1$. But  if
$2m+1>0$ then as $t\rightarrow\infty, \Omega\rightarrow -2m$. So
asymptotically, we have  a quasi-flat open universe.\\

\subsection{General  Asymptotic  Behaviour}
In  this  section, we  consider  general  asymptotic  behaviour
when $2m+1>0$ i.e., for  quasi-flat  open  universe. The equation
(2.44) can be approximated  by
$$
2y'z'-\frac{y z'^{2}}{z}\simeq \Gamma~ z^{2(m+1)}
$$

which  has  a  first  integral
$$
\frac{y}{\sqrt{z}}=\int\frac{\Gamma~
z^{2(m+1)}}{2z'\sqrt{z}}~d\tau
$$

Thus, if we assume $z\sim \tau^{2/\Omega_{\infty}}$, then  $y\sim
\tau^{2/\Omega_{\infty}}$ and

\begin{equation}
\phi\sim\phi_{0}~exp\left(\frac{\tau^{1-2/\Omega_{\infty}}}{1-
\tau^{2/\Omega_{\infty}}}\right)
\end{equation}

For the case of $\Omega_{\infty}<1$, we have
$\phi\rightarrow\phi_{0},~ a(\tau)\sim \tau^{1/\Omega_{\infty}}$
and $b(\tau)\sim \tau^{1/\Omega_{\infty}}$ in the limit as
$\tau\rightarrow\infty$, or equivalently,
$$
\phi\rightarrow\phi_{0},~~~~a\sim
\tau^{1/(\Omega_{\infty}+1)},~~~~b\sim
\tau^{1/(\Omega_{\infty}+1)},~~~~~ as~ t\rightarrow\infty.
$$

In addition, when $\Omega_{\infty}=1$, we get $a\sim t^{1/2},~
b\sim t^{1/2}$, as expected for flat radiation  asymptote.\\

\section{The  Lambda  and  the  Quasi-Lambda  Problems}
In  this  section, we  shall  examine  the  effect  of variation
of  speed  of  light  when  a  cosmological  term  is introduced
in  the  BD  field  equations. In  fact, the incorporation  of  a
cosmological  term  is  equivalent  to introduction of  a  vacuum
stress  which obeys  an  equation  of  state
$$
\rho_{\Lambda}=-\frac{p_{\Lambda}}{c^{2}}
$$
with
$$
\rho_{\Lambda}=\frac{\Lambda c^{2}}{8\pi G}
$$

Thus  equation  of  continuity  can  be  generalized  to

$$
(\dot{\rho}+\dot{\rho}_{\Lambda})+\left(\frac{\dot{a}}{a}+2\frac{\dot{b}}{b}\right)
=\frac{c\dot{c}}{4\pi G b^{2}}
$$

Also  the  field  equation  (2.2)  can  now  be  written  as

\begin{equation}
\frac{\dot{b}^{2}}{b^{2}}+2\frac{\dot{a}}{a}\frac{\dot{b}}{b}=8\pi
G\rho-\frac{c^{2}}{b^{2}}+\Lambda c^{2}
\end{equation}

Now, using  the  above  expression  for  vacuum  stress-energy
and using $\gamma-1=p/\rho c^{2}$ as  the  equation  of  state
for the matter  and assuming

$$
c=c_{0}\left[a^{n}b^{2(n-1)}\right]^{n/(3n-2)},
$$

the  above  equation  of  continuity  can  be  integrated
(assuming $n=\gamma$) to  give

\begin{equation}
\rho=\frac{B}{(ab^{2})^{\gamma}}+\frac{\gamma c_{0}^{2}}{8\pi
G(5\gamma-2)}\frac{\left[a^{\gamma}b^{2(\gamma-1)}\right]^{\frac{5\gamma-2}{3\gamma-2}}}
{(ab^{2})^{\gamma}}-\frac{\Lambda c_{0}^{2}(5\gamma-2)}{8\pi
G}\frac{1}{ab^{2}}\int(ab^{2})^{\gamma}d[a^{\gamma}b^{2(\gamma-1)}]^{\frac{2\gamma}{3\gamma-2}}
\end{equation}

On substituting  this  value of $\rho$ in  the  above  field
equation (2.49), we  have

\begin{eqnarray*}
\frac{\dot{b}^{2}}{b^{2}}+2\frac{\dot{a}}{a}\frac{\dot{b}}{b}=\frac{8\pi
G B}{(ab^{2})^{\gamma}}+\frac{2\gamma
c_{0}^{2}}{(5\gamma-2)}\frac{\left[a^{\gamma}b^{2(\gamma-1)}\right]^{(5\gamma-2)/(3\gamma-2)}}
{(ab^{2})^{\gamma}}+\left(\Lambda-\frac{1}{b^{2}}\right)c_{0}^{2}
\left[a^{\gamma}b^{2(\gamma-1)}\right]^{2\gamma/(3\gamma-2)}
\end{eqnarray*}
\vspace{-9mm}

\begin{equation}
-\Lambda
c_{0}^{2}(5\gamma-2)\frac{1}{(ab^{2})^{\gamma}}\int(ab^{2})^{\gamma}d[a^{\gamma}
b^{2(\gamma-1)}]^{2\gamma/(3\gamma-2)}\hspace{-1.5in}
\end{equation}

If  we  define  the  generalized  density  parameter as

\begin{equation}
\Omega_{1}=\Omega_{m}+\Omega_{\Lambda}=\frac{8\pi
G(\rho+\rho_{\Lambda})b^{2}}{c^{2}}
\end{equation}

then  using  the  density  parameter $\Omega$ from  equation
(2.41) we have

$$
\Omega_{1}=\frac{\Omega}{\Omega-1}=\frac{8\pi
G(\rho+\rho_{\Lambda})b^{2}}{c^{2}}
$$

If  we  now  substitute  the  value  of  $\rho$ from  equation
(2.50) we obtain
\begin{eqnarray*}
\frac{\Omega}{\Omega-1}=\frac{2\gamma}{5\gamma-2}\left[a^{\gamma}b^{2(\gamma-1)}
\right]^{(4\gamma-3)/(3\gamma-2)}+\frac{8\pi G
B}{c_{0}^{2}}\left[a^{\gamma}b^{2(\gamma-1)}
\right]^{-(4\gamma-2)/(3\gamma-2)}+\Lambda b^{2}
\end{eqnarray*}
\vspace{-9mm}

\begin{equation}
-\Lambda(3\gamma-2)\left[a^{\gamma}b^{2(\gamma-1)}
\right]^{-(4\gamma-2)/(3\gamma-2)}\int(ab^{2})^{\gamma}d[a^{\gamma}
b^{2(\gamma-1)}]^{2\gamma/(3\gamma-2)}
\end{equation}

when $\gamma>1$. The 2nd, 3rd  and  4th  terms  are negligible
compare to the first term, hence  we  have
\begin{equation}
\frac{\Omega}{\Omega-1}\sim\frac{2\gamma}{5\gamma-2}\left[a^{\gamma}b^{2(\gamma-1)}
\right]^{(4\gamma-3)/(3\gamma-2)}
\end{equation}

So, for large  $a, b$ (i.e., $a, b\rightarrow\infty$)
$$
\frac{\Omega}{\Omega-1}\rightarrow\infty~~~~i.e.,~~\Omega\rightarrow
1.
$$

Furthermore, if $\Lambda=0$ then

\begin{equation}
\frac{\Omega}{\Omega-1}=\frac{2\gamma}{5\gamma-2}\left[a^{\gamma}b^{2(\gamma-1)}
\right]^{(4\gamma-3)/(3\gamma-2)}+\frac{8\pi G
B}{c_{0}^{2}}\left[a^{\gamma}b^{2(\gamma-1)}
\right]^{-(4\gamma-2)/(3\gamma-2)}
\end{equation}

So for the radiation era  ($\gamma=4/3$)

\begin{equation}
\frac{\Omega}{\Omega-1}=\frac{8}{14}+\frac{8\pi G
B}{c_{0}^{2}}(a^{2}b)^{-14/9}.
\end{equation}

As $a, b\rightarrow\infty,~\Omega/(\Omega-1)\rightarrow 8/14$
i.e., $\Omega=-4/3$. This  asymptotic  behaviour  is  same  as
in  Sec 2.6.1 where  for quasi-flatness  problem  the  BD  scalar
field  is  assumed  to be  constant.

\section{Discussion}

In this chapter, an extensive analysis of the BD solutions for
anisotropic cosmological model has been done where there is a
variation of the velocity of light with time we have assumed the
velocity of light to be in power-law form in the sacle factors.
The flatness problem has been discussed in details. Here
perturbative, non-perturbative and exact solutions of flatness
problem have been obtained. The graphical representation for
non-perturbative solution in the radiation era has some
interesting feature. Quasi-flatness problem with general
asymptotic bahaviour has also been discussed and in most cases we
obtain an open universe. Finally, we have also studied the lambda
and quasi-lambda problem in VSL and the asymptotic behaviour is
very similar to that in the flatness problem.\\

\large \baselineskip .85cm
\chapter{The Cosmology  in  a Perfect or Causal  Viscous Fluid with
Varying Speed of Light} \label{chap3}\markright{\it
CHAPTER~\ref{chap3}. The Cosmology  in  a Perfect or Causal
Viscous Fluid with Varying Speed of Light}

\section{Prelude}
Scalar dissipative in cosmology may be treated via the
relativistic theory of bulk viscosity [Maartens, 1995; Zimdahl,
1996]. The causal and stable thermodynamics of Israel and Stewart
provide a satisfactory replacement of the unstable and non-causal
theories of Eckart and Landau and Lifshitz. If viscosity driven
inflation occurs (leaving aside for the moment various questions
about the hydrodynamical consistency of such models) then this
necessarily involves non-linear bulk viscous pressure [Maartens,
1995]. A non-linear generalization of the Israel-Stewart theory
is more satisfactory model of viscosity driven inflation. We find
exact solutions in a flat and non-flat FRW universe for simple
and consistent thermodynamic co-efficients and equation of state.
Provided the thermodynamic parameters satisfy certain conditions,
the model admits thermodynamically consistent
inflationary solutions.\\

There has been a recent trend of investigation where one
specifies a particular form of the expansion of the universe and
then seeks for the matter or any other field distribution that
would lead to such a temporal behaviour of the scale factor. The
inflationary universe scenario can solve some of outstanding
problems [Guth, 1981] of standard big-bang cosmology. Ellis and
Madsen [1991] have considered an FRW model with a minimally
coupled scalar field along with a potential and perfect fluid in
the form of radiation. For brief but comprehensive reviews we
refer to the papers by Maartens [1995], Hiscock and Lindblom
[1983] and Lindblom [1996]. In recent communication, Zimdahl
[1996] discussed how the truncated theory is indeed a good limit
to the full causal theory in certain physical situations. Mandez
[1996] presented a similar work where attempts had been made to
find $\phi$ and $V$ both functions of time $t$ and thus
$V=V(\phi)$ for an exponentially
expanding universe.\\

In this chapter, we examine the behaviour of $\phi$ and $V$ for
power-law inflation of scale factor $R(t)$ and velocity of light
$c(t)$ which varies with $t$ [Albretch et al, 1999; Barrow et al,
1999]. We generalize the investigation of Ellis and Madsen [1991]
and Banerjee [1998] to include a fluid and we do not ignore the
presence of the fluid in the calculation for the expression for
the scalar field $\phi$ and the potential $V$ in terms of time
$t$ and hence obtained $V=V(\phi)$ for a variety of scale factor
and velocity of light.\\

\section{The Model in Gravitational Field Equations}
The energy momentum tensor for a bulk viscous fluid and minimally
coupled scalar field $\phi$ are given by
\begin{equation}
T^{(F)}_{\mu \nu}=(\rho c^{2}+p+\pi)v_{\mu}v_{\nu}+(p+\pi)g_{\mu
\nu}
\end{equation}
and
\begin{equation}
T^{(\phi)}_{\mu \nu}=\phi_{,\mu}\phi_{,\nu}-g_{\mu
\nu}\left[\frac{1}{2}\phi_{,\alpha}\phi^{,\alpha} +V(\phi)\right]
\end{equation}

where $\rho$ is the energy density, $p$ the thermodynamic
pressure, $\pi$ the bulk viscous pressure, $V(\phi)$
 the scalar potential and $v_{\mu}$ the four velocity satisfying the condition $v^{\mu}=\delta^{\mu}_{0}$ and
 $v^{\mu}v_{\mu}=-1$.

Consider, the line-element in FRW space time,

\begin{equation}
ds^{2}=-c^{2}dt^{2}+R^{2}(t)\left[\frac{dr^{2}}{1-kr^{2}}+r^{2}(d\theta^{2}+sin^{2}\theta
d\phi^{2})\right]
\end{equation}

The gravitational field equations are given by

\begin{equation}
3\left(H^{2}+\frac{kc^{2}}{R^{2}}\right)=\frac{1}{2}\dot{\phi}^{2}+Vc^{2}+\rho
\end{equation}
and
\begin{equation}
2\dot{H}+3H^{2}+\frac{kc^{2}}{R^{2}}=-\frac{1}{2}\dot{\phi}^{2}+Vc^{2}-\frac{(p+\pi)}{c^{2}}
\end{equation}

where, Hubble constant $H=\frac{\dot{R}}{R}$ and $k$ is the
curvature parameter. (We have chosen the unit
 $8\pi G=1$ ).

The `conservation equation' for the fluid is

\begin{equation}
\dot{\rho}+3H\left(\rho+\frac{p+\pi}{c^{2}}
\right)=4\frac{\dot{c}}{c}~\rho
\end{equation}

where we have assumed the velocity of light $c$ is an arbitrary
function of time only.

The equation of state for the fluid distribution is
\begin{equation}
p=(\gamma-1)\rho c^{2}
\end{equation}

where $\gamma$ is the constant adiabatic index of the fluid .

The causal evolution equation for the bulk viscous pressure is
given by

\begin{equation}
\tau\dot{\pi}+\pi=-3\eta
H-\frac{\varepsilon}{2}\tau\pi\left(3H+\frac{\dot{\tau}}{\tau}-\frac{\dot{\eta}}
{\eta}-\frac{\dot{T}}{T}\right)
\end{equation}

where $\tau$ is the ralaxation time, $\eta$ is the co-efficient
of bulk viscosity and $T$ is the temparature of the fluid. For
$\tau=0$, (3.8) reduces to the non-causal equation $\pi=-3\eta
H$. In (3.8), $\varepsilon=0$ gives the truncated theory
(extended irreversible thermodynamics [EIT]), while
$\varepsilon=1$ gives the full Israel-Stewart-Hiscock theory. It
can be noted that in full Israel-Stewart-Hiscock theory the
derivations from equilibrium are small so we can assume $|\pi|\ll
p=(\gamma)\rho c^{2}$ [Zimdahl, 1996; Maartens et al, 1997].

Using equation of state, the field equations (3.4) and (3.5)
becomes

\begin{equation}
\dot{\phi}^{2}=-2\dot{H}+\frac{2kc^{2}}{R^{2}}-\gamma\rho-\frac{\pi}{c^{2}}
\end{equation}
and
\begin{equation}
V=\frac{1}{c^{2}}\left[\dot{H}+3H^{2}-\left(\frac{2-\gamma}{2}\right)\rho+\frac{\pi}{2c^{2}}\right]
+\frac{2k}{R^{2}}
\end{equation}

The power-law form of scale factor $R$ and the velocity of light
are assumed as
\begin{equation}
R=R_{0}t^{n} , c=c_{0}t^{m}
\end{equation}

where, $R_{0} , c_{0} , n$ are positive constants and $m$ is real
constant with the restriction $n>1$ (for accelerating and
expanding universe).

For power-law expansion, (3.9) and (3.10) reduces to

\begin{equation}
\dot{\phi}^{2}=\frac{2n}{t^{2}}+\frac{2kc_{0}^{2}}{R_{0}^{2}}t^{2(m-n)}-\gamma\rho-\frac{\pi}{c_{0}^{2}
t^{2m}}
\end{equation}
and
\begin{equation}
V=\frac{1}{c_{0}^{2}t^{2m}}\left[\frac{3n^{2}-n}{t^{2}}-\left(\frac{2-\gamma}{2}\right)\rho
+\frac{\pi}{2c_{0}^{2}t^{2m}}\right]+\frac{2k}{R_{0}^{2}t^{2n}}
\end{equation}
\\
Since $\dot{\phi}^{2}\ge 0$, this implies,
\begin{equation}
\frac{n}{t^{2}}+\frac{kc_{0}^{2}}{R_{0}^{2}}t^{2(m-n)}-\frac{\gamma\rho}{2}-\frac{\pi}{2c_{0}^{2}
t^{2m}}\ge 0
\end{equation}

for the model to be consistent.\\

\section{Solutions for Perfect Fluid}
For a perfect fluid, $\pi=0$, and the conservation equation (3.6)
gives
\begin{equation}
\rho=\rho_{0}t^{4m-3\gamma n}
\end{equation}

where, $\rho_{0}$ is an integration constant.

In this case, consistent condition (3.14) becomes

\begin{equation}
\frac{n}{t^{2}}+\frac{kc_{0}^{2}}{R_{0}^{2}}t^{2(m-n)}-\frac{\gamma\rho_{0}}{2}t^{4m-3\gamma
n}\ge 0
\end{equation}

For $k=0$, the model is valid for

\begin{center}
$t\le\left(\frac{2n}{\gamma \rho_{0}}\right)^{1/(4m-3\gamma
n+2)}$ , if $4m-3\gamma n+2>0$
\end{center}
OR
\begin{center}
$t\ge\left(\frac{2n}{\gamma \rho_{0}}\right)^{1/(4m-3\gamma
n+2)}$ , if $4m-3\gamma n+2<0$
\end{center}

and for $k=+1 , -1$ the model is also valid for restricted period
of time.

In absence of any fluid ( i.e., $\rho_{0}=0$) the condition (3.16)
satisfies for $k=0,+1$ for all time $t$ and
 for $k=-1$, condition (3.16) satisfies for

\begin{center}
$t\le\left(\frac{nR_{0}^{2}}{c_{0}^{2}}\right)^{1/2(m-n+1)}$ , if
$m>n-1$
\end{center}
OR
\begin{center}
$t\ge\left(\frac{nR_{0}^{2}}{c_{0}^{2}}\right)^{1/2(m-n+1)}$ , if
$m<n-1$
\end{center}

For perfect fluid, (3.12) reduces to

\begin{equation}
\dot{\phi}^{2}=\frac{2n}{t^{2}}+\frac{2kc_{0}^{2}}{R_{0}^{2}}t^{2(m-n)}-\gamma\rho_{0}t^{4m-3\gamma
n}
\end{equation}

Generally, it is very difficult to integrate this equation to
$\phi=\phi(t)$. So we are assuming two cases :

(i) $k=0$ : In this case (3.17) integrates to

\begin{equation}
\frac{(4m+2-3\gamma
n)}{\sqrt{2n}}(\phi-\phi_{0})=2\sqrt{1-\frac{\gamma\rho_{0}}{2n}t^{4m+2-3\gamma
n}}
+\text{log}\left|\frac{\sqrt{1-\frac{\gamma\rho_{0}}{2n}t^{4m+2-3\gamma
n}}-1}{\sqrt{1-\frac{\gamma\rho_{0}}{2n} t^{4m+2-3\gamma n}}+1}
\right|
\end{equation}

where $\phi_{0}$ is an integration constant.

The potential $V$ is given by

\begin{equation}
V=\frac{1}{c_{0}^{2}}\left[\frac{(3n^{2}-n}{t^{2m+2}}+\left(\frac{\gamma-2}{2}\right)\rho_{0}
t^{2m-3\gamma n}\right]
\end{equation}

For simple case, $n=1, m=1/4$ and for radiation $\gamma=4/3$, we
have

\begin{equation}
\phi-\phi_{0}=-2\sqrt{2}\sqrt{1-\frac{2\rho_{0}}{3t}}+\sqrt{2}~\text{log}\left|\frac{2\sqrt{2}
\sqrt{1-\frac{2\rho_{0}}{3t}}+1}{2\sqrt{2}\sqrt{1-\frac{2\rho_{0}}{3t}}-1}\right|
\end{equation}

and
\begin{equation}
V=\frac{2}{c_{0}^{2}t^{5/2}}\left(1-\frac{\rho_{0}}{6t}\right)
\end{equation}

(ii) $k\neq 0$ : We shall discuss for $m=n-1$. In this case (3.17)
yields to

\begin{equation}
[(4-3\gamma)n-2](\phi-\phi_{0})=\sqrt{2B}\left[2\sqrt{1-\frac{\gamma\rho_{0}}{2B}t^{(4-3\gamma)n-2}}
+\text{log}\left|\frac{\sqrt{1-\frac{\gamma\rho_{0}}{2B}t^{(4-3\gamma)n-2}-1}}{\sqrt{1-\frac{\gamma\rho_{0}}{2B}
t^{(4-3\gamma)n-2}+1}}\right|\right]
\end{equation}

and (3.13) gives
\begin{equation}
V=\frac{1}{c_{0}^{2}}[2B+3n(n-1)]\frac{1}{t^{2n}}+\left(\frac{\gamma-2}{2}\right)\rho_{0}
t^{2(n-1)-3\gamma n}
\end{equation}

where $B=n+\frac{kc_{0}^{2}}{R_{0}^{2}}$.

Similarly, we can choose, $2(m-n)=4m-3\gamma n$ OR $4m-3\gamma
n=-2$ and we have same form of $\phi$ and $V$
 as well as (3.22) and (3.23).\\

\section{Solutions for a Fluid with Bulk Viscosity}
A causal viscous fluid with bulk viscous stress is given by
equation (3.8) and we shall discuss the model for
$\varepsilon=1$. For the sake of simplicity, we take the
thermodynamic quantities to be power function of fluid density
$\rho$ i.e.,

\begin{equation}
\eta=\alpha\rho^{\mu},  T=\beta\rho^{r},
\tau=\frac{\eta}{\delta\rho}=\frac{\alpha}{\delta}\rho^{\mu-1}
\end{equation}

where $\alpha$, $\beta$, $\delta$, $\mu$, $r$ are positive
constants.

On differentiating equation (3.6) and using equation of state, we
have

\begin{equation}
\ddot{\rho}+3\dot{H}\left(\gamma\rho+\frac{\pi}{c^{2}}\right)+3H\left(\gamma\dot{\rho}
+\frac{\dot{\pi}}{c^{2}}-\frac{2\pi\dot{c}}{c^{3}}\right)=4\left(\frac{\dot{c}}{c}\dot{\rho}
+\frac{\ddot{c}}{c}\rho-\frac{\dot{c}^{2}}{c^{2}}\rho\right)
\end{equation}

Then equation (3.8) reduces to (using equations (3.6) and (3.25))

\begin{eqnarray*}
\ddot{\rho}+\left[1+3n+\frac{2m(\gamma-1)}{\gamma}\right]\frac{\dot{\rho}}{t}+\left(\frac{9n^{2}\gamma}{2}
-8m^{2}-\frac{6mn}{\gamma}\right)\frac{\rho}{t^{2}}+\frac{\delta}{\alpha}\rho^{1-\mu}\dot{\rho}
\end{eqnarray*}
\vspace{-9mm}

\begin{equation}
+(3n\gamma-4m)\frac{\delta}{\alpha}\frac{\rho^{2-\mu}}{t}-\left(\frac{2\gamma-1}{2\gamma}\right)
\frac{\dot{\rho^{2}}}{\rho}-\frac{9\delta
n^{2}}{c_{0}^{2}}\frac{\rho}{t^{2m+2}}=0
\end{equation}

where we have assumed the standard thermodynamic relation between
temparature and density for barotropic fluid,
\begin{equation}
T\sim\rho^{\frac{\gamma-1}{\gamma}}
\end{equation}

Now comparing $T$ between (3.24) and (3.27) we obtain
\begin{equation}
r=\frac{\gamma-1}{\gamma}
\end{equation}

To complete the solution for $\rho$, we need some cases:\\

{\it Case I} : $\mu\neq 1$ and $\mu\neq\frac{1}{2}-m$.

For $\mu\neq 1$, the solution of the equation (3.26) is of the
form
\begin{equation}
\rho=\rho_{0}t^{\frac{2m+1}{\mu-1}}
\end{equation}

with the condition
\begin{equation}
(2m+1)[6n\gamma+4m(\gamma-1)+1]+(\mu-1)(9n^{2}\gamma^{2}-16m^{2}\gamma-12mn)=0
\end{equation}

where the constant $\rho_{0}$ is given by

\begin{equation}
\rho_{0}=\left[\frac{c_{0}^{2}}{9\alpha
n^{2}}\left(\frac{2m+1}{\mu-1}+3n\gamma-4m\right)\right]
^{\frac{1}{\mu-1}}
\end{equation}

Condition (3.14) alongwith equations (3.6) and (3.29) yields

\begin{equation}
\frac{n}{t^{2}}+\frac{kc_{0}^{2}}{R_{0}^{2}}t^{2(m-n)}+\frac{\rho_{0}}{6n}\left(\frac{2m+1}{\mu-1}-4m\right)
t^{\frac{2m+1}{\mu-1}}\geq 0
\end{equation}

This condition is satisfied for all values of $k$, ($k=0,+1,-1$)
only for restricted period of time. For $k=0$, the model is
consistent for

\begin{equation}
t\leq\left[\frac{6n^{2}}{\rho_{0}[2(a-1)(\mu-1)-a]}\right]^{\frac{1}{a}}
\end{equation}

where $a=\frac{2m+1}{\mu-1}+2$.

Using equation (3.6), the equations (3.12) and (3.13) reduces to

\begin{equation}
\dot{\phi}^{2}=\frac{2n}{t^{2}}+\frac{2kc_{0}^{2}}{R_{0}^{2}}t^{2(m-n)}-\frac{\rho_{0}}{3n}\left(4m
-\frac{2m+1}{\mu-1}\right)t^{\frac{2m+1}{\mu-1}}
\end{equation}

and
\begin{equation}
V=\frac{1}{c_{0}^{2}t^{2m}}\left[\frac{3n^{2}-n}{t^{2}}+\frac{\rho_{0}}{6n}\left(4m-6n+\frac{2m+1}{\mu-1}
\right)t^{\frac{2m+1}{\mu-1}}\right]+\frac{2k}{R_{0}^{2}t^{2n}}
\end{equation}

Two cases are considered as follows:

(i) $k=0$ : Equation (3.34) can be integrated to yield,

\begin{equation}
\phi-\phi_{0}=\frac{\sqrt{2n}}{a}\left[2\sqrt{1-At^{a}}+\text{log}\left|\frac{\sqrt{1-At^{a}}-1}{\sqrt{1-At^{a}}+1}
\right|\right]
\end{equation}

where, $a=\frac{2m+1}{\mu-1}+2$,
$A=\frac{\rho_{0}}{6n^{2}}\left(4m-\frac{2m+1}{\mu-1}\right)$ and
 $\phi_{0}$ is an constant of integration.

Now (3.35) can be written as

\begin{equation}
V=\frac{1}{c_{0}^{2}t^{2m}}\left[\frac{3n^{2}-n}{t^{2}}+\frac{\rho_{0}}{6n}(4m-6n+a-2)t^{a-2}\right]
\end{equation}

It is difficult to write $t=t(\phi)$ from (3.36) and from (3.37)
we cannot write $V=V(\phi)$ analytically.

\begin{figure}
\includegraphics{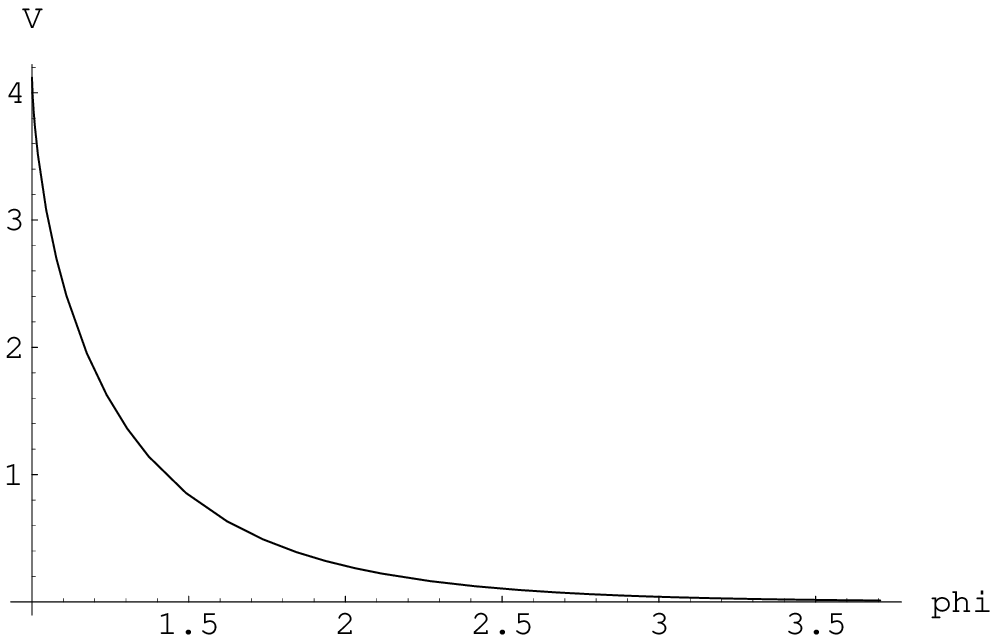}\\

Fig 3.1 : Here the potential function $V$ has been plotted against
$\phi$ from equations (3.36) and (3.37) for radiation era
($\gamma=4/3$)
with $\rho_{0}=1, c_{0}=1$, and $\phi_{0}=1$.\\
\end{figure}

(ii) $k\neq 0$ : In this case, we shall discuss the model for
$m=n-1$. After integration (3.34) we have

\begin{equation}
\phi-\phi_{0}=\frac{\sqrt{2B}}{a}\left[2\sqrt{1-\frac{nA}{B}t^{a}}+\text{log}\left|\frac{\sqrt{1-\frac{nA}{B}t^{a}}
-1}{\sqrt{1-\frac{nA}{B}t^{a}}+1}\right|\right]
\end{equation}

where $B=n+\frac{kc_{0}^{2}}{R_{0}^{2}}$,
$a=\frac{2n-1}{\mu-1}+2$ and $A=\frac{\rho_{0}}{6n^{2}}(4n-a-2)$.

Also (3.35) reduces to

\begin{equation}
V=\frac{1}{c_{0}^{2}}\left[2B+3n(n-1)-\left(nA+\frac{(4-n)\rho_{0}}{3n}\right)t^{a}\right]\frac{1}{t^{2n}}
\end{equation}

Similarly, if we choose $\frac{2m+1}{\mu-1}=2(m-n)$, then we have
same form of $\phi$ and $V$ as well as (3.38) and (3.39)
respectively.\\

\begin{figure}
\includegraphics{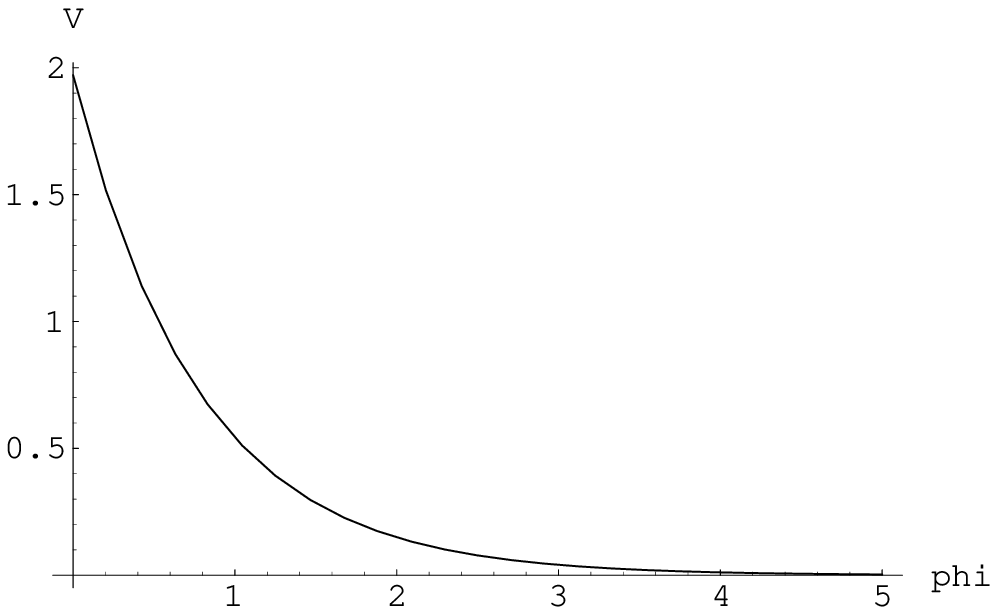}\\

Fig 3.2 : The potential function $V$ in equation (3.42) has been
plotted against $\phi$ for radiation era ($\gamma=4/3$) with $n=1,
m=-3/16, \rho_{0}=1, c_{0}=1$, and $\phi_{0}=1$.\\
\end{figure}

{\it Case II} : $\mu=\frac{1}{2}-m$.

(i) $k=0$ : Equation (3.34) integrates to yields
\begin{equation}
\phi-\phi_{0}=\sqrt{2n(1-A)}~ \text{log} t
\end{equation}

and equation (3.35) becomes
\begin{equation}
V=\frac{V_{0}}{t^{2(m+1)}}
\end{equation}

where
$V_{0}=\frac{1}{c_{0}^{2}}\left[3n^{2}-n+\frac{\rho_{0}}{3n}(2m-3n-1)\right]$
and $A=\frac{\rho_{0}}{3n^{2}}(2m+1)$.

From equations (3.40) and (3.41), we have the potential $V$ is an
exponential function of the scalar field $\phi$ i.e.,
\begin{equation}
V=V_{_{0}}e^{-\frac{2(m+1)(\phi-\phi_{0})}{\sqrt{2n(1-A)}}}
\end{equation}

(ii) $k\neq 0$ : For $k\neq 0$, two cases arises :

($a$) $m=n-1$ : Equation (3.34) integrates to

\begin{equation}
\phi-\phi_{0}=\sqrt{2n(1-A)+\frac{2kc_{0}^{2}}{R_{0}^{2}}}~
\text{log} t
\end{equation}

and equation (3.35) yields

\begin{equation}
V=\left(V_{0}+\frac{2k}{R_{0}^{2}}\right)\frac{1}{t^{2n}}
\end{equation}

Thus $V$ can be expressed as a function of $\phi$ in the form

\begin{equation}
V=\left(V_{0}+\frac{2k}{R_{0}^{2}}\right)\text{Exp}\left[-\frac{2n(\phi-\phi_{0})}{\sqrt{2n(1-A)+\frac{2kc_{0}^{2}}
{R_{0}^{2}}}}\right]
\end{equation}

($b$) $m\neq n-1$ : Equation (3.34) integrates to

\begin{equation}
\phi-\phi_{0}=\frac{\sqrt{2n(1-A)+\frac{2kc_{0}^{2}}{R_{0}^{2}}t^{2(m-n+1)}}}{2(m-n+1)}+\sqrt{2n(1-A)}~
\text{log} t
\end{equation}

and equation (3.35) gives

\begin{equation}
V=\frac{V_{0}}{t^{2(m+1)}}+\frac{2k}{R_{0}^{2}t^{2n}}
\end{equation}

In case ($b$), it is very difficult to eliminate $t$ from (3.46) and (3.47) to the form $V=V(\phi)$.\\

{\it Case III} : $\mu=1$.

In this case relaxation time becomes a constant.

(i) $m=-\frac{1}{2}$ : In this case, the solution of equation
(3.26) is of the form

\begin{equation}
\rho=\rho_{0}t^{-3n\gamma-\frac{9n^{2}\alpha}{c_{0}^{2}}}e^{-\frac{2\delta\gamma}{\alpha}t}
\end{equation}

with the condition :

\begin{equation}
\left(3n\gamma+\frac{9n^{2}\alpha}{c_{0}^{2}}\right)\left(\frac{9n^{2}\alpha}{c_{0}^{2}}+\gamma(2-3n)
-2\right)+9n^{2}\gamma^{2}+6n-4\gamma=0
\end{equation}

OR
\begin{equation}
\rho=\rho_{0}t^{-(2+3n\gamma)+\frac{9n^{2}\alpha}{c_{0}^{2}}}
\end{equation}

with the condition:
\begin{equation}
27n^{3}\alpha^{3}-6n\alpha
c_{0}^{2}(\gamma+1)+2c_{0}^{4}(\gamma^{2}-\gamma+1)=0
\end{equation}

The expression for $\rho$ in (3.48) yields a complicated
expression for $\dot{\phi}^{2}$ from equation (3.12) which cannot
be analytically integrated.

Now we shall discuss for the solution (3.50). Using (3.50) the
equations (3.12) and (3.13) reduces to

\begin{equation}
\dot{\phi}^{2}=\frac{2n}{t^{2}}+\frac{2kc_{0}^{2}}{R_{0}^{2}t^{2n+1}}+\rho_{0}\left(\frac{3n\alpha}{c_{0}^{2}}
-\gamma\right)t^{-(2+3n\gamma)+\frac{9n^{2}\alpha}{c_{0}^{2}}}
\end{equation}

and
\begin{equation}
V=\frac{1}{c_{0}^{2}}\left[\frac{3n^{2}-n}{t}-\rho_{0}\left(1+\frac{1}{2}\left(\frac{3n\alpha}{c_{0}^{2}}
-\gamma\right)\right)t^{-(2+3n\gamma)+\frac{9n^{2}\alpha}{c_{0}^{2}}}\right]+\frac{2k}{R_{0}^{2}t^{2n}}
\end{equation}

($a$) $k=0$ : Equation (3.52) integrates to the form

\begin{equation}
\phi-\phi_{0}=\frac{\sqrt{2n}}{x}\left[2\sqrt{1+wt^{x}}+\text{log}\left|\frac{\sqrt{1+wt^{x}}-1}{\sqrt{1+wt^{x}}+1}
\right|\right]
\end{equation}

where $x=3n\left(\frac{3n\alpha}{c_{0}^{2}}-\gamma\right)$ and
$w=\frac{\rho_{0}}{6n^{2}}x$.

and (3.53) yields

\begin{equation}
V=\frac{1}{c_{0}^{2}}\left[\frac{3n^{2}-n}{t}-(\rho_{0}+nw)t^{x-1}\right]
\end{equation}

It is very difficult to eliminate $t$ from (3.54) and (3.55) to
the form $V=V(\phi)$.\\
($b$) $k\neq 0$ : In this case, we shall discuss for
$3n\alpha=\gamma c_{0}^{2}$. Equation (3.52) integrates to

\begin{equation}
\phi-\phi_{0}=\frac{\sqrt{2n}}{1-2n}\left[2\sqrt{1+\frac{3k\alpha}{\gamma
R_{0}^{2}}t^{1-2n}}
+\text{log}\left|\frac{\sqrt{1+\frac{3k\alpha}{\gamma
R_{0}^{2}}t^{1-2n}}-1}{\sqrt{1+\frac{3k\alpha} {\gamma
R_{0}^{2}}t^{1-2n}}+1}\right|\right]
\end{equation}

and (3.53) reduces to
\begin{equation}
V=\frac{1}{c_{0}^{2}}(3n^{2}-n-\rho_{0})\frac{1}{t}+\frac{2k}{R_{0}^{2}t^{2n}}
\end{equation}

In this case also, we cannot express $V$ in the form $V=V(\phi)$.

(ii) $m=-1$ : The solution of equation (3.26) is of the form
\begin{equation}
\rho=\rho_{0}t^{y}e^{\frac{\gamma\delta}{\alpha}(1\pm z)t}
\end{equation}
with the condition:
\begin{equation}
y^{2}+2y[(3n-2)\gamma-2]+9n^{2}\gamma^{2}-16\gamma+12n=0
\end{equation}

where $z=\sqrt{1+\frac{18n^{2}\alpha}{\gamma\delta c_{0}^{2}}}$
and $\pm yz=(1\mp z)[2+\gamma(3n-1)]
-(3n\gamma+4)$.\\

The expression for $\rho$ yields a complicated expression for
$\dot{\phi}^{2}$ from equation (3.12) which cannot be
analytically integrated.

\section{Discussion}
In this chapter, we have considered perfect fluid distribution
with a scalar field $\phi$ having potential $V(\phi)$ for
isotropic space-time model, considering variation of the velocity
of light. We have considered the power-law inflation and assumed a
polynomial form (in time) for the velocity of light. In this
fluid distribution $V$ cannot be expressed as a function of
$\phi$ explicitely. For simple case ($n=1, m=1/4$) and radiation
dominated era ($\gamma=4/3$), the graphical representation of $V$
with $\phi$ shows that $V$ is monotonically decreasing function
of $\phi$ (Fig 3.1). Also it can be shown that $\phi$ increases
with $c$. In {\it case I} and {\it II} of viscous fluid
distribution we have obtained solutions for different values of
$k$ and arbitrary parameters. In one of the case, where
co-efficient of bulk viscosity $\eta\sim\rho^{\frac{1}{2}-m}$,
the potential function $V$ decreases exponentially with $\phi$
(Fig 3.2), but for other cases,  $V$ cannot be expressed as a
function of $\phi$ analytically. In {\it case III}, where
$\eta\sim\rho$, the relaxation time becomes a constant and
$\frac{\dot{c}}{c}<0$ i.e., velocity of light decreases. In this
case also $V$ cannot be expressed as a function of $\phi$
explicitely, but it can be shown that $V$ is monotonically
decreasing function of $\phi$. So for all cases of perfect and
viscous fluid distribution the potential function $V$ is
monotonically decreasing with $\phi$.\\

\large \baselineskip .85cm
\chapter{Quintessence Problem and Brans-Dicke Theory}
\label{chap4}\markright{\it CHAPTER~\ref{chap4}. Quintessence
Problem and Brans-Dicke Theory}

\section{Prelude}
From theoretical point  of  view, a lot  of  works [Ostriker er
al, 1995; Peebles, 1984; Wang et al, 2000; Caldwell et al, 1998]
has been done to solve this quintessence problem and possible
candidates for Q-matter are Cosmological Constant (or more
generally a variable cosmological term), a scalar field (which
was introduced by Ratra and Peebles [1988]) with  a potential
giving rise to a negative pressure at the present epoch, a
dissipative fluid with an effective negative stress [Cimento et
al,2000] and more exotic matter like  a frustrated network of
non-abelian cosmic strings or domain walls [Bucher et al, 1999;
Battye et al, 1999]. In these works, it  is generally assumed
that Q-matter behaves as a perfect  fluid with barotropic
equation  of state and  most of them  have considered spatially
flat model  of the universe (only the work  of Chimento  etal
[2000] has  been done for  open model of the universe). Recently,
Banerjee etal [2001] have shown that it is possible  to  have an
accelerated universe with BD-theory in  Friedmann  model  without
any matter.\\

In  this  chapter, the quintessence problem has been discussed in
anisotropic cosmological models for Brans-Dicke theory. We have
shown that it is possible to have an accelerated expanding
universe without any quintessence matter due to the Brans-Dicke
field. In order to solve the field equations we have assumed
power-law form of the scale factors. Here we have found that the
anisotropic scalar characterizes the accelerating or decelerating
nature of the universe. Finally flatness problem has been
discussed in this context using conformal transformation in
Jordan frame.\\

\section{Field  Equations  and  Solutions}
Consider  anisotropic  space-time  model described  by  the line
element
\begin{equation}
ds^{2}=-dt^{2}+a^{2}dx^{2}+b^{2}d\Omega_{k}^{2}
\end{equation}

where  $a$  and  $b$  are  functions  of  time  alone : we  note
that
\begin{eqnarray}d\Omega_{k}^{2}= \left\{\begin{array}{lll}
dy^{2}+dz^{2}, ~~~~~~~~~~~~ \text{when} ~~~k=0 ~~~~ (\text{Bianchi ~I ~model})\\
d\theta^{2}+sin^{2}\theta d\phi^{2}, ~~~~~ \text{when} ~~~k=+1~~
(\text{Kantowaski-Sachs~ model})\\
d\theta^{2}+sinh^{2}\theta d\phi^{2}, ~~~ \text{when} ~~~k=-1
~~(\text{Bianchi~ III~ model})\nonumber
\end{array}\right.
\end{eqnarray}

Here  $k$  is  the  curvature  index  of  the  corresponding
2-space, so  that  the  above  three  types  are  described  by
Thorne [1967]  as  Euclidean, open  and  semi  closed
respectively.

Now, in  BD  theory, assuming  a  perfect  fluid  distribution
as  only  matter  field, the  field  equations  for  the  above
space-time  symmetry  are

\begin{equation}
\frac{\ddot{a}}{a}+2\frac{\ddot{b}}{b}=-\frac{1}{(3+2\omega)\phi}\left[
(2+\omega)\rho_{_{f}}+3(1+\omega)p_{_{f}}\right]-\omega\left(\frac{\dot{\phi}}
{\phi}\right)^{2}-\frac{\ddot{\phi}}{\phi}
\end{equation}

\begin{equation}
\left(\frac{\dot{b}}{b}
\right)^{2}+2\frac{\dot{a}}{a}\frac{\dot{b}}{b}=\frac{\rho_{_{f}}}{\phi}-\frac{k}{b^{2}}-\left(\frac{\dot{a}}{a}+2\frac{\dot{b}}{b}
\right)\frac{\dot{\phi}}{\phi}+\frac{\omega}{2}\left(\frac{\dot{\phi}}{\phi}
\right)^{2}
\end{equation}

and  the  wave  equation  for  BD  scalar  field  is

\begin{equation}
\ddot{\phi}+\left(\frac{\dot{a}}{a}+2\frac{\dot{b}}{b}
\right)\dot{\phi}=\frac{1}{3+2\omega}\left(\rho_{_{f}}-3p_{_{f}}
\right)
\end{equation}

Here  $\rho_{_{f}}$  and  $p_{_{f}}$  are  the  density  and
hydrostatic pressure respectively  of  the  fluid  distribution,
obeying  the barotropic  equation  of  state
$$
p_{_{f}}=(\gamma_{_{f}}-1)\rho_{_{f}}
$$
($\gamma_{_{f}}$ being  the  constant  adiabatic  index  of  the
fluid causality  demands  $0\le \gamma_{_{f}}\le 2$) and  as
usual $\omega$ is the BD coupling parameter. Now  the  above
field equations, via the Bianchi identities  lead  to  the energy
conservation equation

\begin{equation}
\dot{\rho}_{_{f}}+\left(\frac{\dot{a}}{a}+2\frac{\dot{b}}{b}
\right)\left(\rho_{_{f}}+p_{_{f}}\right)=0
\end{equation}

At  present, the  universe  is  considered  as  matter dominated
(i.e., filled  with  cold  matter (dust) of negligible  pressure)
so considering $p_{_{f}}= 0$, we  have  from equation (4.5) (after
integration)
\begin{equation}
\rho_{_{f}}=\frac{\rho_{_{0}}}{V}
\end{equation}

where $V=ab^{2}$ is the volume index  at  the present instant and
$\rho_{_{0}}$ is  an  integration  constant. Also  using
$p_{_{f}}=0$  and
 equation (4.6), the  wave equation  has  a  first
integral

\begin{equation}
V\dot{\phi}=\frac{\rho_{_{0}}t}{3+2\omega}+c_{_{0}}
\end{equation}

We  now  assume  a  power-law  form  of  the  scale  factors
keeping  in  mind  that  we  must  have  an  accelerated
universe  to  match  the  recent  observations. So  we  take
\begin{equation}
a(t)=a_{_{0}}t^{\alpha},~~~b(t)=b_{_{0}}t^{\beta}
\end{equation}

and  consequently
\begin{equation}
V=V_{_{0}}t^{\alpha+2\beta},~~(V_{_{0}}=a_{_{0}}b_{_{0}}^{2})
\end{equation}

where  ($a_{_{0}},b_{_{0}}$) are  positive  constants  and
($\alpha,\beta$) are real constants. Thus  the  matter  density
and the BD-scalar field  takes  the  form
\begin{equation}
\rho_{_{f}}=\rho_{_{0}}t^{-(\alpha+2\beta)},~~~(\rho_{_{f}}=\rho_{_{0}}/V_{_{0}})
\end{equation}

and
\begin{equation}
\phi=\frac{\rho_{_{1}}t^{2-\alpha-2\beta}}{(3+2\omega)(2-\alpha-2\beta)}+
\frac{c_{_{0}}t^{1-\alpha-2\beta}}{V_{_{0}}(1-\alpha-2\beta)}
\end{equation}

The  field  equations  will  be  consistent  for  the  above
solutions (eqs.(4.8)-(4.11)) provided  we  have  the  following
restrictions  on  the  parameters

\begin{equation}
c_{_{0}}=0,~~\beta=1~~~\text{and}
\end{equation}
either
\begin{equation}
\alpha=1,~~\omega=-2\left(1+\frac{k}{3b_{_{0}}^{2}}\right)
\end{equation}
or
\begin{equation}
\omega=-2,~~\frac{k}{b_{_{0}}^{2}}=\alpha-1
\end{equation}

It  is  to  be  noted  that  both  the  cases  coincide  for
flat  model  of  the  universe  and  we  have  the  solution
\begin{equation}
a=a_{_{0}}t,~~b=b_{_{0}}t,~~\rho_{_{f}}=\rho_{_{1}}t^{-3},~~
\phi=\rho_{_{1}}t^{-1},~~2\omega+3=-1
\end{equation}

so  for $k\ne 0$, the  solution  in  first  case  is
\begin{equation}
a=a_{_{0}}t,~~b=b_{_{0}}t,~~\rho_{_{f}}=\rho_{_{1}}t^{-3},~~
\phi=-\frac{\rho_{_{1}}t^{-1}}{2\omega+3}
\end{equation}

and
$$
2\omega+3=-\left(1+\frac{4k}{3b_{_{0}}^{2}}\right),
$$

(for $k<0, b_{_{0}}^{2}>4/3$) while  for  the  second  case the
expression for the geometric  and  physical  parameters  are
\begin{equation}
a=a_{_{0}}t,~~b=b_{_{0}}t,~~\rho_{_{f}}=\rho_{_{1}}t^{-(\alpha+2)},~~
\phi=\frac{\rho_{_{1}}}{\alpha}~t^{-\alpha},~~2\omega+3=-1
\end{equation}

and  we  have  a  restriction
$$
\frac{k}{b_{_{0}}^{2}}=\alpha-1
$$

The  deceleration  parameter  has  the  expression
\begin{equation}
q=-\left(\frac{\alpha-1}{\alpha+2}\right),
\end{equation}

(Note  that $\alpha=1$ corresponds  to  the  first  two  case)

Thus  we  always  have  an  accelerated  model  of  the  universe
(in  the  third  case  i.e., eq.(4.17)) (except  for $-2<\alpha
\le 1 $) as predicted  by  recent  Supernova  observation. Hence
for the solutions  represented  by  equations  (4.15)  and  (4.16)
$q=0$ i.e., at  present  the  universe  is  in  a  state  of
uniform expansion  and  the  conclusion  is  identical  to  that
of Banerjee  et al [2001]. Furthermore, the  solution  corresponds
to equation  (4.17)  is  new  as  we  have  a  negative
deceleration parameter. This  solution  is  valid  for  closed
(or  open) model  of  the  universe  for  $\alpha>1$ (or
$\alpha<1$). Therefore, it  is possible  to  have  an
accelerated  universe today  by considering  anisotropy  model
of  the  universe  and hence the  quintessence  problem  may  be
solved  for  closed, open or  flat  type  of  model  of  the
universe.\\

\section{Conformal  Transformations : Flatness  Problem}
In  cosmology, the  technique  of  conformal transformation is
often  used  (for  mathematical simplification)  to transform a
non-minimally  coupled  scalar field  to  a minimally coupled
one  [Faraoni et al, 1999]. Usually, in  `Jordan Conformal frame'
the scalar field (also  BD  scalar  field) couples non-minimally
to the `Einstein  frame'  in  which  the transformed  scalar field
is minimally  coupled. In the  last section, we  have  developed
the  BD  theory  in Jordan  frame and  to  introduce  the
Einstein  frame  we make the  following transformations:\\
\begin{equation}
d\eta=\sqrt{\phi}~dt,~ \bar{a}=\sqrt{\phi}~a,~
\bar{b}=\sqrt{\phi}~b,~ \psi=ln~\phi,~
\bar{\rho}_{_{f}}=\phi^{-2}\rho_{_{f}},~
\bar{p}_{_{f}}=\phi^{-2}p_{_{f}}
\end{equation}

As  a  result  the  field  equations  (4.2)-(4.4) transformed  to

\begin{equation}
\frac{\bar{a}''}{\bar{a}}+2\frac{\bar{b}''}{\bar{b}}=-\frac{1}{2}\left(\bar{\rho}_{_{f}}+
3\bar{p}_{_{f}}\right)-\frac{(3+2\omega)}{2}\psi'^{2}
\end{equation}

\begin{equation}
\left(\frac{\bar{b}'}{\bar{b}}\right)^{2}+2\frac{\bar{a}'}{\bar{a}}\frac{\bar{b}'}{\bar{b}}+
\frac{k}{\bar{b}^{^{2}}}=\bar{\rho}_{_{f}}+\frac{(3+2\omega)}{4}\psi'^{2}
\end{equation}
and
\begin{equation}
\psi''+\left(\frac{\bar{a}'}{\bar{a}}+2\frac{\bar{b}'}{\bar{b}}\right)\psi'=\frac{1}
{3+2\omega}\left(\bar{\rho}_{_{f}}-3\bar{p}_{_{f}}\right)
\end{equation}

where~~  $' \equiv \frac{d}{d\eta}$.

These  equations  are  the  well  known  fields  equations  for
the  anisotropic  cosmological  models (described here)  with a
minimally  coupled  scalar  field  $\psi$ (massless). This scalar
field  behaves  like  a  `stiff'  perfect  fluid  with equation of
state\\
\begin{equation}
\bar{p}_{_{\psi}}=\bar{\rho}_{_{\psi}}=\frac{\psi'^{2}}{16\pi G}
\end{equation}

In  Einstein  frame, the  total  stress-energy  tensor  is
conserved, but  the  scalar  field  and  normal  matter  change
energy  according  to\\
\begin{equation}
\bar{\rho}_{_{f}}+\left(\frac{\bar{a}'}{\bar{a}}+2\frac{\bar{b}'}{\bar{b}}\right)
\left(\bar{\rho}_{_{f}}+\bar{p}_{_{f}}\right)=-\left[\bar{\rho}_{_{\psi}}+
\left(\frac{\bar{a}'}{\bar{a}}+2\frac{\bar{b}'}{\bar{b}}\right)\left(\bar{\rho}_{_{\psi}}+
\bar{p}_{_{\psi}}\right)\right]=-\frac{\psi'}{2}\left(\bar{\rho}_{_{f}}-
3\bar{p}_{_{f}}\right)
\end{equation}

Thus  combining  the  two  energy  densities  we  have  from the
above  equation
\begin{equation}
\bar{\rho}'+3\gamma H\bar{\rho}=0
\end{equation}

Here
$H=\frac{1}{3}\left(\frac{\bar{a}'}{\bar{a}}+2\frac{\bar{b}'}{\bar{b}}\right)$
is  the  Hubble  parameter  in  the Einstein frame and  $\gamma$
is the  average  barotropic  index  defined by the relation
\begin{equation}
\gamma\Omega=\gamma_{_{f}}\Omega_{_{f}}+\gamma_{_{\psi}}\Omega_{_{\psi}}
\end{equation}
where
\begin{equation}
\Omega=\Omega_{_{f}}+\Omega_{_{\psi}}=\frac{\bar{\rho}}{3H^{^{2}}}
\end{equation}

is  the density  parameter.

From  equations  (4.21)  and  (4.25)  after  some  algebra, we
have
\begin{equation}
\Omega'=\Omega(\Omega-1)[\gamma H_{a}+2(\gamma-1)H_{b}]
\end{equation}

where ~~ $H_{a}=\frac{\bar{a}'}{\bar{a}}$    and
$H_{b}=\frac{\bar{b}'}{\bar{b}}$ .

This evolution in $\Omega$ shows that  $\Omega=1$ is a possible
solution  of it and  for stability of  this solution, we  have
\begin{equation}
\gamma<\frac{2}{3},
\end{equation}
for  the  solutions  given  in  equations  (4.15)  and  (4.16)
and the  restriction  is
\begin{equation}
\gamma<\frac{2}{\alpha+3},
\end{equation}
for  the  solution  (4.17).

Since  the  adiabatic  indices  do  not  change  due  to
conformal  transformation  so  we  take  $\gamma_{_{f}}=1$ (since
$p_{_{f}}=0$) and $\gamma_{_{\psi}}=2$. Hence  from  (4.26)  and
(4.27), we have

\begin{equation}
\gamma=\frac{\Omega_{_{f}}+2\Omega_{_{\psi}}}{\Omega_{_{f}}+\Omega_{_{\psi}}}
\end{equation}
Now  due  to  upper  limit  of  $\gamma$ (given above) we  must
have the inequalities
$$
\Omega_{_{f}}<4|\Omega_{_{\psi}}|
$$
$$
\text{OR}
$$
\begin{equation}
\Omega_{_{f}}<\frac{2(\alpha+1)}{\alpha}|\Omega_{_{\psi}}|
\end{equation}

according  as  $\gamma$ is  restricted  by  (4.29)  or  (4.30).
Also from the  field  equation  (4.21), the  curvature  parameter
$\Omega_{_{k}}=-k/\bar{b}^{2}$ vanishes for  the  solution
$\Omega=1$, provided  we are restricted to $\alpha=1$. Therefore,
depending  on the relative magnitude  of the energies  of matter
and  the BD-scalar  field (as  in  isotropic  case) it  is
possible  to  have  a stable  solution  corresponding  to
$\Omega=1$ and hence  the flatness  problem  can  be solved.\\

\section{Discussion}
In  this  chapter, we  have  considered  three  anisotropic
cosmological  models  namely, Bianchi III ($k<0$), axially
symmetric  Bianchi I ($k=0$)  and  Kantowski-Sachs ($k>0$)
space-time. We  have shown  that  the  anisotropic character is
responsible  for  getting  an  accelerated  model of  the
universe. In  fact, for  the  present  model  the  shear scalar
is given  by $\sigma^{2}=\frac{2}{3}(\alpha-1)^{2}$, so  the
deceleration parameter (eq. (4.18)) is  proportional  to
$\sqrt{\sigma}$ and it shows  how anisotropy characterizes  the
accelerating  or decelerating universe. In other  words, we can
say  that anisotropic nature  of the universe  has  an effect on
the quintessence problem. The problem  for negative coupling
constant $\omega$ is same  as in isotropic  case and  there is
problem in big-bang nucleosynthesis  scenario as  claimed  by
Banerjee etal [2001]. In  this  case the  modified version of
BD-theory (where  the  coupling parameter $\omega$ is a function
of the scalar  field) is  very similar  to  that for isotropic
case, so  we  have  not presented  in this chapter.

\large \baselineskip .85cm
\chapter{A  Quintessence  Problem  in  Self-interacting
Brans-Dicke  Theory } \label{chap5}\markright{\it
CHAPTER~\ref{chap5}. A  Quintessence  Problem  in Self-interacting
Brans-Dicke  Theory }

\section{Prelude}
The  quintessence  proposal faces  two  types  of problems
[Steinhardt et al, 1999]. One  of  these  problems (referred to as
fine tuning problem, eliminate classically by Ratra and Peebles
[1988]), is the smallness  of the  energy  density compared  to
other typical particle physics  scales. The other problem known
as the  cosmic coincidence  problem  is that although the missing
energy density  and  matter density decrease at different  rates
as the universe  expands, it appears that the initial condition
has to be  set  so precisely that the two densities become
comparable today. Quintessence  has been proposed  as that
missing  energy density  component that along with  the matter
and baryonic density  makes the density parameter  equal to  1.
A  special form  of quintessence  field called  the `tracker
field' has been proposed by Ratra and Peebles [1988] to  tackle
this problem [Also Zlatev et al, 1999]. Since in a variety of
inflationary  models scalar fields have been used in describing
the  transition from the quasi-exponential expansion of  the
early  universe to  a power law  expansion, it  is natural  to
try  to understand the present acceleration of  the universe
which has  an exponential behaviour  too, by constructing models
where the matter responsible  for  such behaviour  is also
represented by a scalar  field [Strarobinsky, 1998; Saini et al,
2000]. Inverse power law is the other potential [Peebles et al,
1998; Ferreira et al, 1987] that has been studied extensively for
quintessence models, particularly, for solving the cosmic
coincidence problem. Recently, Bertolami and Martins [2000]
obtained  an accelerated expansion  for  the universe in  a
modified Brans-Dicke (BD) theory  by  introducing a potential
which is a  function  of BD scalar  field  itself.\\

Similar to the previous chapter, we have studied quintessence
problem in this chapter for Brans-Dicke theory but with a scalar
field which is self-interacting. Assuming the power law form for
the scale factors and Brans-Dicke scalar field, we have studied
the accelerating or decelerating nature of the universe for
various values of the parameters involved. Though for the present
universe we have considered dust models only but it has been
shown in principle that perfect fluid with barotropic equation of
state can be considered to study the quintessence problem.
Lastly, in addition to the non-decelerating solution, flatness
problem has been solved potentially.\\

\section{Field  Equations  and  Solutions}
The  Brans-Dicke  theory  is  given  by  the  action [Brans and
Dicke, 1961]\\
\begin{equation}
S=\int\sqrt{-g}\left[\phi
R-\frac{\omega}{\phi}~\phi_{,\alpha}\phi^{,\alpha}+{\cal
L_{m}}\right]d^{4}x
\end{equation}

where  $\phi$  is  the  BD  scalar  field, $\omega$ is  the
dimensionless constant  BD  parameter  and  ${\cal L_{m}}$  is
the Lagrangian for all other  matter  fields. We  have  chosen
the units   $8\pi G= c =1$.

The  matter  content  of  the  universe  is  composed  of
perfect  fluid  and  a  scalar  field  $\psi$ as  the
quintessence matter. We  assume that  the  universe  is
homogeneous  and  we consider  an  anisotropic  space-time  with
line-element which is given in eq.(4.1).

Now the  BD-field  equations are

\begin{equation}
2\frac{\ddot{b}}{b}+\left(\frac{\dot{b}}{b}
\right)^{2}+\frac{k}{b^{2}}=-\frac{(p_{_{m}}+p_{_{\psi}})}{\phi}-\frac{1}{2}\omega
\left(\frac{\dot{\phi}}{\phi}\right)^{2}-2\frac{\dot{b}}{b}\frac{\dot{\phi}}{\phi}
-\frac{\ddot{\phi}}{\phi}
\end{equation}

\begin{equation}
\frac{\ddot{a}}{a}+\frac{\ddot{b}}{b}+\frac{\dot{a}}{a}\frac{\dot{b}}{b}
=-\frac{(p_{_{m}}+p_{_{\psi}})}{\phi}-\frac{1}{2}\omega
\left(\frac{\dot{\phi}}{\phi}\right)^{2}-\left(\frac{\dot{a}}{a}+\frac{\dot{b}}{b}\right)
\frac{\dot{\phi}}{\phi}-\frac{\ddot{\phi}}{\phi}
\end{equation}

\begin{equation}
\left(\frac{\dot{b}}{b}
\right)^{2}+2\frac{\dot{a}}{a}\frac{\dot{b}}{b}+\frac{k}{b^{2}}=\frac{(\rho_{_{m}}+\rho_{_{\psi}})}{\phi}-\left(\frac{\dot{a}}{a}+2\frac{\dot{b}}{b}
\right)\frac{\dot{\phi}}{\phi}+\frac{\omega}{2}\left(\frac{\dot{\phi}}{\phi}
\right)^{2}
\end{equation}

and  the  wave  equation  for  the  BD  scalar  field  $\phi$  is

\begin{equation}
\ddot{\phi}+\left(\frac{\dot{a}}{a}+2\frac{\dot{b}}{b}
\right)\dot{\phi}=\frac{1}{3+2\omega}\left[(\rho_{_{m}}-3p_{_{m}})+(\rho_{_{\psi}}-3p_{_{\psi}})
\right]
\end{equation}

$\rho_{_{m}}$  and  $p_{_{m}}$  are  the  density  and  the
pressure of normal matter, $\rho_{_{\psi}}$ and  $p_{_{\psi}}$
are  those due to the quintessence field  given  by

\begin{equation}
\rho_{_{\psi}}=\frac{1}{2}\dot{\psi}^{2}+V(\psi),~~p_{_{\psi}}=\frac{1}{2}\dot{\psi}^{2}-V(\psi)
\end{equation}

where  $V(\psi)$  is  the  relevant  potential.

The wave equation  for  the  quintessence  scalar  field $\psi$ is

\begin{equation}
\ddot{\psi}+\left(\frac{\dot{a}}{a}+2\frac{\dot{b}}{b}
\right)\dot{\psi}=-\frac{dV(\psi)}{d\psi}
\end{equation}

From  the  above  field  equations, we  have  the  matter {\it
conservation}  equation

\begin{equation}
\dot{\rho}_{m}+\left(\frac{\dot{a}}{a}+2\frac{\dot{b}}{b}
\right)(\rho_{m}+p_{m})=0
\end{equation}

Assuming  that  at  the  present  epoch, the  universe  is
filled  with  cold  matter (dust)  with  negligible  pressure,
so  using  $p_{m}=0$, the  conservation  equation (5.8)  gives
\begin{equation}
\rho_{m}=\frac{\rho_{1}}{ab^{2}}
\end{equation}

where  $\rho_{1}$ is  an  integration  constant.

Now  we  assume, the  power  law  form  of  scale  factors $a, b$
and  the  BD  scalar  field  $\phi$  are\\
\begin{equation}
a=a_{1}t^{\alpha},~~ b=b_{1}t^{\beta},~~ \phi=\phi_{1}t^{\delta}
\end{equation}

where  $a_{1},~ b_{1},~ \phi_{1}$ are  positive constants and
$\alpha, \beta, \delta$ are real  constants  with
$\alpha+2\beta\ge 3$ (for accelerating universe).

From  the  field  equations (5.2), (5.3) and (5.4) using (5.10) we
have the  expression  for $\dot{\psi}^{2}$  as
\begin{equation}
\dot{\psi}^{2}=\frac{2k\phi_{1}}{b_{1}^{2}}t^{\delta-2\beta}-\frac{\rho_{1}}{a_{1}b_{1}^{2}}
t^{-\alpha-2\beta}+\phi_{1}[2\beta^{2}-2\alpha(\alpha-1)-\omega\delta^{2}-\delta(\delta-1)
-(\alpha-2\beta)\delta]t^{\delta-2}
\end{equation}

From (5.5) the potential $V$ is given  by

\begin{eqnarray*}
V=\frac{k\phi_{1}}{2b_{1}^{2}}t^{\delta-2\beta}-\frac{\rho_{1}}{2a_{1}b_{1}^{2}}
t^{-\alpha-2\beta}+\frac{1}{4}\phi_{1}[(2\omega+3)(\alpha+2\beta+\delta-1)\delta+
2\beta^{2}-2\alpha(\alpha-1)
\end{eqnarray*}
\vspace{-13mm}

\begin{equation}
-\omega\delta^{2}-\delta(\delta-1)-(\alpha-2\beta)\delta]
t^{\delta-2}  \hspace{-2in}
\end{equation}

The wave equation (5.7) for the quintessence scalar field $\psi$
can be written in the form

\begin{equation}
-\frac{dV}{dt}=\dot{\psi}\ddot{\psi}+\left(\frac{\dot{a}}{a}+
2\frac{\dot{b}}{b}\right)\dot{\psi}^{2}
\end{equation}

After integration (5.13) the expression for  $V$  is

\begin{eqnarray*}
V=-\frac{k\phi_{1}}{b_{1}^{2}}\frac{(2\alpha+2\beta+\delta)}{(\delta-2\beta)}~t^{\delta-2\beta}
-\frac{\rho_{1}}{2a_{1}b_{1}^{2}}~t^{-\alpha-2\beta}+\frac{\phi_{1}(2\alpha+4\beta+\delta-2)}
{2(\delta-2)}[2\alpha(\alpha-1)
\end{eqnarray*}
\vspace{-14mm}

\begin{equation}
+\delta(\delta-1)+\omega\delta^{2}+(\alpha-2\beta)\delta-
2\beta^{2}]~t^{\delta-2}\hspace{-2in}
\end{equation}

Now, the  consistency  relations  of  the  constants  coming
from  the  two  identical  equations (5.12)  and (5.14)  for  the
potential  V  are
\begin{equation}
4\alpha+2\beta+3\delta=0
\end{equation}
\begin{eqnarray*}
\delta(\delta-2)(\alpha+2\beta+\delta-1)(2\omega+3)=(4\alpha+8\beta+3\delta-6)
[2\alpha(\alpha-1)+\delta(\delta-1)
\end{eqnarray*}
\vspace{-14mm}

\begin{equation}
+\omega\delta^{2}+(\alpha-2\beta)\delta-2\beta^{2}] \hspace{-3in}
\end{equation}

From  these  relations  for  an  accelerating  universe
($\alpha+2\beta\ge 3$), there are  two  possibilities: one  in
which  $k = 0$ and  the second where  $k\ne 0$.\\

{\it Case I} : $ k = 0 $

In  this  case, consistency  conditions  are  (5.16)  and
\begin{equation}
\alpha+2\beta=c
\end{equation}

where  $c$  is  an  constant  ($\ge 3$).

For  solving  (5.16)  and  (5.17), we  may  choose  $\delta= - 2$
and we have  two  possible  solutions:

(i) ~~~~ $\alpha=\beta=\frac{c}{3}$,~~~  $\delta=-2$

 ~~~~~~~~~~~~~~~~~~~~~~~~~~~~~ or

(ii) ~~~  $\alpha=4-c,~~ \beta=-2+c,~~ \delta=-2$

For  both  solutions  (i)  and  (ii)  the  expression  for
$\dot{\psi}^{2}$  (eq.(5.11))  becomes

\begin{equation}
\dot{\psi}^{2}=-\frac{\rho_{1}}{a_{1}b_{1}^{2}}~t^{-c}-2(2\omega+3)\phi_{1}t^{-4}
\end{equation}

This  indicates  that  $\omega<-3/2$  as  $\dot{\psi^{2}}$
cannot  be negative.

For  $c = 4$, one  has $2|2\omega+3|\phi_{1}\ge
\frac{\rho_{1}}{a_{1}b_{1}^{2}}$ and the deceleration parameter
$q=-\frac{1}{4}$.

In  this  case, equation (5.18) integrates  to
\begin{equation}
\psi=\pm \frac{A}{t}
\end{equation}

where
$A^{2}=-2(2\omega+3)\phi_{1}-\frac{\rho_{1}}{a_{1}b_{1}^{2}}$~~
and  the  relation  between  $V$  and  $\psi$  becomes
\begin{equation}
V=V_{1}\psi^{4}
\end{equation}

where   $V_{1}$  being  a  constant, related  to  the  constants
$a_{1}, \rho_{1}$  etc. The  model  works  for  all  time
$0<t<\infty$ with the condition  $2|2\omega+3|\phi_{1}\ge
\frac{\rho_{1}}{a_{1}b_{1}^{2}}$  is satisfied. The  value  of
$\omega$ is related to the other constants as follows:

$$2(2\omega+3)\phi_{1}+\frac{\rho_{1}}{a_{1}b_{1}^{2}}=
\frac{-1\pm \sqrt{1+\frac{64}{9}V_{1}}}{4V_{1}},~~\text{for~
solution~ (i)}$$ and
$$2(2\omega+3)\phi_{1}=\frac{\rho_{1}}{a_{1}b_{1}^{2}}-\frac{\epsilon}{2V_{1}},
~~(\epsilon=0,1),~~\text{for ~solution~ (ii)}$$

For  $c \ne 4$, the  model  does  not  work  for  the  whole
range of  time $0<t<\infty$. In  this  case, the  deceleration
parameter  is  $q=\frac{3}{c}-1$.

If  $c>4$, then  the  rate  of  acceleration  is  faster than
$q=-\frac{1}{4}$  and  from  equation (5.18), $\dot{\psi}^{2}>0$
restricts the validity of the model  for  $t>t_{1}$  where

\begin{equation}
t_{1}=\left[\frac{\rho_{1}}{2|2\omega+
3|a_{1}b_{1}^{2}\phi_{1}}\right]^{\frac{1}{c-4}}
\end{equation}

Further  if  $3<c<4$, then  the  universe  will  expands with an
acceleration  but  with  a  rate  less  than  $q=-\frac{1}{4}$ and
as before from  equation (5.18)  for  real $\dot{\psi}$,  the
model works upto  the time  $t_{2}$  where
\begin{equation}
t_{2}=\left[\frac{\rho_{1}}{2|2\omega+
3|a_{1}b_{1}^{2}\phi_{1}}\right]^{\frac{1}{c-4}}
\end{equation}

For  $q=-\frac{1}{4}$, the  present  age  of  the  universe  can
be calculated  from  (5.4)  as

$$
t_{0}=\left[2-2\omega-\frac{A^{2}}{2\phi_{1}}+\frac{V_{1}A^{4}}{\phi_{1}}\right]^{1/2}
\frac{1}{\left[(H_{b}^{2})_{0}+2(H_{a})_{0}(H_{b})_{0}\right]^{1/2}}
$$

where  $H_{a}=\frac{\dot{a}}{a}$  and  $H_{b}=\frac{\dot{b}}{b}$.

For  large $\omega$ limit,
$$
t_{0}\cong
\frac{\sqrt{-2\omega}}{\left[(H_{b}^{2})_{0}+2(H_{a})_{0}(H_{b})_{0}\right]^{1/2}}
$$

where $\omega$  is  obviously  a  negative  quantity.\\

{\it Case II :}  $k\ne 0$.

In  this  case, consistency  conditions  are  (5.15), (5.16)  and
\begin{equation}
\alpha+2\beta=3
\end{equation}

Solving  these  three  equations  we  have  the  following
solutions:

(i)~~~ $\alpha=1,~ \beta=1,~ \delta=-2$ ~~ for  all  values  of
$\omega$.

(ii)~~ $\alpha=1\mp\sqrt{\frac{2\omega+3}{2\omega-3}},~
\beta=1\pm\sqrt{\frac{2\omega+3}{2\omega-3}},~
\delta=-2\pm\sqrt{\frac{2\omega+3}{2\omega-3}} $

provided  for $\omega>3/2$ or $\omega\le -3/2$.

For  the  solution (i),  the  model  works  for  a  limited
period  of  time, $0<t<t_{1}$  where

\begin{equation}
t_{1}=2\phi_{1}\left[\frac{k}{b_{1}^{2}}-(2\omega+3)\right]\frac{a_{1}b_{1}^{2}}{\rho_{1}}
\end{equation}

For  an  open  universe  i.e., for  $k = -1, (2\omega+3)<0 $  and
$|2\omega+3|>\frac{1}{b_{1}^{2}}$.

For a closed  universe  i.e., for  $k = +1 ,(2\omega+3)>0 $  and
$|2\omega+3|<\frac{1}{b_{1}^{2}}$.

For the solution (ii) the model  works  for  the  time, where $t$
satisfies the equation\\
\begin{equation}
2(2\omega+3)\phi_{1}\left[\left(1\mp\sqrt{\frac{2\omega+3}{2\omega-3}}\right)^{2}-
\frac{2}{2\omega-3}\right]~t^{\pm
2\sqrt{\frac{2\omega+3}{2\omega-3}}}+\frac{\rho_{1}}{a_{1}b_{1}^{2}}~t
\le \frac{2k\phi_{1}}{b_{1}^{2}}
\end{equation}

provided  for  $\omega > 3/2$ or  $\omega\le -3/2 $.\\

\section{Flatness  Problems and its Solutions}
One important  aspect  of  this model  is  that potentially it
can solve the flatness  problem. Now we make  a  conformal
transformation [Faraoni et al, 1999] as
\begin{equation}
\bar{g}_{\mu\nu}=\phi~g_{\mu\nu}
\end{equation}

In  this  section, we  make the  following  transformations:
\begin{eqnarray*}
d\eta=\sqrt{\phi}~dt,~~ \bar{a}=\sqrt{\phi}~a,~~
\bar{b}=\sqrt{\phi}~b,~~ \psi=\text{ln}~\phi,~~
\bar{\rho}_{_{m}}=\phi^{-2}\rho_{_{m}},~~
\bar{\rho}_{_{\psi}}=\phi^{-2}\rho_{_{\psi}},
\end{eqnarray*}
\vspace{-14mm}

\begin{equation}
\bar{p}_{_{m}}=\phi^{-2}p_{_{m}},~~\bar{p}_{_{\psi}}=\phi^{-2}p_{_{\psi}}
\hspace{-2in}
\end{equation}

So  the  field  equations  (5.2) - (5.4)  transformed  to (after
some manipulations)

\begin{equation}
\left(\frac{\bar{b}'}{\bar{b}}\right)^{2}-2\frac{\bar{a}''}{\bar{a}}-2\frac{\bar{a}'}{\bar{a}}
\frac{\bar{b}'}{\bar{b}}+\frac{k}{\bar{b}^{2}}=(\bar{p}_{_{m}}+\bar{p}_{_{\psi}})+
\frac{(3+2\omega)}{4}\left(\frac{\phi'}{\phi}\right)^{2}
\end{equation}

\begin{equation}
\left(\frac{\bar{b}'}{\bar{b}}\right)^{2}+2\frac{\bar{a}'}{\bar{a}}\frac{\bar{b}'}{\bar{b}}+
\frac{k}{\bar{b}^{^{2}}}=(\bar{\rho}_{_{m}}+\bar{\rho}_{_{\psi}})+
\frac{(3+2\omega)}{4}\left(\frac{\phi'}{\phi}\right)^{2}
\end{equation}
and
\begin{equation}
\frac{\bar{b}''}{\bar{b}}-\frac{\bar{a}''}{\bar{a}}+\frac{k}{\bar{b}^{2}}=
\frac{\bar{a}'}{\bar{a}}\frac{\bar{b}'}{\bar{b}}-\left(\frac{\bar{b}'}{\bar{b}}\right)^{2}
\end{equation}

where~~  $' \equiv \frac{d}{d\eta}$.

The BD scalar field in new version $\bar{\rho}_{_{\phi}}$ is given
by\\
\begin{equation}
\bar{\rho}_{_{\phi}}=\frac{(3+2\omega)}{4}\left(\frac{\phi'}{\phi}\right)^{2}=\bar{p}_{_{\phi}}
\end{equation}

We define the dimensionless density parameter $\bar{\Omega}$ as
\begin{equation}
\bar{\Omega}=\bar{\Omega}_{_{m}}+\bar{\Omega}_{_{\phi}}+\bar{\Omega}_{_{\psi}}=
\frac{\bar{\rho}}{3\bar{H}^{2}}
\end{equation}

where
$\bar{\rho}=\bar{\rho}_{_{m}}+\bar{\rho}_{_{\phi}}+\bar{\rho}_{_{\psi}}$
is the total density and $\bar{\Omega}_{i}$ are defined
accordingly.

Using (5.28)-(5.31), and combining the energy densities, we have
the equation for the conservation for the total energy,
\begin{equation}
\bar{\rho}'+3\bar{H}(\bar{\rho}+\bar{p})=0
\end{equation}

Here
$\bar{H}=\frac{1}{3}\left(\frac{\bar{a}'}{\bar{a}}+2\frac{\bar{b}'}{\bar{b}}\right)$
is  the  Hubble  parameter  in  the Einstein frame and  $\gamma$
is the  net  barotropic  index  defined  as
\begin{equation}
\gamma~\bar{\Omega}=\gamma_{_{m}}~\bar{\Omega}_{_{m}}+\gamma_{_{\phi}}~\bar{\Omega}_{_{\phi}}
+\gamma_{_{\psi}}~\bar{\Omega}_{_{\psi}}
\end{equation}

From  equations (5.29)  and  (5.33), we  have  the  evolution
equation for  the  density  parameter  as
\begin{equation}
\bar{\Omega}'=\bar{\Omega}(\bar{\Omega}-1)[\gamma\bar{H}_{a}+2(\gamma-1)\bar{H}_{b}]
\end{equation}

where ~~ $\bar{H}_{a}=\frac{\bar{a}'}{\bar{a}}$    and
$\bar{H}_{b}=\frac{\bar{b}'}{\bar{b}}$ .

The individual $\gamma_{i}$~'s are defined by the relation
$p_{i}=(\gamma_{i}-1)\rho_{i}$. So the ratios
$\frac{p_{i}}{\rho_{i}}$ remain same in both frames. For our
choices of matter, $\bar{p}_{_{m}}=0$ and
$\bar{p}_{_{\phi}}=\bar{\rho}_{_{\phi}}$, so we have
$\gamma_{_{m}}=1$ and $\gamma_{_{\phi}}=2$. The other index
$\gamma_{_{\psi}}$ is related by the equation

\begin{equation}
\gamma_{_{\psi}}=\frac{p_{_{\psi}}+\rho_{_{\psi}}}{\rho_{_{\psi}}}=\frac{\dot{\psi}^{2}}
{\frac{1}{2}\dot{\psi}^{2}+V}
\end{equation}

It has been shown that $\gamma_{_{\psi}}$ is varies with time.
The equation (5.35) indicates that $\bar{\Omega}=1$ is a possible
solution and this solution determines that
$\left(\frac{\partial\dot{\bar{\Omega}}}{\partial\bar{\Omega}}\right)_{\bar{H}}<0$.
This solution is stable for expanding universe $(\bar{H}>0)$ if
$\gamma<\frac{2}{3}$ with the relevant condition\\
\begin{equation}
\bar{\Omega}_{_{m}}+4\bar{\Omega}_{_{\phi}}<(2-3\gamma_{_{\psi}})\bar{\Omega}_{_{\psi}}
\end{equation}

From  the  field equation  (5.29), the  curvature  parameter
$\bar{\Omega}_{_{k}}=-k /~\bar{b}^{2}$ vanishes  for  the
solution $\bar{\Omega}=1$. So for BD-scalar field it  is possible
to have a stable  solution corresponding  to $\bar{\Omega}=1$ and
hence  the flatness problem  can  be solved.\\

\section{Discussion}
We  have  investigated  the  nature  of  the potential  relevant
to  the  power  law expansion  of  the universe  in  a
self-interacting  Brans-Dicke (BD) cosmology with  a  perfect
fluid  distribution  for  anisotropic cosmological  models. We
have  considered  a non-gravitational quintessence  scalar field
$\psi$  with  a potential $V = V(\psi)$. This scalar field  in
BD-theory  is shown  to give rise  to an accelerated expansion
for  the present  dust universe (where we have taken $p_{m}=0$).
It  is  to  be noted that at early stages of  the evolution  of
the universe $p_{m}$ is non-zero. But if  we take barotropic
equation of  state $p_{m}=(\gamma-1)\rho_{m}$, then equation (5.9)
is modified  to $\rho_{m}=\frac{\rho_{1}}{(ab^{2})^{\gamma}}$.

We  have  presented  a  class  of  solutions  describing
non-decelerating  universe  for  both  flat ($k=0$) and  curved
space-time ($k\ne 0$). For  $k=0$, there  are  two  possible
solutions for  different choice  of  the  parameters. In  both
the  solutions, the  parameter $\omega$ must  be  negative (in
fact $2\omega+3$ is  negative) to  make  the  quintessence  field
real. The validity (on  the  time  scale) of  the  solutions
depends on the  parameter  $c$  (defined  in  eq.(5.17)). For
$c=4$, the model  works  for  all  time $0<t<\infty$, while  for
$c>4$, the model works  for ($t_{1}, \infty$) where $t_{1}$ is
given by (5.21). From this model  we  cannot  predict  the
geometry of the universe before $t_{1}$. Similarly  for  $3<c<4$
the model is valid for  $t<t_{2}$  i.e., from  the  begining  to
the time $t_{2}$. Here also we cannot  predict  the  nature  of
space-time of  the universe after $t=t_{2}$.

In  curved  space-time ($k\ne 0$)  there  are  also  two  possible
solutions  for  different  choice  of  the  parameters.  The
time  interval  over  which  the  solutions  are  valid  are
given  in  equations (5.24)  and  (5.25). For  close  model
($k>0$) the  coupling  parameter  $\omega$  is  restricted  by
the inequality $0<2\omega+3<1/b_{1}^{2}$ while   for   open  model
$\omega$ satisfies $-1/b_{1}^{2}<2\omega+3<0$. From   the   above
solutions, we note that the coupling parameter  $\omega$  may  be
positive  or negative (with some restrictions). Hence  for  all
solutions the choice of $\omega$ is not in  agreement  with  the
observations.

Finally, for  non-decelerating  solution  it  can  also
potentially  solve  the  flatness  problem (without  any
restriction  on  the  parameters)  and  it  has  been  shown that
$\bar{\Omega}=1$ could  be  a  stable  solution  in  this  model.\\

\large \baselineskip .85cm
\chapter{A  Quintessence  Problem  in  Brans-Dicke Theory  with
Varying  Speed  of  Light } \label{chap6}\markright{\it
CHAPTER~\ref{chap6}. A  Quintessence  Problem  in  Brans-Dicke
Theory  with Varying  Speed  of  Light }

\section{Prelude}
In this chapter, quintessence problem has been studied as in the
last two chapters for the Brans-Dicke theory but considering time
variation of the speed of light. Here also without any
quintessence matter, an accelerated form of the universe is
possible. For power law expansion of the universe, we have shown
that the time variation of the velocity of light is also in the
power law form. Further assuming exponential form for the scale
factors and the Brans-Dicke scalar field, the velocity of light
is also in the exponential form and the deceleration is negative
definite without any dependence of the parameters involved.
Similar to the previous chapters the flatness problem can be
solved automatically.\\

\section{Field  Equations  and  Solutions}
We  consider  the  line-element  of  anisotropic  space-time
model\\
\begin{equation}
ds^{2}=-c^{2}dt^{2}+a^{2}dx^{2}+b^{2}d\Omega_{k}^{2}
\end{equation}

where  $a , b$ are  functions  of  time  only  and
$d\Omega_{k}^{2}$ is given in chapter 4.

Now, the  BD-field  equations  with  varying  speed  of  light are

\begin{equation}
\frac{\ddot{a}}{a}+2\frac{\ddot{b}}{b}=-\frac{8\pi}{(3+2\omega)\phi}\left[
(2+\omega)\rho_{_{f}}+3(1+\omega)\frac{p_{_{f}}}{c^{2}}\right]-\omega\left(\frac{\dot{\phi}}{\phi}
\right)^{2}-\frac{\ddot{\phi}}{\phi}
\end{equation}

\begin{equation}
\left(\frac{\dot{b}}{b}
\right)^{2}+2\frac{\dot{a}}{a}\frac{\dot{b}}{b}=\frac{8\pi\rho_{_{f}}}{\phi}-\frac{kc^{2}}{b^{2}}-\left(\frac{\dot{a}}{a}+2\frac{\dot{b}}{b}
\right)\frac{\dot{\phi}}{\phi}+\frac{\omega}{2}\left(\frac{\dot{\phi}}{\phi}
\right)^{2}
\end{equation}

and  the  wave  equation  for  the  BD  scalar  field  $\phi$  is

\begin{equation}
\ddot{\phi}+\left(\frac{\dot{a}}{a}+2\frac{\dot{b}}{b}
\right)\dot{\phi}=\frac{8\pi}{3+2\omega}\left(\rho_{_{f}}-\frac{3p_{_{f}}}{c^{2}}
\right)
\end{equation}

Here  the  velocity  of  light  $c$  is  an  arbitrary  function
of  time, $\omega$  is  the  BD  coupling  parameter.
$\rho_{_{f}}$ and $p_{_{f}}$ are density  and  hydrostatic
pressure respectively  of the fluid distribution  with  barotropic
equation  of  state

\begin{center}
$p_{_{f}}=(\gamma_{_{f}}-1)\rho_{_{f}}$
\end{center}

( $\gamma_{_{f}}$   being  the  constant  adiabatic  index  of
the  fluid with  $0\leq \gamma_{_{f}}\leq 2$ ).

From  the  above  field  equations, we  have  the
`non-conservation'  equation

\begin{equation}
\dot{\rho_{_{f}}}+\left(\frac{\dot{a}}{a}+2\frac{\dot{b}}{b}
\right)\left(\rho_{_{f}}+\frac{p_{_{f}}}{c^{2}}
\right)=\frac{kc\dot{c}}{4\pi b^{2}}\phi
\end{equation}

As  at  the present  epoch  the  universe  is  filled  with cold
matter (dust)  with  negligible pressure, so  using $p_{_{f}}=0$,
the equation  (6.4)  and  (6.5)  gives

\begin{equation}
\frac{d}{dt}(ab^{2}\dot{\phi})=\frac{8\pi}{3+2\omega}ab^{2}\rho_{_{f}}
\end{equation}
and
\begin{equation}
\frac{d}{dt}(ab^{2}\rho_{_{f}})=\frac{ka\phi}{8\pi}\frac{d}{dt}(c^{2})
\end{equation}

Now  eliminating $\rho_{_{f}}$  between  (6.6)  and  (6.7), we
have

\begin{equation}
\frac{d^{2}}{dt^{2}}(ab^{2}\dot{\phi})=\frac{k}{3+2\omega}a\phi\frac{d}{dt}(c^{2})
\end{equation}

Since  the  scale  factors, scalar  field  and  velocity  of
light  are  time  dependent, so we  shall  consider  the
following  cases  to  obtain  exact  analytic  form  for  the
variables  assuming  some  of  them  in  polynomial  form  or in
exponential  form.\\

{\it Case I} :  The  power  law  form  of  scale  factors  $a , b$
and BD  scalar  field  $\phi $  are  assumed  as\\
\begin{equation}
a(t)=a_{_{0}}t^{\alpha},~ b(t)=b_{_{0}}t^{\beta},~
\phi(t)=\phi_{_{0}}t^{\mu}
\end{equation}

where $a_{_{0}}, b_{_{0}},\phi_{_{0}}$ are   positive   constants
and  $\alpha, \beta, \mu$ are real   constants  with   the
restriction $\alpha+2\beta\geq 3$ (for accelerating  universe).

If  we  use  the  relations  (6.9)  in  (6.8), we  have
\begin{equation}
c=c_{_{0}}t^{\beta-1}
\end{equation}

where  the  positive  constant  $c_{_{0}}$  is  restricted as
\begin{equation}
b_{_{0}}^{2}\mu(\alpha+2\beta+\mu-1)(\alpha+2\beta+\mu-2)=\frac{2kc_{_{0}}^{2}(\beta-1)}{3+2\omega}
\end{equation}

Also  from  (6.6)  the  expression  for  $\rho_{_{f}}$ is
\begin{equation}
\rho_{_{f}}=\rho_{_{0}}t^{\mu-2}
\end{equation}
with
$$
\rho_{_{0}}=\frac{3+2\omega}{8\pi}~\phi_{_{0}}\mu(\alpha+2\beta+\mu-1)
$$

For  consistency  of  the  field  equations  the  restriction
between  the  parameters  is\\
\begin{equation}
\left(\alpha+2\beta+\mu-\frac{1}{2}\right)^{2}=\frac{\mu(\alpha+2\beta+\mu-1)}{\beta-1}[(2+\omega)
(1-\beta)-(\alpha-\mu)(3+2\omega)]+\frac{1}{4}
\end{equation}
\\

{\it Case II} :  The  exponential  form  of  scale  factors  and
BD scalar  field  are  assumed  as
\begin{equation}
a(t)=a_{_{0}}e^{\alpha t},~ b(t)=b_{_{0}}e^{\beta t},~
\phi(t)=\phi_{_{0}}e^{\mu t}
\end{equation}

where  $a_{_{0}}, b_{_{0}},\phi_{_{0}}$  are  positive constants
and $\alpha, \beta, \mu$  are  real  constants.

From  (6.8), we have
\begin{equation}
c=c_{_{0}}e^{\beta t}
\end{equation}

where  the  positive  constant  $c_{_{0}}$ will  satisfy
\begin{equation}
(3+2\omega)\mu(\alpha+2\beta+\mu)^{2}=\frac{4k\beta
c_{_{0}}^{2}}{b_{_{0}}^{2}}
\end{equation}

From  (6.6), the  expression  for  $\rho_{_{f}}$  is
\begin{equation}
\rho_{_{f}}=\rho_{_{0}}e^{\mu t}
\end{equation}

with~~~$\rho_{_{0}}=\frac{(3+2\omega)}{8\pi}\phi_{_{0}}\mu(\alpha+2\beta+\mu)$.

From  field  equations  the  relation  between  the  parameters
becomes
\begin{equation} \text{either}~~~~ \alpha+2\beta+\mu=0
\end{equation}
\begin{equation}
\text{or}~~~~(\alpha+2\beta+\mu)\left\{1+\frac{(3+2\omega)\mu}{\beta^{2}}\right\}=\mu(4+3\omega)
\end{equation}

In  both  the  cases  we  have  the  deceleration  parameter
\begin{center}
$q=-1\le 0$
\end{center}

\begin{figure}
\includegraphics{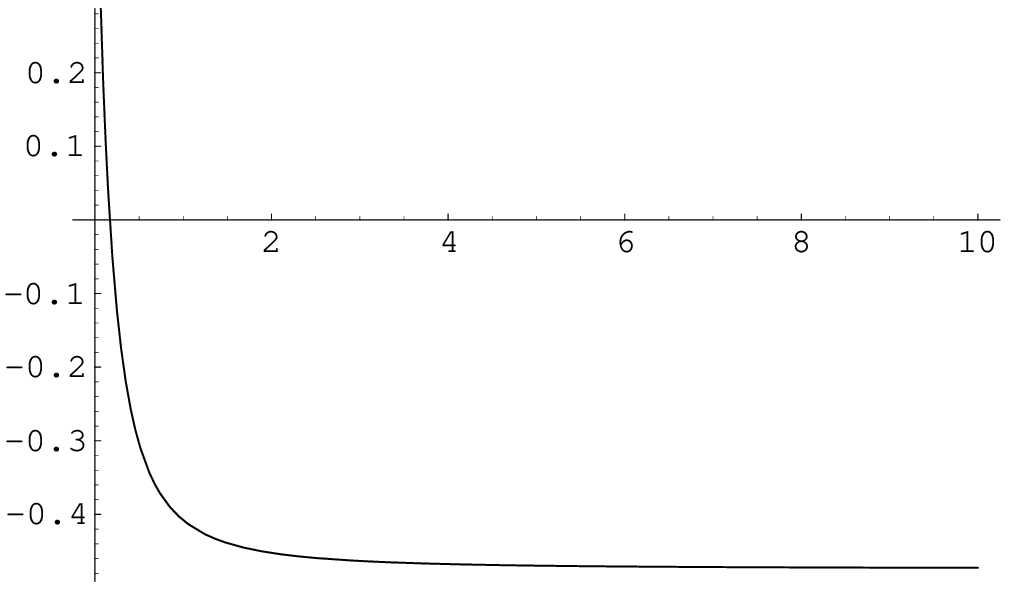}\\
$$\text{Fig}~ 6.1$$\\
\includegraphics{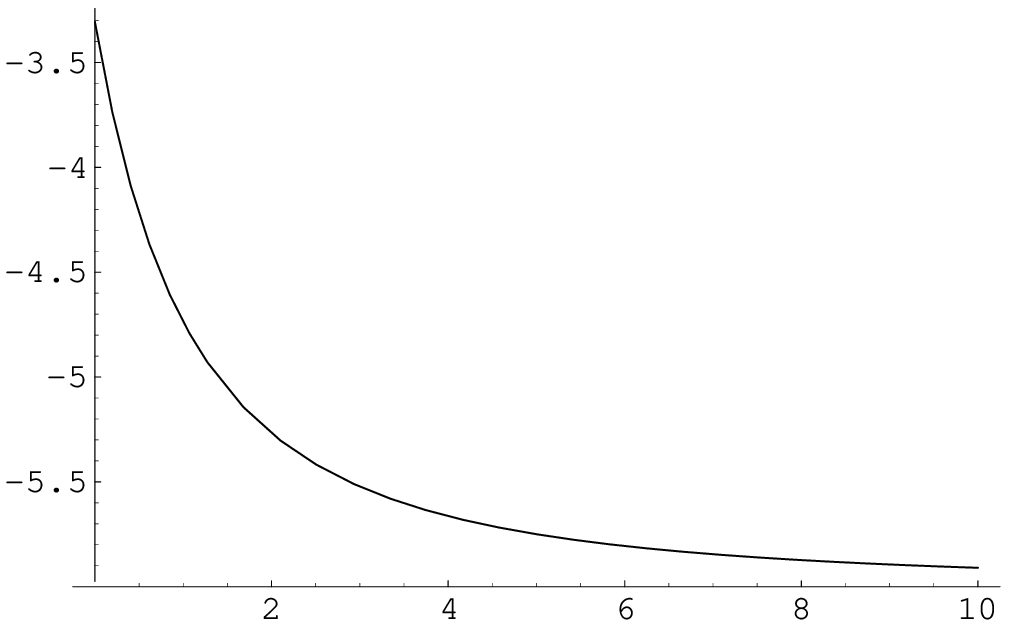}
\\
$$\text{Fig}~ 6.2$$\\
\includegraphics{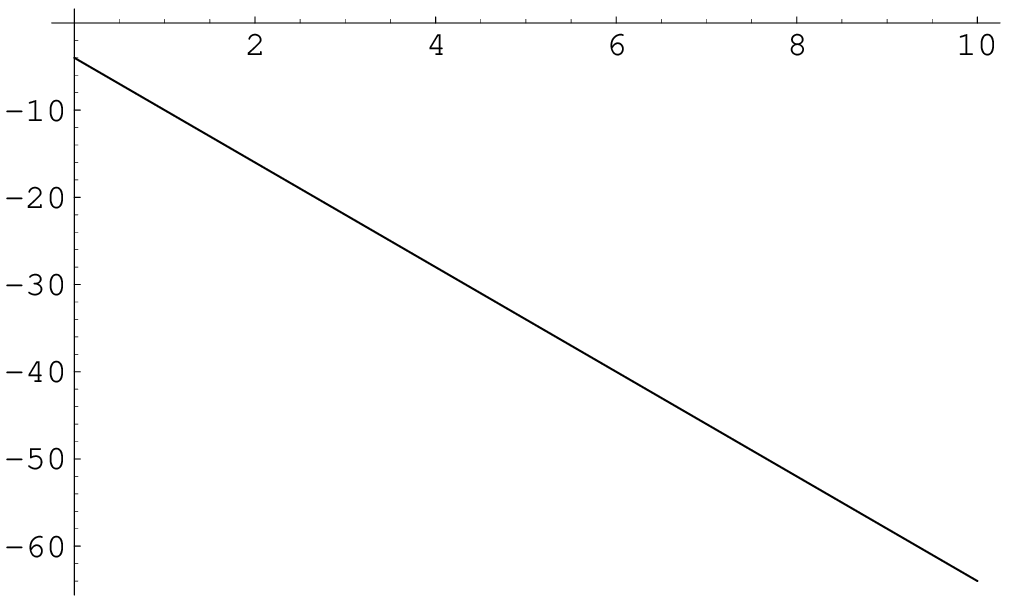}~~~~\\
\\
$$\text{Fig}~ 6.3$$\\
\end{figure}
\begin{figure}
\includegraphics{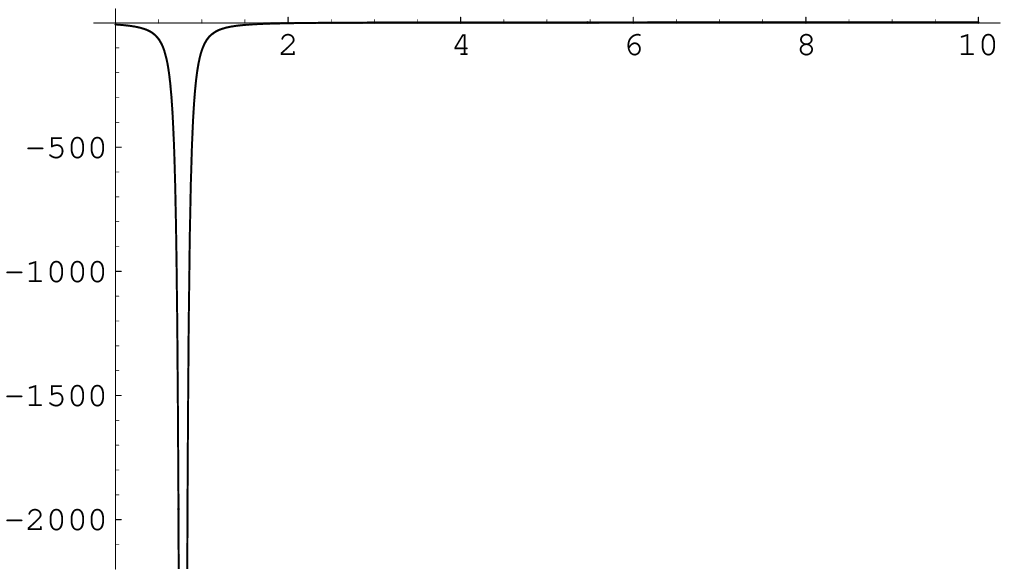}
\\$$\text{Fig}~ 6.4$$
\\
\includegraphics{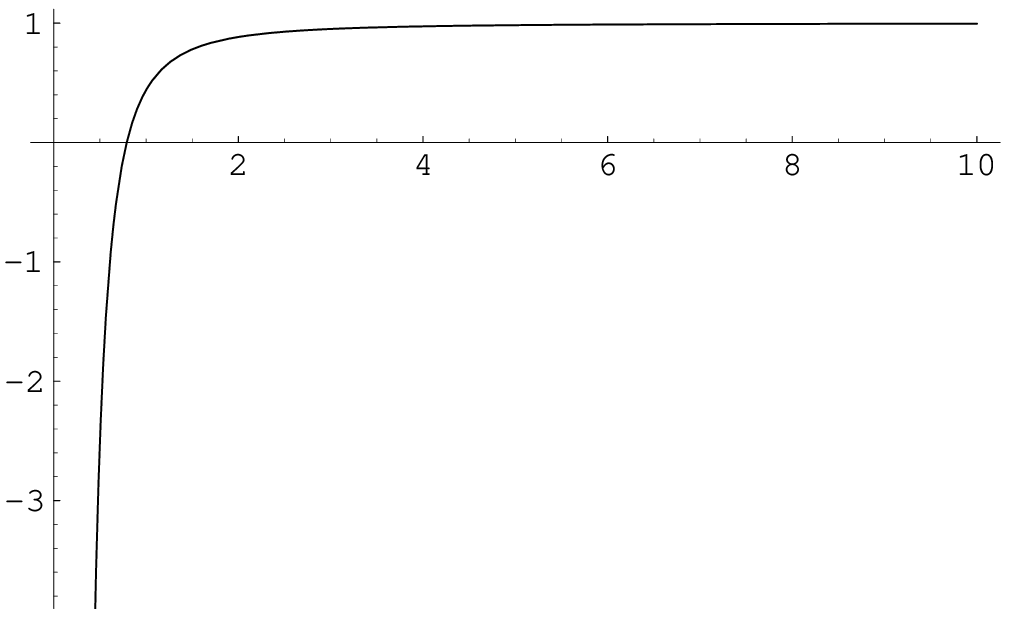}
\\
$$\text{Fig}~ 6.5$$\\

In Figs 6.1 - 6.5, we have shown the variations of $q$ over $t$
for various values of $\omega$ and the values of parameters
$a_{0}=a_{1}=1$. We have taken $\omega=-1.8, -1.5, -1, -0.87,
-0.75 $ respectively in Figs 6.1 - 6.5.
\end{figure}
\begin{figure}
\includegraphics{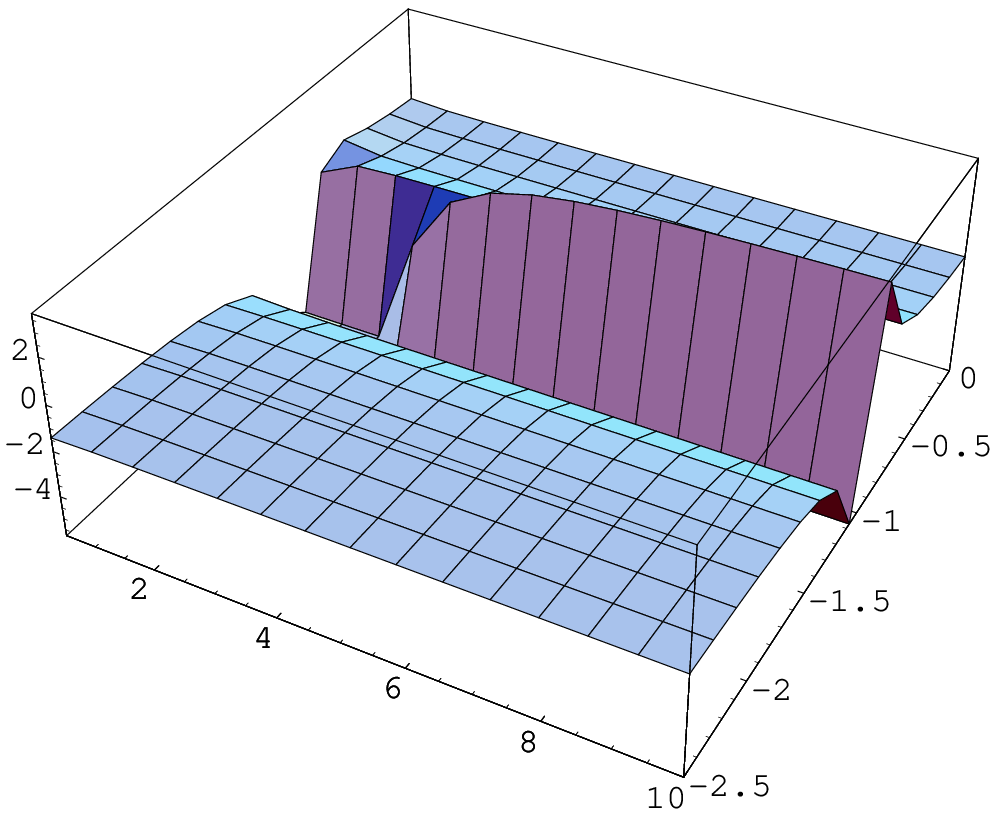}
\\$$\text{Fig}~ 6.6$$\\

In Fig 6.6, we have shown the variations of $q$ over $t$ and
$\omega$ in the range
$0\le t\le 10$ and $-2.5\le \omega \le 0$.\\
\end{figure}

{\it Case III} :   In  this  case, we  have  assumed the power
law form  of  scale  factor  $b$, velocity  of  light $c$ and BD
scalar  field $\phi$  to be
\begin{equation}
b(t)=b_{_{0}}t^{\beta},~ c(t)=c_{_{0}}t^{\delta},~
\phi(t)=\phi_{_{0}}t^{\mu}
\end{equation}

Using  (6.20), we  have  from  (6.8),

\begin{equation}
\frac{d^{2}}{dt^{2}}\left(a t^{2\beta+\mu-1}
\right)=\frac{2\delta k c_{_{0}}^{2}}{(3+2\omega)\mu
b_{_{0}}^{2}}\left(a t^{\mu+2\delta-1} \right)
\end{equation}

To  solve  the  differential  equation  we  assume
$\delta=\beta$. It is to  be  noted  that  the  above
differential  equation  can be solved  for  all  $k$. But  for
$k=\pm 1$ the  solutions  are  not consistent  to  the  field
equations. So  we  consider  only  $k = 0$. In  this  case  the
explicit solution  is
\begin{eqnarray}
\left.\begin{array}{llll}
a(t)=a_{_{0}}t^{-\frac{\omega+2}{3\omega+4}}+a_{_{1}}t^{\frac{2(\omega+1)}{3\omega+4}}\\\\
b(t)=b_{_{0}}t^{\frac{2(\omega+1)}{3\omega+4}}\\\\
c(t)=c_{_{0}}t^{\frac{2(\omega+1)}{3\omega+4}}\\\\
\phi(t)=\phi_{_{0}}t^{\frac{2}{3\omega+4}}\\\\
\rho_{_{f}}=\frac{1}{4\pi}\frac{2\omega+3}{3\omega+4}\frac{\phi_{_{0}}a_{_{1}}}{(a_{_{0}}
+a_{_{1}}t)}~t^{-\frac{3\omega+2}{3\omega+4}}
\end{array}\right\}
\end{eqnarray}

with $a_{_{0}}, a_{_{1}}$   as  integration  constants.

The  deceleration  parameter  has  the  expression

\begin{center}
$q=-1+\frac{3[(3\omega+2)(a_{_{0}}+a_{_{1}}t)^{2}+a_{_{1}}^{2}(3\omega+4)t^{2}]}
{[(3\omega+2)(a_{_{0}}+a_{_{1}}t)+a_{_{1}}(3\omega+4)t]^{2}}$ ,
\end{center}

which  is  finite  for  all  $t$  and $\omega$  except
$t=-\frac{a_{_{0}}(3\omega+2)}{6a_{_{1}}(\omega+1)}$ and we have
the non-decelerated universe for $-2\le \omega\le-2/3$. Also  we
have the non-decelerated  universe asymptotically except  for
$-1\le\omega\le-1/2$. The variation  of   $q$  over  time has
been shown graphically for different  values  of  $\omega$   in
figures 6.1 - 6.5.\\

\section{Conformal  Transformation and Flatness  Problem}
One important  aspect  of  this model  is  that  it  can solve the
flatness  problem  as  well. To  see  this  we  make  a
conformal  transformation [Faraoni et al, 1999]  as
\begin{equation}
\bar{g}_{\mu\nu}=\phi g_{\mu\nu}
\end{equation}

which  enables  us  to  identify  the  energy  contributions
from  different  components  of  matter  very  clearly.

In  this  section, we  have  developed  the  BD  theory  in
Jordan  frame  and  to  introduce  the  Einstein  frame, we make
the  following  transformations :
\begin{eqnarray*}
d\eta=\sqrt{\phi}~a,~ \bar{a}=\sqrt{\phi}~a,~
\bar{b}=\sqrt{\phi}~b,~ \psi=\text{ln}~\phi,~
\bar{\rho}_{_{f}}=\phi^{-2}\rho_{_{f}},
\end{eqnarray*}
\begin{equation}
\bar{\rho}_{_{\psi}}=\phi^{-2}\rho_{_{\psi}},~
\bar{p}_{_{f}}=\phi^{-2}p_{_{f}},~\bar{p}_{_{\psi}}=\phi^{-2}p_{_{\psi}}
\end{equation}

So  the  field  equations  (6.2) - (6.4)  transformed  to

\begin{equation}
\frac{\bar{a}''}{\bar{a}}+2\frac{\bar{b}''}{\bar{b}}=-4\pi\left(\bar{\rho}_{_{f}}+\frac{3\bar{p}_{_{f}}}{c^{2}}\right)-\frac{(3+2\omega)}{2}\psi'^{2}
\end{equation}

\begin{equation}
\left(\frac{\bar{b}'}{\bar{b}}\right)^{2}+2\frac{\bar{a}'}{\bar{a}}\frac{\bar{b}'}{\bar{b}}+\frac{k
c^{2}}{\bar{b}^{^{2}}}=8\pi\bar{\rho}_{_{f}}+\frac{(3+2\omega)}{4}\psi'^{2}
\end{equation}
and
\begin{equation}
\psi''+\left(\frac{\bar{a}'}{\bar{a}}+2\frac{\bar{b}'}{\bar{b}}\right)\psi'=\frac{8\pi}
{3+2\omega}\left(\bar{\rho}_{_{f}}-\frac{3\bar{p}_{_{f}}}{c^{2}}\right)
\end{equation}

where~~  $' \equiv \frac{d}{d\eta}$.

The  scalar  field  $\psi$ (massless)  behaves  like  a  `stiff'
perfect  fluid  with  equation  of  state
\begin{equation}
\bar{p}_{_{\psi}}=\bar{\rho}_{_{\psi}}=\frac{\psi'^{2}}{16\pi G}
\end{equation}

If  the  velocity  of  light  is  constant, then  in  Einstein
frame  total  stress-energy  tensor  is  conserved  but  there
is  an  exchange  of  energy  between  the  scalar  field  and
normal  matter  according  to  the  following  equation\\
\begin{equation}
\bar{\rho}_{_{f}}+\left(\frac{\bar{a}'}{\bar{a}}+2\frac{\bar{b}'}{\bar{b}}\right)
\left(\bar{\rho}_{_{f}}+\frac{\bar{p}_{_{f}}}{c^{2}}\right)=-\left[\bar{\rho}_{_{\psi}}+
\left(\frac{\bar{a}'}{\bar{a}}+2\frac{\bar{b}'}{\bar{b}}\right)\left(\bar{\rho}_{_{\psi}}+
\bar{p}_{_{\psi}}\right)\right]=-\frac{\psi'}{2}\left(\bar{\rho}_{_{f}}-
\frac{3\bar{p}_{_{f}}}{c^{2}}\right)
\end{equation}

On  the  other  hand, if  the  velocity  of  light  varies then
we  have  separate  `non-conservation'  equations
\begin{equation}
\bar{\rho}_{_{f}}+\left(\frac{\bar{a}'}{\bar{a}}+2\frac{\bar{b}'}{\bar{b}}\right)\left(\bar{\rho}_{_{f}}+\frac{\bar{p}_{_{f}}}{c^{2}}\right)=-\frac{\psi'}{2}\left(\bar{\rho}_{_{f}}-\frac{3\bar{p}_{_{f}}}{c^{2}}\right)+\frac{k
c c'}{4\pi G\bar{b}^{2}}
\end{equation}
and
\begin{equation}
\bar{\rho}_{_{\psi}}+\left(\frac{\bar{a}'}{\bar{a}}+2\frac{\bar{b}'}{\bar{b}}\right)\left(\bar{\rho}_{_{\psi}}+\bar{p}_{_{\psi}}\right)=\frac{\psi'}{2}\left(\bar{\rho}_{_{\psi}}-\frac{3\bar{p}_{_{\psi}}}{c^{2}}\right)
\end{equation}

Thus  combining  the  two  energy  densities, we  have  from the
above  equations, the  equation  for  the  conservation for the
total  energy  is
\begin{equation}
\bar{\rho}'+3\gamma\bar{H}\bar{\rho}=0
\end{equation}

Here
$\bar{H}=\frac{1}{3}\left(\frac{\bar{a}'}{\bar{a}}+2\frac{\bar{b}'}{\bar{b}}\right)$
is  the  Hubble  parameter  in  the Einstein frame and  $\gamma$
is the  net  barotropic  index  defined  as
\begin{equation}
\gamma\bar{\Omega}=\gamma_{_{f}}\bar{\Omega}_{_{f}}+\gamma_{_{\psi}}\bar{\Omega}_{_{\psi}}
\end{equation}
where
\begin{equation}
\bar{\Omega}=\bar{\Omega}_{_{f}}+\bar{\Omega}_{_{\psi}}=\frac{\bar{\rho}}{3\bar{H}^{^{2}}}
\end{equation}

is  the  dimensionless  density  parameter.

From  equations (6.26)  and  (6.32), we  have  the  evolution
equation  for  the  density  parameter  as
\begin{equation}
\bar{\Omega}'=\bar{\Omega}(\bar{\Omega}-1)[\gamma\bar{H}_{a}+2(\gamma-1)\bar{H}_{b}]
\end{equation}

where ~~ $\bar{H}_{a}=\frac{\bar{a}'}{\bar{a}}$    and
$\bar{H}_{b}=\frac{\bar{b}'}{\bar{b}}$ .\\

This equation in $\bar{\Omega}$ shows that  $\bar{\Omega}=1$ is a
possible solution  of it and  for stability of  this solution,
we  have
\begin{equation}
\gamma<\frac{2}{3}
\end{equation}

Since  the  adiabatic  indices  do  not  change  due  to
conformal  transformation  so  we  take  $\gamma_{_{f}}=1$ (since
$ p_{_{f}}=0$) and $\gamma_{_{\psi}}=2$. Hence from (6.33) and
(6.34), we have\\
\begin{equation}
\gamma=\frac{\bar{\Omega}_{_{f}}+2\bar{\Omega}_{_{\psi}}}{\bar{\Omega}_{_{f}}+\bar{\Omega}_{_{\psi}}}
\end{equation}

Now  due  to  upper  limit  of  $\gamma$ , we  must  have  the
inequality
\begin{equation}
\bar{\Omega}_{_{f}}<4|\bar{\Omega}_{_{\psi}}|
\end{equation}

according  as  $\gamma$  is  restricted  by  (6.36).

From  the  field equation  (6.26), the  curvature  parameter
$\bar{\Omega}_{_{k}}=-k c^{2}/~\bar{b}^{2}$ vanishes  for  the
solution $\bar{\Omega}=1$ . So for BD-scalar field it  is possible
to have a stable  solution corresponding  to $\bar{\Omega}=1$ and
hence  the flatness problem  can  be solved.\\

\section{Discussion}
We  have  performed  an  extensive  analysis  of  solutions  to
Brans-Dicke  theories  with  varying  speed  of  light. For the
power-law  forms  and  exponential  forms  of  the cosmological
scale  factors  and  scalar  field  in  {\it Cases I} and {\it II}
respectively, we  have  the  velocity  of  light  in the  same
form  and  we  identified  the  cases  where  the quintessence
problem  can  be  solved  for  some  restrictions on  parameters.
In  the  {\it Case II}, we  always  have  the accelerated
universe. For Figs 6.1  and  6.2, $q$  decreases to a  fixed
negative value asymptotically  for  $\omega=-1.8$   and
$\omega=-1.5$ . For Fig 6.3, $q$ is linear with time  and always
negative for $\omega=-1$ . For Fig 6.4, $q$ decreases and then
there  is a singularity and after that $q$ increases. In Fig 6.5,
$q$ increases  to a fixed positive value asymptotically. From Fig
6.6, we conclude that the decelerated parameter  is negative
whenever  $-2<\omega<-2/3$ . In this range $q$ decreases till
$t=-\frac{a_{_{0}}(3\omega+2)}{6a_{_{1}}(\omega+1)}$ and then
increases  to  a  fixed  positive  value asymptotically.

Along  with  providing  a  non-decelerating  solution, it  can
solve  the  flatness  problem  also. In  fact, it  has  been
shown  that  $\bar{\Omega}=1$   could  be  a  stable  solution  in
this model.\\

\large \baselineskip .85cm
\chapter{Does Cosmic No-Hair Conjecture in Brane Scenarios Follow from
General Relativity?}\label{chap7} \markright{\it
CHAPTER~\ref{chap7}.~Does Cosmic No-Hair Conjecture in Brane
Scenarios Follow from General Relativity?}

\section{Prelude}
The idea of brane world scenarios may resolve the challenging
problem in theoretical physics namely the unification of all
forces and particles in nature. It is suggested that we live in a
four dimensional brane embedded in a higher dimensional
space-time. As a result, the fundamental higher dimensional
Planck mass could be of same order as the electro weak scale and
thereby one of the hierarchy problems in the current standard
model of high-energy physics are resolved [Maartens, 2001; Chakraborty
et al, 2002; Randall et al, 1999].\\

According to Randall and Sundrum [1999] it is possible to have a
single massless bound state confined to a domain wall or 3-brane
in five dimensional non-factorizable geometries. They have shown
that this bound state corresponds to the zero mode of the
Kaluza-Klein dimensional reduction and is related to the four
dimensional gravitation [Chakraborty et al, 2002]. Hence all
matter and gauge fields (except gravity) are confined to a 3-brane
embedded in a five dimensional space-time (bulk) while gravity can
propagate in the bulk. As a consequence, the gravity on the brane
can be described by the Einstein's equations modified by two
additional terms, namely (i) quadratic in matter variables and
(ii) the electric part of the five dimensional Weyl tensor [Maartens,
2000; Campos et al, 2001].\\

In general terminology, the CNHC [Gibbons et al, 1977; Hawking et
al, 1982] states that ``{\it all expanding universe models with a
positive cosmological constant asymptotically approach the
de-Sitter solution}''. To address the question whether the
universe evolves to a homogeneous and isotropic state during an
inflationary epoch, Gibbons and Hawking [1977] and then Hawking
and Moss [1982] developed this conjecture. Subsequently, Wald
[1983] gave a formal proof of it for homogeneous cosmological
models (Bianchi models) with a positive cosmological constant. He
assumed that the matter field should satisfy strong and weak
energy conditions.\\

In this chapter, we wish to extend Wald's [1983] result in the
brane world scenario and examine whether the new conditions can be
minimize using those in general relativity.\\

\section{Cosmic No Hair Conjecture in Brane World}
According to Roy Maartens [2001], the Einstein equations on the
brane can be written as
\begin{equation}
G_{\mu\nu}=-\Lambda
g_{\mu\nu}+\kappa_{4}^{2}T_{\mu\nu}+\kappa_{5}^{4}S_{\mu\nu}-\cal{E}_{\mu\nu}
\end{equation}

where $S_{\mu\nu}$ and $T_{\mu\nu}$ are the two correction terms
(local and non-local) in the energy momentum tensor. The local
correction term $S_{\mu\nu}$ has the expression

\begin{equation}
4S_{\mu\nu}=\frac{1}{3}T
T_{\mu\nu}-T_{\mu\rho}T_{\nu}^{\rho}-\frac{1}{2}g_{\mu\nu}\left(\frac{1}{3}T^{2}
-T_{\rho\sigma}T^{\rho\sigma}\right)
\end{equation}

while $\cal{E}_{\mu\nu}$ is the electric part of the 5D Weyl
tensor in the bulk. Now the scalar constraint (initial value
constraint) equation and the Roychoudhuri equation on the brane
has the form
\begin{equation}
G_{\mu\nu}n^{\mu}n^{\nu}=\Lambda+\kappa_{4}^{2}T_{\mu\nu}n^{\mu}n^{\nu}
+\kappa_{5}^{4}S_{\mu\nu}n^{\mu}n^{\nu}-\cal{E}_{\mu\nu}n^{\mu}n^{\nu}
\end{equation}
and
\begin{equation}
R_{\mu\nu}n^{\mu}n^{\nu}=-\Lambda+\kappa_{4}^{2}\left(T_{\mu\nu}-\frac{1}{2}
g_{\mu\nu}T\right)n^{\mu}n^{\nu}+\kappa_{5}^{4}\left(S_{\mu\nu}-\frac{1}{2}
g_{\mu\nu}S\right)n^{\mu}n^{\nu}-\cal{E}_{\mu\nu}n^{\mu}n^{\nu}
\end{equation}

where $n^{\mu}$ is the unit normal to the spatial homogeneous
hypersurfaces. In terms of the homogeneous hypersurface elements
namely, the projected metric
$h_{\mu\nu}(=g_{\mu\nu}+n_{\mu}n_{\nu})$ and the extrinsic
curvature $K_{\mu\nu}(=\nabla_{\nu}n_{\mu})$ and using the
Gauss-Codazzi equations the above two equations namely equations
(7.3) and (7.4) become
\begin{equation}
K^{2}=3\Lambda+\frac{3}{2}\sigma_{\mu\nu}\sigma^{\mu\nu}-\frac{3}{2}~^{(3)}R
+3\kappa_{4}^{2}T_{\mu\nu}n^{\mu}n^{\nu}+3\kappa_{5}^{4}S_{\mu\nu}n^{\mu}n^{\nu}
-3\cal{E}_{\mu\nu}n^{\mu}n^{\nu}
\end{equation}
and
\begin{equation}
\dot{K}=\Lambda-\frac{1}{3}K^{2}-\sigma_{\mu\nu}\sigma^{\mu\nu}-
\kappa_{4}^{2}\left(T_{\mu\nu}-\frac{1}{2}g_{\mu\nu}T\right)n^{\mu}n^{\nu}
-\kappa_{5}^{4}\left(S_{\mu\nu}-\frac{1}{2}g_{\mu\nu}S\right)n^{\mu}n^{\nu}
+\cal{E}_{\mu\nu}n^{\mu}n^{\nu}
\end{equation}

where the dot denotes the Lie derivative with respect to proper
time, $K$ is the trace of the extrinsic curvature,
$\sigma_{\mu\nu}$ is the shear of the time like geodesic
congruence orthogonal to the homogeneous hypersurfaces and
~$^{(3)}R$ is the scalar curvature of the homogeneous
hypersurfaces.

Using the idea of Wald and proceeding along his approach (for
details, Wald et al [1983] and Chakraborty et al [2001]) one can
find that for CNHC we must have
\begin{equation}
(a)~~~~~~~~~~~~~~~~~~~~~ S_{\mu\nu}n^{\mu}n^{\nu} \ge 0 ~~~~
\text{and}
~~~~\left(S_{\mu\nu}-\frac{1}{2}g_{\mu\nu}S\right)n^{\mu}n^{\nu}
\ge 0 \hspace{1in}
\end{equation}
and
\begin{equation}
(b)~~~~~~~~~~~~~~~~~~~~~~~~~~~~~~~~~~~~~~~~~~~~~~\cal{E}_{\mu\nu}n^{\mu}n^{\nu}
\le 0 \hspace{2in}
\end{equation}

in addition to the weak and strong energy conditions for the
matter field

\begin{equation}
T_{\mu\nu}n^{\mu}n^{\nu} \ge 0
~~~\text{and}~~~\left(T_{\mu\nu}-\frac{1}{2}g_{\mu\nu}T\right)n^{\mu}n^{\nu}
\ge 0
\end{equation}

Now if we use the expression (7.2) for $S_{\mu\nu}$ in (7.7) then
we get

\begin{equation}
\frac{1}{3}T
b-\frac{1}{2}T_{\rho\sigma}T^{\rho\sigma}-(T_{\mu\rho}n^{\mu})(T_{\nu}^{\rho}n^{\nu})
\ge 0
\end{equation}
and
\begin{equation}
\frac{1}{3}T a-(T_{\mu\rho}n^{\mu})(T_{\nu}^{\rho}n^{\nu}) \ge 0
\end{equation}

where $a=T_{\mu\nu}n^{\mu}n^{\nu}$ and
$b=\left(T_{\mu\nu}-\frac{1}{2}g_{\mu\nu}T\right)n^{\mu}n^{\nu}$
are positive due to (7.9).

Also using the symmetry properties of $\cal{E}_{\mu\nu}$, it is
possible to decompose it with respect to any time like observer
$\vec{u}~(u^{\alpha}u_{\alpha}=-1)$ as [Maartens, 2000; Campos et
al, 2001]

$$
\cal{E}_{\mu\nu}=-\left(\frac{\kappa_{5}}{\kappa_{4}}\right)^{4}\left[\left(u_{\mu}u_{\nu}+
\frac{1}{3}h_{\mu\nu}\right)\cal{U}+2(u_{\mu}Q_{\nu})+P_{\mu\nu}\right]
$$

with the properties
$$
Q_{\mu}u^{\mu}=0,~~ P_{(\mu\nu)}=P_{\mu\nu},~~ P_{\mu}^{\mu}=0,~~
P_{\mu\nu}u^{\nu}=0
$$

If we consider the Bianchi models then due to the symmetry of the
spatial geometry we may choose
$$
Q_{\mu}=P_{\mu\nu}=0,
$$

and the scalar part namely $\cal{U}$ is termed as dark energy
density as it has energy-momentum tensor that of a radiation
perfect fluid. So the restriction (7.8) implies that dark energy
density should be always positive i.e.,
\begin{equation}
\cal{U}\ge 0
\end{equation}

As it is not possible to make any restriction on $T_{\mu\nu}$ to
satisfy inequations (7.10) and (7.11), so let us examine with some
realistic model for the matter field.\\

\section{Examples}

\subsection{Perfect Fluid Model}
In this case the energy-momentum tensor has the form
$$
T_{\mu\nu}=(\rho+p)n_{\mu}n_{\nu}+p
g_{\mu\nu},~~~n_{\mu}n^{\mu}=-1
$$

with $\rho$ and $p$ as the energy density and isotropic pressure
respectively.

The weak and strong energy conditions demand
\begin{equation}
a=\rho \ge 0 ~~~\text{and}~~~ 2b=\rho+3p \ge 0
\end{equation}

Hence the inequations (7.7) (i.e., inequations (7.10) and (7.11))
take the form
\begin{equation}
\rho^{2} \ge 0 ~~~\text{and} ~~~ \rho(3p+2\rho) \ge 0
\end{equation}

which are always true. Thus for perfect fluid model CNHC is
automatically satisfied in brane scenarios if it is valid in
general relativity.\\

\subsection{General Form of Energy-Momentum Tensor}
The general form of the brane energy momentum tensor for any
matter fields (scalar field, perfect fluids, kinematic gases,
dissipative fluids, etc.) including a combination of different
fields can be covariently written as [Maartens, 2001]
\begin{equation}
T_{\mu\nu}=\rho n_{\mu}n_{\nu}+p
h_{\mu\nu}+\Pi_{\mu\nu}+q_{\mu}n_{\nu}+q_{\nu}n_{\mu}
\end{equation}

Here the energy flux $q_{\mu}$ and the anisotropic stress
$\Pi_{\mu\nu}$ are projected, symmetric and traceless that is
$$
q_{\mu}n^{\mu}=0,~~\Pi_{\mu\nu}n^{\mu}=0,~~\Pi_{\mu\nu}=\Pi_{\nu\mu},
~~\Pi_{\mu\nu}g^{\mu\nu}=0
$$

Thus for this form of energy-momentum tensor, the restrictions on
$S_{\mu\nu}$ now result
\begin{equation}
\frac{1}{3}\rho^{2}-\frac{1}{2}\Pi_{\mu\nu}\Pi^{\mu\nu} \ge 0
\end{equation}
and
\begin{equation}
\frac{1}{3}\rho(3p+2\rho)-q_{\mu}q^{\mu} \ge 0
\end{equation}

As for realistic matter, the energy density should be larger than
the anisotropic stress and heat flux is very small in magnitude
so the inequalities (7.16) and (7.17) are automatically satisfied.
Hence the CNHC is satisfied for the above form of general
energy-momentum tensor.\\

\addcontentsline{toc}{part}{Part B~: Study
of Some Collapsing Models}
 \baselineskip
.81cm

\markright{ }

\vspace*{8cm}
\begin{center}
{\Huge{{\bf Part B} \\ \vspace{1cm} Study of Some Collapsing
Models }}
\end{center}

\large \baselineskip .85cm

\chapter{General Introduction}
\label{chap8}\markright{\it CHAPTER~\ref{chap8}. General
Introduction}

An outstanding problem in relativistic astrophysics and
gravitation theory today is the final outcome of an endless
gravitational collapse. What will be the end state of such a
continual collapse which is entirely dominated by the force of
gravity? According to the general theory of relativity, which is
a classical gravity theory, after undergoing supernova explosion
in the last state of its evolution, a star with a mass much
larger than the solar mass will contracted without limit, due to
its strong gravity and form a `domain' called {\it space-time
singularity}. Space-time singularity is a very general phenomena,
not only the gravitational collapse of stars of very large but
also in physical processes in which the general theory of
relativity plays an important role. In fact it was proved by
Hawking and Penrose [1965, 1967, 1970] that the appearance of
space-time singularity is generic i.e., space-time singularity
appears for any space-time symmetry. However the singularity
theorems of Hawking and Penrose only prove the causally geodesic
incompleteness of space-time and say nothing about the detailed
features of the singularities themselves. For example, one can not
obtain information about how the space-time curvature and the
energy density diverge in a space-time singularity from these
theorems.\\

Space-time singularities can be classified into two kinds,
according to whether or not they can observed. A singularity in
space-time is said to be {\it naked singularity} when non-space
like geodesics can escape from it in future direction that means
the singularity is visible to the local or distant observer. If
such singularity can reach the neighbouring or asymptotic regions
of space-time, the singularity becomes {\it locally} or {\it
globally} naked respectively. A space-time singularity that can
not observed is called {\it black hole}. Is such naked
singularity formed in our universe? With regard to this question,
Penrose [1969, 1979] proposed the so-called {\it Cosmic Censorship
Conjecture}. There are two versions of this conjecture. The {\it
Weak Cosmic Censorship Conjecture} states that the singularities
produced by gravitational collapse must always hidden in a black
hole [Penrose, 1969]. If weak cosmic censorship conjecture fails
then gravitational collapse can (generically) result in a naked
singularity rather than black hole. If so, then it is quite
possible that the formation of black hole would be a non-generic
outcome of collapse . This conjecture implies the future
predictability of the space-time outside the event horizon. A
closely related conjecture, known as the {\it Strong Cosmic
Censorship Conjecture} [Penrose, 1979], asserts that, generically,
time-like singularities never occur, so that even an observer who
falls into a black hole will never experience the singularity. It
states that all physically reasonable space-times are globally
hyperbolic.\\

Most of the counter examples of Cosmic Censorship Conjecture
present in the literature either belong to Vaidya space-time
[Vaidya, 1943, 1951] or to the Tolman-Bondi space-time [Tolman,
1934; Bondi, 1947]. The Vaidya metric describes the external
gravitational field of a radiating star, whereas the Tolman-Bondi
metric describes the gravitational field inside a spherically
symmetric inhomogeneous distribution of dust matter. It has been
shown by Lemos [1992] that the nature of naked singularities are
same in Vaidya and Tolman-Bondi space-times. When the binding
energy function in Tolman-Bondi metric goes to infinity, the
Vaidya metric and Tolman-Bondi metric represent the
same collapse. \\

\section{Spherical collapse}
Before restricting matter fields, one can present the Einstein
equation for a general spherically symmetric space-time. In
spherically symmetric space-time, without loss of generality, the
line element can be written as

\begin{equation}
ds^{2}=-e^{2\nu(t,r)}dt^{2}+e^{2\lambda(t,r)}dr^{2}+R^{2}(t,r)(d\theta^{2}+sin^{2}\theta
d\phi^{2} )
\end{equation}

One can choose a reference system which is comoving for matter
fields. In this co-ordinate system the stress-energy tensor
$T^{\mu}_{\nu}$ that is the source of spherically symmetric
gravitational field must be of the form

\begin{equation}
T^{\mu}_{\nu}=\left(
\begin{array}{c}
-\rho~~~~0~~~~0~~~~0\\
~~0~~~~\Sigma~~~~0~~~~0\\
~~0~~~~0~~~~\Pi~~~~0\\
~~0~~~~0~~~~0~~~~\Pi
\end{array}\right)
\end{equation}

where $\rho(t,r), ~\Sigma(t,r)$ and $\Pi(t,r)$ are the energy
density, radial stress and tangential stress respectively. If one
considers a perfect fluid, which is described by

\begin{equation}
T_{\mu\nu}=(\rho+p)u_{\mu}u_{\nu}+pg_{\mu\nu}
\end{equation}

then the stress is isotropic i.e., $\Sigma=\Pi=p$.

From the Einstein equation and the equation of motion for the
matter fields, one can obtain [Harada et al, 2002] (using units,
$8\pi G=c=1$)

\begin{equation}
m'=4\pi\rho R^{2}R'
\end{equation}
\vspace{-5mm}
\begin{equation}
\dot{m}=-4\pi\Sigma R^{2}\dot{R}
\end{equation}
\vspace{-5mm}
\begin{equation}
\dot{R}'=\dot{R}\nu'+R'\dot{\lambda}
\end{equation}
\vspace{-5mm}
\begin{equation}
\Sigma'=-(\rho+\Sigma)\nu'-2(\Sigma-\Pi)~\frac{R'}{R}
\end{equation}
\begin{equation}
m=\frac{1}{2}~R\left(1-R'^{2}e^{-2\lambda}+\dot{R}^{2}e^{-2\nu}\right)
\end{equation}

where $m=m(t,r)$ is the Misner-Sharp mass [Misner et al, 1964] and
the prime and dot denote partial derivatives w.r.t. $r$ and $t$ respectively.\\

\subsection{Dust Collapse}
Tolman-Bondi-Lema\^{\i}tre (TBL) [Tolman, 1934; Bondi, 1947;
Lema\^{\i}tre, 1933] class of solutions of Einstein's equations
represents inhomogeneous spherically symmetric dust models. One
can restrict the matter content of the model to a dust fluid
i.e., a pressure less fluid. Therefore $\Sigma=\Pi=0$. So from
equations (8.4) to (8.8), one obtains the solutions
\begin{equation}
m(r)=\frac{F(r)}{2}
\end{equation}
\vspace{-5mm}
\begin{equation}
\nu=0
\end{equation}
\vspace{-5mm}
\begin{equation}
e^{2\lambda}=\frac{R'^{2}}{1+f(r)}
\end{equation}
\vspace{-5mm}
\begin{equation}
\dot{R}^{2}=f(r)+\frac{F(r)}{R}
\end{equation}
\vspace{-5mm}
\begin{equation}
\rho(t,r)=\frac{F'}{R^{2}R'}
\end{equation}

where the arbitrary functions $F(r)$ and $f(r)(>-1)$ are twice the
conserved Misner-Sharp mass and the specific energy respectively.
In equation (8.10) one can use the rescaling freedom of the time
co-ordinate. This means that a synchronous comoving co-ordinate
system is possible.\\

For gravitational collapse of dust model, $\dot{R}(t,r)<0$ and
without loss of generality rescale $R$ such that $R(0,r)=r$. So
the solution of $R$ (from eq. (8.12)) has the implicit form
[Dwivedi et al, 1992]

\begin{equation}
t=\frac{1}{\sqrt{F}}\left[r^{3/2}G\left(-\frac{fr}{F}\right)-
R^{3/2}G\left(-\frac{fR}{F}\right)\right]
\end{equation}

Here $G(y)$ is a strictly real positive and bounded function which
has the range $-\infty\le y\le 1$ and is given by

\begin{eqnarray}G(y)=\left\{\begin{array}{lll}
\frac{sin^{-1}\sqrt{y}}{y^{3/2}}-\frac{\sqrt{1-y}}{y}~~~~\text{for}~~~ 0<y\le 1\\\\
~~~~\frac{2}{3}~~~~\text{for}~~~ y=0\\\\
-\frac{sinh^{-1}\sqrt{-y}}{(-y)^{3/2}}-\frac{\sqrt{1-y}}{y}~~~~\text{for}~~~y<0
\end{array}\right.
\end{eqnarray}

One should note that the cases $f<0,~f=0$ and $f>0$ correspond
respectively to the ranges in terms of $y$ given by $0<y\le
1,~y=0$ and $y<0$. Also it should be noted that $f<0,~f=0$ and
$f>0$ represents bound (elliptic), marginally bound (parabolic)
and unbound (hyperbolic) models.\\

The solution for marginally bound collapse ($f=0$) is explicitly
given by

\begin{equation}
R=\left(\frac{9F}{4}\right)^{1/3}\left\{t_{sf}(r)-t\right\}^{2/3}
\end{equation}

where $t_{sf}(r)$ is given by
\begin{equation}
t_{sf}(r)=\frac{2}{3\sqrt{F}}~r^{3/2}
\end{equation}

which represents the shell focusing singularity ($R=0$) time and
the singularity occurs at different instants of time for
different shells characterized by $r$-constant hypersurface. The
central shell focusing singularity occurs at $r=0$ and the
corresponding time being $t_{0}=t_{sf}(0)$. This brings in the
essential difference from the analogous situation in the
homogeneous space-time where the function $t_{sf}(r)$ is a
constant and all the particles reach the centre simultaneously.\\

The total mass within radius $r$ is given by the integral

\begin{equation}
m(r)=4\pi\int_{0}^{r}\rho R^{2}dR=4\pi\int_{0}^{r}F'(r)dr
\end{equation}

or $2m(r)=F(r)$, since $F(0)=0$. This is because $R=0$ when
$r=0$. The function $F(r)$ thus represents the effective
gravitational mass within radius $r$. For $F=$constant ($\ne 0$),
we find $\rho=0$, so that the solution applies to empty space
outside the boundary of the spherical distribution. In addition
if $f=0$ and $t_{sf}(r)=r$ are chosen then one may obtain
Schwarzchild metric in a different co-ordinate system. The case
$F=0$ corresponds to the absence of any gravitational field and
by a suitable transformation the metric can be brought to
Minkowskian form.\\

An important property of the TBL solution is that the assignment
of the arbitrary functions appearing in it over the interval from
0 to some $r_{0}$ completely determines the solution of the
interior problem for the sphere of this radius [Landau et al,
1975]. It does not depend on how the functions are assigned for
$r>r_{0}$. Physical interpretation of the arbitrary functions
becomes more clear when we write the equation (8.12) in the form

\begin{equation}
\frac{1}{2}\dot{R}^{2}-\frac{\frac{1}{2}F(r)}{R}=\frac{1}{2}f(r)
\end{equation}

which invites the interpretation of the quantities

\begin{equation}
E(r)=\frac{1}{2}f(r)
\end{equation}
\vspace{-5mm}
\begin{equation}
T(t,r)=\frac{1}{2}\dot{R}^{2}(t,r)
\end{equation}
\vspace{-5mm}
\begin{equation}
V(t,r)=-\frac{\frac{1}{2}F(r)}{R(t,r)}=-\frac{m(r)}{R(t,r)}
\end{equation}

as a conserved total energy at a co-ordinate radius $r$, the
kinetic energy and potential energy respectively.\\

The TBL model has two kinds of singularities. The expression
(8.13) shows that the density $\rho$ diverges when either $R=0$ or
$R'=0$. There are known as shell focusing and shell crossing
singularities respectively. A shell focusing singularity ($R=0$)
on a shell of dust occurs when it collapses at or starts
expanding from the centre of matter distribution. Now the moment
$t=t_{sf}(r)$ corresponds to the arrival of the shell with a given
radial co-ordinate $r$ at the centre of matter distribution.
Since the different shells of dust arrive at the centre at
different times, there is always a possibility that any two
shells of dust cross each other at a finite radius in course of
their collapse. This is a situation when the comoving system
breaks down and one encounters the shell crossing singularity
($R'=0$). On the hypersurface $R'=0$ [Seifert, 1979; Hellaby et
al, 1984], both the matter density and Kretchmann scalar diverges.
But beyond the boundary surface it has negative density. So the
region beyond it is unacceptable. It is therefore of interest to
find condition which guarantee that no shell crossing will occur.
A simple solution is to assure that the greater the radius of a
shell, the latter it reaches the centre. In such a situation one
must have $t'_{sf}(r)>0$ to avoid shell crossing.\\

In TBL model there are some important distinction between shell
crossings where only one metric component goes to zero ($g_{rr}$)
and the true big bang and recollapse surfaces where the angular
components go to zero ($g_{\theta\theta}$ and $g_{\phi\phi}$).
The shell crossing surfaces are always timelike whereas the bang
or recollapse surfaces are always spacelike. The frequency shift
of light coming from the shell crossing surfaces is finite (red
or blue) [Hellaby et al, 1984] and quite regular while that coming
from the big bang surface is always infinite. The surface density
$\sigma$ on shell crossing remains finite [Lake, 1984] while on
big bang or recollapse surfaces $\sigma$ diverges. For these
reasons it is believed that shell crossings are not serious
physical singularities but they indicate the breakdown of the
basic choice of comoving co-ordinate system in the Tolman-Bondi model.\\

To investigate whether the naked singularities found by Eardley
and Smarr [1979] and Christodoulou [1984] go against the Cosmic
Censorship Conjecture, Newman [1986] proposes an alternative
formulation to the Cosmic Censorship Conjecture. He replaces the
stability condition present in the original conjecture by the
curvature strength criteria for the singularity. According to
Tipler [1977] a curvature singularity is {\it strong} if any
object hitting it is crushed to zero volume. Newman proposes that
all naked singularities are in some sense gravitationally {\it
weak}. The curvature strength of a singularity may be estimated in
terms of the following {\it limiting focusing conditions} (LFC)
[Clarke and Krolak, 1986]. Now it is convenient to study the
strength of naked singularity, consider a null geodesic affinely
parameterized by $\mu$ which terminates at the singularity where
$\mu=0$. If the Ricci curvature along this geodesic grows as
$\mu^{-2}$ or more rapidly near the singularity the later is said
to be strong curvature in nature. Analytically, define
\begin{equation}
\Psi(\mu)\equiv R_{ab}K^{a}K^{b}
\end{equation}

where $R_{ab}$ is the Ricci tensor and $K^{a}$ is the four
tangent to the geodesic. Then the following focusing conditions:

the LFC
\begin{eqnarray}\begin{array}{ll}
~lim~~\mu \Psi > 0\\
\mu\rightarrow 0
\end{array}
\end{eqnarray}
and the strong LFC
\begin{eqnarray}\begin{array}{ll}
~lim~~\mu^{2} \Psi > 0\\
\mu\rightarrow 0
\end{array}
\end{eqnarray}

may be used to estimate the strength of the singularity. If only
LFC is satisfied the singularity is essentially weak in nature.
Fulfillment of strong LFC is equivalent to the termination of the
geodesic in a strong curvature singularity in the sense of Tipler
[1987]. Newman [1986] has simplified and generalized
Christodoulou's analysis to cover a wide class of Tolman-Bondi
models including both the marginally bound and the time symmetric
cases. He has shown that both the shell focusing and shell
crossing dust naked singularities that were found by Eardley and
Smarr and by Christodoulou are not strong curvature singularities
as defined by Tipler, Clarke and Ellis [1980].\\

For dust, Newman has shown that naked shell focusing
singularities are not strong curvature singularities, whereas
Lake [1988] and independently Ori and Piran [1990] point out that
the singularities found by Ori and Piran [1987] are strong
curvature singularities in the sense of Tipler. The addition of
pressure in the spherical collapse of a perfect fluid should not
lead to a stronger singularity than in the pressureless case. The
problem is resolved by Waugh and Lake [1988] by considering the
singularity structure of self-similar space-time, specifically
the marginally bound Tolman-Bondi case. By a clear and simple
demonstration they showed that when the inhomogeneity parameter
is greater than a critical value, the apparent horizon will
always occur after the Cauchy horizon resulting a globally naked
strong curvature shell focusing singularity.\\

It may be interesting to point out that most of the space-times
that have been studied to investigate the nature of singularity
following a gravitational collapse are self-similar. It becomes
important therefore to give a clear definition of self-similarity
and also to discuss its advantages and disadvantages.
Self-similarity is defined by the existence of a homothetic
killing vector field $\xi^{a}$. By a homothetic killing vector
field we mean that in a local co-ordinate system
\begin{equation}
\xi_{\alpha;\beta}+\xi_{\beta;\alpha}=\frac{1}{2}~\theta
g_{\alpha\beta}
\end{equation}

So along each integral curve of that vector field all the points
are similar to each other, apart from the scale factor that
evolves linearly with the proper length along the integral curve
in question. In specific co-ordinate systems, self-similarity is
manifested by a simple scaling relation for the metric functions.
For instance, a spherical space-time is {\it self-similar} if
there exists a radial area co-ordinate $r$ and an orthogonal time
co-ordinate $t$ for which the metric functions satisfy
\begin{equation}
g_{tt}(ct, cr)=g_{tt}(t, r)
\end{equation}
\begin{equation}
g_{rr}(ct, cr)=g_{rr}(t, r)
\end{equation}
\begin{equation}
R(ct, cr)=c R(t, r)
\end{equation}

for every $c>0$. It is well known that Einstein's equations admit
self-similar solutions. In a spherical self-similar space-time it
is easy to solve the equations of motion for radial null
geodesics and this enables us to study the causal structure of
the space-time. The main difficulty arises when one tries to
employ self-similarity to describe gravitational collapse in the
asymptotic behaviour at infinity. Self-similar solutions are not
asymptotically flat. In other words, in a self-similar solution
the mass of the collapsing object is infinite.\\

Now consider TBL model for bound, marginally bound and unbound
cases ($f<0, f=0$ and $f>0$). For gravitational collapse situation
of dust, one require $\dot{R}(t,r)<0$ and without any loss of
generality rescale $R$ such that
\begin{equation}
R(0,r)=0
\end{equation}

Now partial derivatives of $R$ like $R'$ and $\dot{R}'$ are of
crucial importance in the analysis. From eq.(8.14) solution of $R$
using the initial condition (8.30) have the form

\begin{equation}
t=\frac{r^{3/2}G\left(-\frac{fr}{F}\right)}{\sqrt{F}}\left[1-\left(\frac{R}{r}\right)^{3/2}
\frac{G\left(-\frac{fR}{F}\right)}{G\left(-\frac{fr}{F}\right)}
\right]
\end{equation}

So the shell focusing singularity occurs at the time

\begin{equation}
t_{sf}(r)=\frac{r^{3/2}G\left(-\frac{fr}{F}\right)}{\sqrt{F}}
\end{equation}

The central singularity occurs at $r=0$, the corresponding time
being $t_{0}=t_{sf}(0)$. For the above three cases the expression
for $R',~ \dot{R}'$ have the form

\begin{equation}
R'=(\eta-\beta)P-\left[\frac{1+\beta-\eta}{\sqrt{\lambda+f}}+\left(\eta-\frac{3}{2}\beta\right)\frac{t}{r}
\right]\dot{R}
\end{equation}

\begin{equation}
\dot{R}'=\frac{\beta}{2r}\dot{R}+\frac{\lambda}{2rP^{2}}\left[\frac{1+\beta-\eta}{\sqrt{\lambda+f}}+
\left(\eta-\frac{3}{2}\beta\right)\frac{t}{r}\right]
\end{equation}

where
\begin{equation}
R(t,r)=rP(t,r),~ \eta\equiv\eta(r)=\frac{rF'}{F},~
\beta\equiv\beta(r)=\frac{rf'}{f},~ F(r)=r\lambda(r)
\end{equation}

At this point, the functions $f(r),~ \lambda(r)$ are analytic at
$r=0$ such that $\lambda(0)\ne 0$. It should be pointed out that
when $\lambda(r)=$constant and $f(r)=$constant, the space-time
becomes self-similar in the sense of equations (8.27), (8.28) and
(8.29) for any $c>0$, whereas in general it is non-self-similar.
The condition that $\lambda(0)\ne 0$ implies that $t_{sf}(0)=0$.
It follows that the point $r=0,~t=0$ corresponds to the central
singularity on the $t=0$ hypersurface where the energy density $\rho$ blows up.\\

Now introducing a new variable
\begin{equation}
X=\frac{t}{r}
\end{equation}

the function $P(t,r)=P(X,r)$ is given with the help of (8.31) and
(8.32) by

\begin{equation}
X-\Theta=-\frac{P^{3/2}}{\sqrt{\lambda}}~G\left(-\frac{fP}{\lambda}\right)
\end{equation}

where
$t_{sf}(r)=r\Theta(r)=rG\left(-\frac{f}{\lambda}\right)\sqrt{\lambda}$~.\\

The tangents $K^{r}=\frac{dr}{d\mu}$ and $K^{t}=\frac{dt}{d\mu}$
to the outgoing radial null geodesic with $\mu$ as the affine
parameter satisfy
\begin{equation}
\frac{dK^{t}}{d\mu}+\frac{\dot{R}'}{\sqrt{1+f}}K^{r}K^{t}=0
\end{equation}

\begin{equation}
\frac{dt}{dr}=\frac{K^{t}}{K^{r}}=\frac{R'}{\sqrt{1+f}}
\end{equation}

The purpose is to find whether these geodesics terminate at the
central singularity formed at $r=0,~t=t_{sf}(0)$ in the past. The
exact nature of this singularity $t=0,~r=0$ could be analyzed by
the limiting value of $X=\frac{t}{r}$ at $t=0,~r=0$. If the
geodesics meet the singularity with a definite value of the
tangent then using equation (8.39) and L'Hospitals rule, one can
have

\begin{eqnarray}
\begin{array}{c}
X_{0}~=\\
{}\\
{}
\end{array}
\begin{array}{c}
~~~lim~~~X\\
t\rightarrow 0\\ r\rightarrow 0
\end{array}
\begin{array}{c}
=lim~~~\frac{t}{r}\\
t\rightarrow 0\\ r\rightarrow 0
\end{array}
\begin{array}{c}
=lim~~~\frac{dt}{dr}\\
t\rightarrow 0\\r\rightarrow 0
\end{array}
\begin{array}{c}
=~~~~~lim~~~\frac{R'}{\sqrt{1+f}}\\
t\rightarrow 0\\r\rightarrow 0
\end{array}
\end{eqnarray}

Now use the notation $\lambda_{0}=\lambda(0),~
\beta_{0}=\beta(0),~ f_{0}=f(0)$ and $Q=Q(X)=P(X,0)$, which is a
function of $X$ alone given by equation (8.37)

\begin{equation}
X-\Theta_{0}=-\frac{Q^{3/2}}{\sqrt{\lambda_{0}}}~G\left(-\frac{f_{0}Q}{\lambda_{0}}\right)
\end{equation}

with $\Theta_{0}=\Theta(0)$ and $Q_{0}=Q(X_{0})$. One can
simplify equation (8.40) with the help of equations (8.31),
(8.32), (8.33) and (8.41) as

\begin{equation}
(1-\beta_{0})Q_{0}+
\left[\frac{\beta_{0}}{\sqrt{\lambda_{0}+f_{0}}}+\left(1-\frac{3}{2}\beta_{0}\right)X_{0}
\right]\sqrt{\frac{\lambda_{0}}{Q_{0}}+f_{0}}
-X_{0}\sqrt{1+f_{0}}=0
\end{equation}

If the equation (8.42) has real positive root then the singularity
could be naked. In order to be the end point of null geodesics at
least one real positive value of $X_{0}$ should satisfy equation
(8.42). Clearly if no real positive root of the above is found the
singularity $t=0,~r=0$ is not naked. The existence of a real
positive root of equation (8.42) is a necessary and sufficient
condition for the singularity to be at least locally naked. Such
a singularity could be globally naked as well. The details of
this will depend on the global features of $\lambda(r)$.\\

Now for simplicity, consider $f(r)=0$ and $\lambda_{0}\ne 0$.
From equations (8.32) and (8.41) one may have
\begin{equation}
\Theta_{0}=\frac{2}{3\sqrt{\lambda}},~~
Q(X)=Q_{0}^{2/3}(X-\Theta_{0})^{2/3}
\end{equation}

Equations (8.42) and (8.43) imply that

\begin{equation}
(3\Theta_{0}-X_{0})^{3}=27X_{0}^{3}\Theta_{0}^{2}(\Theta_{0}-X_{0})
\end{equation}

This has real and positive roots if

\begin{equation}
\Theta_{0}\ge \frac{2}{9}\left(\frac{25}{3}+5\sqrt{3}\right)
\end{equation}

and therefore the singularity is naked for which equation (8.45)
is satisfied. The equation (8.44) has two real positive roots.
Such a situation is analyzed by Joshi and Dwivedi [1992] and it is
shown that in this case the singularity will be globally naked.\\

For the TBL model under consideration, using equations (8.12) and
(8.15) the expression of $\Psi$ in (8.23) becomes
\begin{equation}
\Psi\equiv R_{ab}K^{a}K^{b}=\frac{F'~
\left(K^{t}\right)^{2}}{r^{2}P^{2}R'}
\end{equation}

Using L'Hospital's rule and equations (8.33), (8.34), (8.38) and
(8.39), one can see that
\begin{equation}
\begin{array}{c}
{}\\
lim\\
\mu\rightarrow 0\\
\end{array}
\begin{array}{c}
{}\\
\mu^{2}\Psi=\\
{}
\end{array}
\frac{\lambda_{0}X_{0}}{(1+f_{0})^{1/2}(h_{0}-1)^{2}Q_{0}^{2}}>0
\end{equation}

where $h_{0}$ is a related with the constants $f_{0},~
\lambda_{0},~ \beta_{0},~ Q_{0}$ and $X_{0}$.\\

So equation (8.47) satisfies the strong limiting focusing
condition (8.25). For this reasons, the naked singularity at
$t=0,~ r=0$ is strong curvature singularity.\\

Now to determine naked singularity in TBL model, one should
discuss another approach of geodesic study by Barve, Singh, Vaz
and Witten [1999]. From equations (8.16) and (8.17) the solution
of $R$ for marginally bound case ($f=0$) is

\begin{equation}
R^{3/2}=r^{3/2}-\frac{3}{2}\sqrt{F}~t
\end{equation}

It follows from the equation (8.13) that the function $F(r)$
becomes fixed once the initial density distribution
$\rho(0,r)=\rho(r)$ is given i.e.,

\begin{equation}
F(r)=\int \rho(r) r^{2}dr
\end{equation}

If the initial density $\rho(r)$ has a series expansion

\begin{equation}
\rho(r)=\rho_{0}+\rho_{1}r+\rho_{2}r^{2}+\rho_{3}r^{3}+......
\end{equation}

near the centre $r=0$, the resulting series expansion for the mass
function $F(r)$ is

\begin{equation}
F(r)=F_{0}r^{3}+F_{1}r^{4}+F_{2}r^{5}+F_{3}r^{6}+......
\end{equation}

where $F_{j}=\frac{\rho_{j}}{j+3}$ and $j=0,1,2,3,...$.\\

The central singularity at $r=0$ forms at the time

\begin{equation}
t_{0}=t_{sf}(r)=\frac{2}{3\sqrt{F_{0}}}
\end{equation}

From equation (8.17) the singularity time at $r$, near $r=0$
approximately written as

\begin{equation}
t_{sf}(r)=t_{0}-\frac{F_{m}}{3F_{0}^{3/2}}~r^{m}
\end{equation}

Here $F_{m}$ is the first non-vanishing term beyond $F_{0}$ in
equation (8.51). The first non-vanishing derivative in the series
expansion in (8.50) should be negative and the density functions
decreases as one moves out from the centre. So $F_{m}<0$. From
equation (8.53) it is to be seen that $t_{sf}(r)>t_{0}$\\

To examine the singularity at $t=t_{0},~ r=0$ is naked i.e.,
there are one or more outgoing null geodesics which terminate in
the past at the central singularity, only radial null geodesic is
considered. Let the geodesic have the form

\begin{equation}
t=t_{0}+a~r^{\alpha}
\end{equation}

to leading order in ($t,~r$) plane where $a,~\alpha>0$. Now
comparing (8.53) and (8.54) one may have $\alpha\ge m$ and if
$\alpha=m$ then $a<-\frac{F_{m}}{3F_{0}^{3/2}}$.\\

Now for TBL metric, an outgoing null geodesic must satisfy the
equation

\begin{equation}
\frac{dt}{dr}=R'
\end{equation}

Consider first $\alpha>m$: In this case, equations (8.48), (8.52),
(8.54) and (8.55) near $r=0$ give the results

\begin{equation}
a\alpha
r^{\alpha-1}=\left(1+\frac{2m}{3}\right)\left(-\frac{F_{m}}{2F_{0}}\right)^{2/3}
r^{\frac{2m}{3}}
\end{equation}

which implies $\alpha=1+\frac{2m}{3},~
a=\left(-\frac{F_{m}}{2F_{0}}\right)^{2/3}$. So for $\alpha>m$, ~
$m$ takes values 1, 2 only. Hence the singularity is naked for
$m=1$ and 2 i.e., for the models $\rho_{1}<0$ and
$\rho_{1}=0,~\rho_{2}<0$. There is at least one outgoing geodesic
given by (8.54) which terminates in the central singularity in the
past. If $m>3$ then the condition $\alpha>m$ cannot be satisfied
and the singularity is covered.\\

Consider next that $\alpha=m$. In this case also equations (8.48),
(8.52), (8.54) and (8.55) give the result

\begin{equation}
mar^{m-1}=\left(-\frac{3}{2}~a\sqrt{F_{0}}-\frac{(2m+3)}{6}\frac{F_{m}}{\sqrt{F_{0}}}
\right)\left(-\frac{F_{m}}{2F_{0}}-\frac{3}{2}~a\sqrt{F_{0}}\right)^{-\frac{1}{3}}r^{\frac{2m}{3}}
\end{equation}

which implies $m=3$ and $a$ satisfies the biquadratic equation

\begin{equation}
12\sqrt{F_{0}}~a^{4}-\left(-\frac{4F_{3}}{F_{0}}+F_{0}^{3/2}\right)a^{3}-3F_{3}a^{2}-
\frac{3F_{3}^{2}}{F_{0}^{3/2}}~a-\frac{F_{3}^{3}}{F_{0}^{3}}=0
\end{equation}

By substituting $b=\frac{a}{F_{0}},~
\xi=\frac{F_{3}}{F_{0}^{5/2}}$, this equation can be written as

\begin{equation}
4b^{3}(3b+\xi)-(b+\xi)^{3}=0
\end{equation}

The singularity will be naked if this equation admits one or more
positive roots for $b$ which satisfies the constraint
$b<-\frac{\xi}{3}$ and also $\xi$ satisfies the condition $\xi\le
-25.9904$. Now to investigate whether there is an entire family
of radial null geodesics which terminate at the naked
singularity, a solution for the geodesics correct to one order
beyond the solution (8.54), is assumed
\begin{equation}
t=t_{0}+a~r^{\alpha}+d~r^{\alpha+\beta}
\end{equation}

where $d,~\beta>0$. In this case, $\alpha,~a$ are same as in
before and $d$ is totally arbitrary. There will be an entire
family of outgoing null geodesics terminating at the singularity
provided $\beta$ is non-negative. It is essential that $\beta$ be
non-negative, otherwise these geodesics will not lie in the
space-time.\\

The normal vector to the boundary of any hypersphere $R-R_{0}=0$
is given by $l_{\mu}=(\dot{R},R',0,0)$. In order that $l_{\mu}$
is a null vector and the boundary of this hypersphere is a null
surface, so
\begin{equation}
R(t_{ah}(r),r)=F(r)
\end{equation}

where $t=t_{ah}(r)$ is the equation of apparent horizon which
marks the boundary of the trapped region. If the apparent horizon
starts developing earlier than the epoch of singularity formation
then the event horizon can fully cover the strong gravity regions
including the final singularity which will thus be hidden within
a black hole. On the other hand, if trapped surfaces form
sufficiently later during the evolution of collapse then it is
possible for the singularity to communicate with outside
observers. That means, naked singularity or black hole appears if
$t_{ah}(r)\ge t_{0}$ ~or~ $t_{ah(r)}<t_{0}$ respectively near
$r=0$.\\

From equations (8.48) to (8.53) and (8.61), apparent horizon time
is given by

\begin{equation}
t_{ah}(r)=t_{0}-\frac{2}{3}F_{0}r^{3}-\frac{F_{m}}{2F_{0}^{3/2}}~r^{m}+......
\end{equation}

For $m<3$, equation (8.62) shows that $t_{ah}(r)>t_{0}$ and so
naked singularity forms. For $m=3$, naked singularity or black
hole forms if $F_{3}<-2F_{0}^{5/2}$ or $F_{3}\ge -2F_{0}^{5/2}$
respectively. For $m\ge 4$, equation (8.62) shows that
$t_{ah}(r)\le t_{0}$ and so a black hole forms.\\

The expression for shear scalar is given by

\begin{equation}
\sigma^{2}=\frac{1}{3}\left(\frac{\dot{R}'}{R'}-\frac{\dot{R}}{R}\right)^{2}=
\frac{m^{2}F_{m}^{2}}{12F_{0}}\left[1-3F_{0}^{1/2}t+\frac{9}{4}F_{0}t^{2}
\right]^{-1}~r^{2m}+......
\end{equation}

As expected, the shear diverges as one approach to the
singularity.\\

\subsection{Perfect Fluid Collapse}

The perfect fluid matter model is one of the most natural ways to
introduce matter pressure. If the pressure is bounded from above,
there can appear a shell crossing naked singularity [Muller Zum
Hagen et al, 1974] which is gravitationally weak. There can
appear a strong curvature naked singularity even with unbounded
pressure.\\

The examples presented before are all based on collapse of
pressureless matter based on TBL model and one may reasonably
think that these naked singularities will not be covered by an
event horizon when pressure is introduced. Also from physical
point of view the assumption that matter is dust like is never
admissible near the singularity, since $\rho\rightarrow\infty$
and one should use the ultra-relativistic equation of state
$p=\frac{1}{3}~\rho$. However, it appears that the general
character of the limiting laws of compression are to a large
extent independent of the equation of state of the matter
[Lifshitz et al, 1960, 1961]. Ori and Piran [1987] have
investigated the self-similar collapse of a perfect fluid with an
equation of state $p=(\gamma-1)\rho$. They have shown that if the
equation of state is soft enough i.e., $\gamma-1\ll 1$, a naked
singularity may appear. In another work, Ori and Piran [1990]
used the self-similarity to model spherically symmetric
gravitational collapse of compact objects with a reasonable
equation of state. It is shown that a $p=(\gamma-1)\rho$ equation
of state is compatible with self-similarity. They show that in a
significant part of the space of self-similar solutions, there is
a globally naked central singularity from which null geodesics
emerge to infinity.\\

For $1<\gamma\le 1.4$, they found numerically that there is a
discrete set of self-similar solutions that allow analytic
initial data beyond a `sonic point'. These solutions can be
labeled by the number of zeros in the velocity field of world
lines of constant circumferential radius relative to the fluid
element. There exists a pure collapse solution among these
self-similar solutions which they call the general relativistic
Larson-Penston solution. They showed that a central naked
singularity forms in this Larson-Penston solution for
$1<\gamma\lesssim 1.0105$. They also showed that there are naked
singular solutions with oscillations in the velocity field for
$1<\gamma\le 1.4$. The results of this work were confirmed and
extended to $1<\gamma\le 25/16$ [Foglizzo et al, 1993]. This naked
singularity is ingoing null. It was shown that this naked
singularity satisfies both the limiting focusing condition and
the strong curvature condition for the first null ray [Waugh et
al, 1988, 1989].\\

In order to judge the necessity of self-similarity assumption,
Harada [1998] numerically simulated the spherically symmetric and
adiabatic gravitational collapse of perfect fluid with the same
equation of state $p=(\gamma-1)\rho$ without this assumption.
Since null co-ordinates were used in these simulations, he could
detect naked singularities not relying upon the absence of an
apparent horizon. The result was that naked singularities develop
from rather generic initial data sets for $1<\gamma\lesssim
1.01$, which is consistent with the result of Ori and Piran
obtained using the self-similarity assumption. In fact, through
numerical simulations by Harada and Maeda [2001], it was found
that generic spherical collapse approaches the Larson-Penston
self-similar solution in the region around the centre, at least
for $1<\gamma\le 1.03$. This finding is supported by a linear
stability analysis of self-similar solutions. Although the final
fate of the generic spherical collapse of a perfect fluid with
larger values of $\gamma$ is not known, Harada [2001]
analytically showed that the Larson-Penston self-similar solution
is no longer stable. Recently Ghosh et al [2002, 2003], Joshi et
al [2002] and Goswami et al [2002, 2003] gave the solutions for
perfect fluid and they have shown that perfect fluid halts
collapse.

\subsection{Other Examples on Spherical Collapse}

Critical behaviour in gravitational collapse was discovered by
Choptuik [1993]. He found so-called critical behaviour, such as a
scaling law for the formed black hole mass. He also found that
there is a discrete self-similar solution that sits at the
threshold of black hole formation which is called a {\it critical
solution}. Similar phenomena have been observed in the collapse
of various kinds of matter fields, for example, axisymmetric
gravitational waves [Abrahams et al, 1993], radiation fluid
[Evans et al, 1994] and more general perfect fluids [Neilsen et
al, 2000, 2001]. One interesting example of spherical collapse is
a spherical self-gravitating system of counter rotating particles
i.e., an Einstein cluster. One should mention several additional
examples of naked singularities in type I and type II matter. If
one consider imploding ``null dust'' into the centre, the
space-time is given by the Vaidya metric [Vaidya, 1943, 1951]. For
this space-time, it has been shown that naked singularity
formation is possible from regular initial data [Ghosh et al,
2000, 2001; Beesham et al, 2003]. There have been many analysis
with a set of assumptions and the conditions for the appearance
of a naked singularity are written down in terms of the energy
density, radial stress and tangential stress [Goncalves et al,
2001, 2002]. In these analysis, in general the conclusion is that
naked singularities are possible for generic matter fields that
satisfy some energy conditions. In this approach, this conclusion
is very natural because there remains great freedom in the choice
of matter fields, even if some energy condition is imposed.\\

\section{Non-Spherical Collapse}

There is very little progress in studying non-spherical collapse.
The {\it quasi-spherical} dust collapse models given by Szekeres
metric [Szekeres, 1975] were analyzed by Szekeres himself [1975].
It has been shown that shell focusing singularities are possible
in the solution and the conditions necessary for the appearance of
naked singularities are very similar to those in the case of
spherical dust collapse [Joshi and Krolak, 1996; Goncalves, 2001].
The global visibility of this singularity was also examined
[Deshingkar et al, 1998]. Now the basic difficulty is the
ambiguity of horizon formation in non-spherical geometries and the
influence of gravitational radiation. Schoen and Yau [1983]
proposed a sufficient criterion for the formation of trapped
surfaces in an arbitrary space-time but it fails to say anything
about the conditions which lead to the formation of naked
singularities. This problem has been restated by Thorne [1972] in
the form of a conjecture known as {\it hoop conjecture} which
states that {\it ``horizon form when and only when a
gravitational mass $M$ gets compacted into a region whose
circumference in every direction is $C \lesssim 4\pi M$''}. He
analyzed the causal structure of cylindrically symmetric
space-times and found remarkably different nature from
spherically symmetric space-times. The event horizon cannot exist
in cylindrically symmetric space-time. Therefore, if an
infinitely long cylindrical object gravitationally collapses to a
singularity, it becomes a naked singularity, not a black hole
covered by an event horizon. Then it is natural to ask what
happens if a very long but finite object collapses. The hoop
conjecture was derived from such a thought experiment. It should
be noted that the hoop conjecture itself does not assert that a
naked singularity will not appear. It is a conjecture regarding
the necessary and sufficient condition for black hole formation.\\

Many examples of naked singularity formation are obtained from
the analysis of spherically symmetric space-time. These examples
are consistent with the hoop conjecture. The circumference $C$
corresponding to a radius $R$ centered at the symmetric centre is
$2\pi R$. For a central naked singularity of a spherically
symmetric space-time, the ratio of $C$ to the gravitational mass
$M$ within $R$ can be estimated as

\begin{eqnarray}
\begin{array}{c}
~~~~~~lim~~~\frac{2\pi R}{4\pi M}\\
r\rightarrow 0
\end{array}
\begin{array}{c}
=~~lim~~~\frac{R}{2M}\\
r\rightarrow 0
\end{array}
\begin{array}{c}
=~~~~lim~~~\frac{r^{\alpha}}{2M_{c}r^{3}}\\
r\rightarrow 0
\end{array}
\begin{array}{c}
>1\\
{}
\end{array}
\nonumber
\end{eqnarray}

where $M_{c}$ is a constant and $1<\alpha\le 3$. In this case the
condition $C>4\pi M$ coincides with the condition that the centre
is not trapped. Therefore the appearance of a central naked
singularity in spherical gravitational collapse does not
constitute a counter example to the hoop conjecture.\\

As for the non-spherically symmetric case several studies support
the hoop conjecture. A pioneering investigation is the numerical
simulation of general relativistic collapse of axially symmetric
stars [Nakamura et al, 1982]. From initial conditions that
correspond to a very elongated prolate fluid with sufficiently
low thermal energy, the fluid collapses with this elongated form
maintained. In the numerical simulation, black hole was not
found. This result suggests that such an elongated object might
collapse to a naked singularity and in this case it would provide
evidence supporting the hoop conjecture.\\

Shapiro and Teukolsky [1991, 1992] numerically studied the
evolution of a collisionless gas spheroid with fully general
relativistic simulations. They found some evidence that a prolate
spheroid with a sufficiently elongated initial configuration and
even with a small angular momentum, might form naked singularity.
They also found that when a spheroid is highly prolate, a spindle
singularity forms at the pole and the spindle might be a naked
singularity. Subsequently, there were attempts namely,
gravitational radiation emission in aspherical collapse [Nakamura
et al, 1993], analytical studies of prolate collapsing spheroids
[Barrabes et al, 1991; Pelath et al, 1998] and others [Harada et
al, 1998; Iguchi et al, 1999, 2000] to prove or disprove the
conjecture. Interestingly, all of them either confirmed or failed
to refute the conjecture.\\

It should be noted that there are problems in the hoop conjecture
to prove it as a mathematically unambiguous theorem, such as how
one should define the mass of the object and the length of the
hoop. Although nobody have found a counter example to the hoop
conjecture, it is necessary to obtain a complete solution to
these problems.

\large \baselineskip .85cm

\chapter{Naked Singularities in Higher Dimensional Spherical Gravitational Collapse}
\label{chap9}\markright{\it CHAPTER~\ref{chap9}. Naked
Singularities in Higher Dimensional Spherical Gravitational
Collapse}

\section{Prelude}
The Cosmic Censorship Conjecture [Penrose, 1969, 1979] is yet one
of the unresolved problems in General Relativity. It states that
the space-time singularity produced by gravitational collapse
must be covered by the horizon. However the singularity theorems
as such do not state anything about the visibility of the
singularity to an outside observer. In fact the conjecture has
not yet any precise mathematical proof. Several models related to
the gravitational collapse of matter has so far been constructed
where one encounters a naked singularity [Waugh et al, 1988;
Lemos, 1991, 1992; Joshi et al, 1999; IIha et al, 1997,
1999; Rocha et al, 2000].\\

It has recently been pointed out by Joshi et al [2002] that the
physical feature which is responsible for the formation of naked
singularity is nothing but the presence of shear. It is the shear
developing in the gravitational collapse, which delays the
formation of the apparent horizon so that the communication is
possible from the very strong gravity region to observers
situated outside. Joshi et al [2002] have analyzed in details how
the presence of shear determines the growth and evolution of an
inhomogeneous dust distribution represented by four dimensional
Tolman-Bondi metric and have attempted to clarify the nature of
singularities as the final outcome.\\

The objective of this chapter is to fully investigate the
situation in the background of higher dimensional space-time with
both non-marginally and marginally bound collapse [Banerjee et al,
1994; Ghosh et al, 2000, 2001; Deshingkar et al, 1999]. As it is
only to be expected, in one way or another, these works all deal
with propagation of null geodesics in the space-time of a
collapsing dust [Barve et al, 1999]. In this context we mention
that Ghosh and Beesham [2000, 2001] have also studied dust
collapse for (n+2) dimensional Tolman-Bondi space-time for
marginally bound case ($f=0$), considering the self-similar
solutions. They have concluded that higher dimensions are
favourable for black holes rather than naked singularities. Also
recently, Ghosh and Banerjee [2002] have considered non-marginal
case ($f\ne 0$) for dust collapse in 5D Tolman-Bondi model and
have shown that the degree of inhomogeneity of the collapsing
matter
is necessary to form a naked singularity. \\

In this chapter, we show that in non-marginally bound collapse
(i.e., for $f\ne 0$), the naked singularity may appear in any
dimensional space-time, but for marginally bound collapse the
naked singularity may appear only when the space-time has
dimensions upto five. In more than five dimensions and marginally
bound case $t_{ah}<t_{0}$ that is the apparent horizon at any
point in the region forms earlier than the central shell focusing
singularity, which indicates that the shell focusing singularity
first appearing at $r=0$ remains hidden behind the apparent
horizon and so gives rise to a black hole always. Also we study
gravitational collapse for non-marginally bound case considering
non self-similar solutions and it is generalization to higher
dimension of the Dwivedi and Joshi [1992]. The geodesic equations
can not completely solved due to the presence of the complicated
hypergeometric function. So we present numerical results which
favour the formation of naked singularity in any dimension.\\

\section{Basic Equations in Higher Dimensional TBL Model}
The ($n+2$) dimensional Tolman-Bondi type metric is given by

\begin{equation}
ds^{2}=e^{\nu}dt^{2}-e^{\lambda}dr^{2}-R^{2}d\Omega^{2}_{n}
\end{equation}

where $\nu,\lambda,R$ are functions of the radial co-ordinate $r$
and time $t$ and $d\Omega^{2}_{n}$ represents the metric on the
$n$-sphere. Since we assume the matter in the form of dust, the
motion of particles will be geodesic allowing us to write
\begin{equation}
e^{\nu}=1
\end{equation}
Using comoving co-ordinates one can in view of the field
equations [Banerjee et al, 1994] arrive at the following relations
in ($n+2$) dimensional space-time

\begin{equation}
e^{\lambda}=\frac{R'^{2}}{1+f(r)}
\end{equation}
and
\begin{equation}
\dot{R}^{2}=f(r)+\frac{F(r)}{R^{n-1}},
\end{equation}

where the function $f(r)$ classifies the space-time [Joshi, 1993]
as bound, marginally bound and unbound depending on the range of
its values which are respectively
$$
f(r)<0,~~~f(r)=0,~~~f(r)>0.
$$

The function $F(r)$ can be interpreted as the mass function which
is related with the mass contained within the comoving radius $r$.

The energy density $\rho(t,r)$ is therefore given by

\begin{equation}
\rho(t,r)=\frac{nF'(r)}{2R^{n}R'}
\end{equation}

The above models are characterized by the initial data specified
on the initial hypersurface $t=t_{i}$, from which the collapse
develops. As it is possible to make an arbitrary relabeling of
spherical dust shells by $r\rightarrow g(r)$, without loss of
generality we fix the labeling by the choice $R(t_{i},r)=r$, so
that initial density distribution is given by
\begin{equation}
\rho_{i}(r)=\frac{n F'}{2r^{n}}
\end{equation}

Now the curve $t=t_{sf}(r)$ defines the shell-focusing singularity
and is characterized by

\begin{equation}
R(t_{sf}(r),r)=0
\end{equation}

It indicates that the shell focusing singularity occurs at
different $r$ at different epochs. So the collapse in this case
is not simultaneous in comoving co-ordinates and is in variance
with that in the homogeneous dust model [IIha et al, 1997, 1999;
Chatterjee et al, 1990].

Further, within the collapsing cloud, the trapped surfaces will
be formed due to unbounded growth of the density and these
trapped surfaces are characterized by the outgoing null
geodesics. We observe that the normal vector to the boundary of
any hypersphere $R-R_{0}=0$ is given by
$l_{\mu}=(\dot{R},R',0,0,0,......)$. In order that $l_{\mu}$ is a
null vector and the boundary of this hypersphere is a null surface
we must have

\begin{equation}
R(t_{ah}(r),r)=[F(r)]^{\frac{1}{n-1}}
\end{equation}

where $t=t_{ah}(r)$ is the equation of apparent horizon which
marks the boundary of the trapped region.

To characterize the nature of the singularity we shall discuss
the two possibilities namely, $(i)$ $t_{ah}<t_{sf}(0)$ and $(ii)$
$t_{ah}>t_{sf}(0)$. The first case may correspond to formation of
black hole while the second one may lead to naked singularity. If
the apparent horizon will form earlier than the instant of the
formation singularity, then the event horizon can fully cover the
strong gravity region and also the singularity. As a result, no
light signal from the singularity can reach to any outside
observer and the singularity is totally hidden within a black
hole. On the other hand, in the second case the trapped surfaces
will form much later during the evolution of the collapse and it
is possible to have a communication between the singularity and
external observers.\\

\section{Marginally Bound Case}
In this case (i.e., $f(r)=0$) equation (9.4) can be integrated out
easily to give

\begin{equation}
R^{\frac{n+1}{2}}=r^{\frac{n+1}{2}}-\frac{n+1}{2}\sqrt{F(r)}~(t-t_{i})
\end{equation}

where we have used the initial condition $R(t_{i},r)=r$.

Combining (9.8) and (9.9) along with the initial condition
$R(t_{i},r)=r$ we arrive at the relation
\begin{equation}
t_{ah}(r)-t_{i}=\frac{2}{n+1}r^{\frac{n+1}{2}}F^{^{-1/2}}-\frac{2}{n+1}F^{\frac{1}{n-1}}
\end{equation}

Since our study is restricted to the region near $r=0$, the mass
function $F(r)$ should vanishes exactly at $r=0$ and it is
possible to express $F(r)$ as a polynomial function of $r$ near
the origin. One should at the same time keep in mind that
initially the central density is regular. Further from physical
considerations one may argue that $\rho_{i}'(r)$ should vanish
exactly at $r=0$, but is negative in the neighbouring region. All
these consideration lead to the following expressions for $F(r)$
and the initial energy density $\rho_{i}(r)$,
\begin{equation}
F(r)=F_{0}r^{n+1}+F_{2}r^{n+3}+F_{3}r^{n+4}.........
\end{equation}
and
\begin{equation}
\rho_{i}(r)=\rho_{0}+\rho_{1}r+\rho_{2}r^{2}+.........
\end{equation}

with ~~$\rho_{j}=\frac{(n+j-1)(n-2)}{2}F_{j},~~~j=0,1,2,...$~.
Obviously the initial central density is given by
$\rho_{c}=\frac{n(n+1)}{2}F_{0}$. Since $\rho_{i}'(r)<0$ in the
region $r\approx 0$ one must have $F_{2}<0$.

Now in view of (9.7) and (9.9) one can write

\begin{equation}
t_{sf}(r)-t_{i}=\frac{2}{(n+1)}\frac{r^{(n+1)/2}}{\left[F_{0}r^{n+1}+F_{2}r^{n+3}+...\right]^{1/2}}
\end{equation}

If we now denote $t_{0}=t_{sf}(0)$ as the instant of first shell
focusing singularity to occur at the centre $r=0$, the relation
(9.13) can be used to obtain the following relation

\begin{equation}
t_{0}-t_{i}=\frac{2}{(n+1)F_{0}^{1/2}},
\end{equation}

which in combination with the relation (9.10) yields

\begin{eqnarray*}
t_{ah}(r)-t_{0}=-\left[\frac{1}{n+1}\frac{F_{2}}{F_{0}^{3/2}}r^{2}+
\frac{1}{n+1}\frac{F_{3}}{F_{0}^{3/2}}r^{3}+\frac{1}{n+1}\frac{F_{4}}{F_{0}^{3/2}}r^{4}+...\right]
\end{eqnarray*}
\vspace{-3mm}
\begin{equation}
-\left[\frac{2}{n+1}F_{0}^{\frac{1}{n-1}}r^{\frac{n+1}{n-1}}+
\frac{2}{n^{2}-1}\frac{F_{2}}{F_{0}^{\frac{n-2}{n-1}}}r^{\frac{3n-1}{n-1}}+
\frac{2}{n^{2}-1}\frac{F_{3}}{F_{0}^{\frac{n-2}{n-1}}}r^{\frac{4n-2}{n-1}}+...\right]
\end{equation}
\\
The relation (9.15) is valid in general $(n+2)$ dimensional
space-time and reduces to the 4 dimensional expression $(n=2)$ in
the region near the centre as given in the paper of Joshi et al
[2002]. In $5D$ space-time the naked singularity appears only if
$F_{2}\ne 0$ and further $|F_{2}|>2F_{0}^{2}$. Otherwise for
either $F_{2}=0$ or $|F_{2}|<2F_{0}^{2}$ there must occur a black
hole, because in that case $t_{ah}(r)<t_{0}$ that is, the horizon
appears earlier than the shell focusing singularity at $r=0$.
Only in a very special case of $|F_{2}|=2F_{0}^{2}$ the occurrence
of naked singularity or black hole will depend on the
co-efficients of higher powers of $r$. For general $(n+2)$
dimensions the close examination of (9.15) reveals that if there
is to exist a naked singularity we must have

\begin{equation}
\frac{n+1}{n-1}\geq 2,
\end{equation}

which means $n\leq 3$. It is interesting to note that $\left(
\frac{n+1}{n-1}\right)$ has the value 2 when $n=3$ and then it
decreases monotonically with the increasing number of dimensions.
So for space-time with dimension larger than five the term
$\frac{2}{n+1}F_{0}^{\frac{1}{n-1}}r^{\frac{n+1}{n-1}}$ dominates
leading to a black hole. Our conclusion is that for the space-time
having larger than five dimensions the existence of the naked
singularity is prohibited or in other words the shell focusing
singularity is fully covered by the horizon in such cases.
However, if we relax the restriction namely, $\rho'_{i}(r)=0$ at
$r=0$ then $F_{1}$ will be non-zero and this will leads to the
possibility of naked singularity in any dimension.\\

\section{Non-Marginally Bound Case}
In this case (i.e., $f(r)\ne 0$), integrating equation (9.4) with
the help of initial condition $R(t_{i},r)=r$, we have the form of
$R$ as

\begin{equation}
t-t_{i}=\frac{2}{(n+1)\sqrt{F}}\left[r^{\frac{n+1}{2}}~_{2}F_{1}[\frac{1}{2},a,a+1,-\frac{f
r^{n-1}}{F}]-R^{\frac{n+1}{2}}~_{2}F_{1}[\frac{1}{2},a,a+1,-\frac{f
R^{n-1} }{F}]\right]
\end{equation}

where $_{2}F_{1}$ is the usual hypergeometric function with
$a=\frac{1}{2}+\frac{1}{n-1}$. Using equations (9.7) and (9.8)
separately in equation (9.17) we have

\begin{equation}
t_{sf}(r)-t_{i}=\frac{2}{(n+1)\sqrt{F}}r^{\frac{n+1}{2}}~_{2}F_{1}[\frac{1}{2},a,a+1,-\frac{f
r^{n-1}}{F}]
\end{equation}
and
\begin{equation}
t_{ah}(r)-t_{i}=\frac{2r^{\frac{n+1}{2}}}{(n+1)\sqrt{F}}~_{2}F_{1}[\frac{1}{2},a,a+1,-\frac{f
r^{n-1}}{F}]-\frac{2F^{\frac{1}{n-1}}}{n+1}~_{2}F_{1}[\frac{1}{2},a,a+1,-f]
\end{equation}
\\
The only new input in the present case is the function $f(r)$,
which may be expressed as a power series in $r$ near the centre
$r=0$.

We assume
\begin{equation}
f(r)=f_{0}r^{2}+f_{1}r^{3}+f_{2}r^{4}+.........
\end{equation}

The form (9.20) is chosen because $f(r)$ vanishes as $r\rightarrow
0$, which is demanded by the regularity condition at $r=0$
[Harada et al, 2000, 2002].

From equation (9.18) we have the central shell focusing
singularity time

\begin{equation}
t_{0}=t_{i}+\frac{2}{(n-1)\sqrt{F_{0}}}~_{2}F_{1}[\frac{1}{2},a,a+1,-\frac{f_{0}
}{F_{0}}]
\end{equation}

Therefore, equation (9.19) and (9.21) give (after using (9.11) and
(9.20) in (9.19))

\begin{eqnarray*}
t_{ah}(r)-t_{0}=-\frac{2}{n+1}F_{0}^{\frac{1}{n-1}}r^{\frac{n+1}{n-1}}
-\frac{f_{1}}{(3n-4)F_{0}^{3/2}}~_{2}F_{1}[\frac{3}{2},a+1,a+2,-\frac{f_{0}
}{F_{0}}]r+\left[-\frac{F_{2}}{(n+1)F_{0}^{3/2}}\right.
\end{eqnarray*}
\vspace{-4mm}
\begin{eqnarray*}
\left._{2}F_{1}[\frac{1}{2},a,a+1,-\frac{f_{0}
}{F_{0}}]+\frac{1}{4(n-1)(3n-4)f_{0}F_{0}^{5/2}(f_{0}+F_{0})}\left\{(4-3n)f_{0}^{2}F_{0}^{2}
~_{2}F_{1}[\frac{1}{2},a,a+1,-\frac{f_{0} }{F_{0}}]\right.\right.
\end{eqnarray*}
\vspace{-4mm}
\begin{eqnarray*}
\left.\left.+\{
-2f_{0}F_{0}((3-2n)f_{1}^{2}+2(n-1)f_{2}F_{0})-4(n-1)f_{0}^{2}F_{0}(f_{2}-F_{2})
+4(n-1)f_{0}^{3}F_{2}\right.\right.
\end{eqnarray*}
\vspace{-4mm}
\begin{equation}
\left.\left.+(3n-4)f_{1}^{2}F_{0}^{2} \}
~_{2}F_{1}[\frac{3}{2},a+1,a+2,-\frac{f_{0} }{F_{0}}] \right\}
\right]r^{2}+O(r^{3})
\end{equation}
\\
From the above relation we note that near $r=0$, $t_{ah}(r)> or
\le 0$ according as
$$f_{1}~_{2}F_{1}[\frac{3}{2},a+1,a+2,-\frac{f_{0}}{F_{0}}]< or \ge
0 .$$ Though the hypergeometric function has arguments depending
on $n$ yet $_{2}F_{1}$ is always positive. So the above
restriction simply implies $f_{1}< or \ge 0$. Now according to
Joshi etal, if a comoving observer (at fixed $r$) does not
encounter any trapped surfaces until the time of singularity
formation then it is possible to visualize the singularity
otherwise the singularity is covered by trapped surfaces, leading
to a black hole. Thus $f_{1}<0$ leads to naked singularity while
$f_{1}\ge 0$ is the restriction for black hole. However, the
above conditions are not sufficient, particularly for non-smooth
initial density profile. Therefore, we may have locally naked
singularity or black hole depending on the sign of $f_{1}$.\\

{\bf Calculation of shear near $r=0$:}\\

We still consider the non-marginally bound case $f(r)\ne 0$. The
shear scalar in the $(n+2)$ dimensional spherically symmetric dust
metric so far discussed in this section may be estimated by the
factor
$\sigma=\sqrt{\frac{n}{2(n+1)}}\left(\frac{\dot{\lambda}}{2}-
\frac{\dot{R}}{R}\right)$. Using (9.3) and (9.4) one gets in turn

\begin{equation}
\sigma=\sqrt{\frac{n}{2(n+1)}}\left(\frac{\dot{R}'}{R'}-\frac{\dot{R}}{R}\right)
=\sqrt{\frac{n}{8(n+1)}}~\frac{\left[\{R F'-(n+1)R' F \}+R^{n-1}(R
f'-2R' f)\right]}{R^{\frac{n}{2}}R'\left(F+f R^{n-1}
\right)^{1/2}}
\end{equation}

Since at the initial hypersurface ($t=t_{i}$) we have chosen
$R(t_{i},r)=r$, so the initial shear $\sigma_{i}$ is of the form

\begin{equation}
\sigma=\sqrt{\frac{n-2}{8(n-1)}}~\frac{\left[\{r F'-(n-1)F
\}+r^{n-3}(r f'-2f)\right]}{r^{\frac{n-1}{2}}\left(F+f r^{n-3}
\right)^{1/2}}
\end{equation}

Thus using the power series expansions (9.11) and (9.20) for
$F(r)$ and $f(r)$ in the above expression for $\sigma_{i}$ we have

\begin{eqnarray*}
\sigma_{i}=\sqrt{\frac{n-2}{8(n-1)}}~\frac{f_{1}r+\sum_{m=2}^{\infty}m(f_{m}+F_{m})r^{m}}
{\sqrt{(f_{0}+F_{0})+f_{1}r+\sum_{m=2}^{\infty}m(f_{m}+F_{m})r^{m}}}
\end{eqnarray*}
\vspace{1mm}
\begin{equation}
\hspace{1in}=\sqrt{\frac{n-2}{8(n-1)}}~\frac{1}{\sqrt{f_{0}+F_{0}}}~\left[f_{1}r+
\left\{2(f_{2}+F_{2})-\frac{f_{1}^{2}}{2(f_{0}+F_{0})}\right\}r^{2}+O(r^{3})\right]
\end{equation}

For marginally bound case $f(r)=0$ i.e.,
$f_{0}=f_{1}=f_{2}=...=0$. So the shear in this case becomes
\begin{equation}
\sigma=\sqrt{\frac{n}{8(n+1)}}\frac{\sum_{m=2}^{\infty}m
F_{m}r^{m}}{F_{0}^{1/2}\left[1+\frac{(n+1)^{2}}{4}F_{0}(t-t_{i})^{2}
-(n+1)F_{0}^{1/2}(t-t_{i})\right]}
\end{equation}
\\
and the initial shear at $t=t_{i}$ is given by

\begin{equation}
\sigma_{i}=\sqrt{\frac{n}{8(n+1)}}\frac{\sum_{m=2}^{\infty}m
F_{m}r^{m}}{F_{0}^{1/2}}
\end{equation}

We note that the initial shear (9.25) vanishes when $f_{1}=0,
(F_{2}+f_{2})=(F_{3}+f_{3})=......=0$ and hence even if the
initial shear is zero the dust distribution may be inhomogeneous.
It is interesting to note that  the dependence on $r$ of the
initial shear does not depend on the number of dimensions of the
space-time. In fact the expression (9.27) coincides exactly with
the value of the initial shear $\sigma_{i}$ calculated earlier by
Joshi et al [2002] for $4D$ space-time $(n=2)$. In view of what
has been discussed so far the expression (9.27)  reveals that the
existence of the naked singularity is directly related with the
non-vanishing shear in four and five dimensions. But in larger
dimensions the central singularity seems to be covered  by the
appearance of apparent horizons. Also the above expression (9.25)
for $\sigma_{i}$ reveals that the existence of the naked
singularity is not directly related to the non-vanishing of the
shear as it does in the marginally bound case. In fact the shear
in view of (9.26) increases and goes to infinity at this epoch,
which is expected.\\

\section{Geodesic Study in Marginally Bound Case}
Since the radius of the spherical shell $R$ shrinks to zero at the
time $t_{sf}(r)$ so from (9.13) we have (chosen $t_{i}=0$)
\begin{equation}
t_{sf}(r)=\frac{2}{n+1}\frac{r^{\frac{n+1}{2}}}{\sqrt{F(r)}}
\end{equation}

Now the Kretchmann scalar\\
\begin{equation}
K=[n(n-1)+1]\frac{F'^{2}}{R^{2n}R'^{2}}-2
n^{2}(n-1)\frac{FF'}{R^{2n+1}R'^{2}}+
n^{2}(n^{2}-1)\frac{F^{2}}{R^{2n+2}}
\end{equation}

diverges at $t=t_{sf}(r)$ i.e., $R=0$. Thus it represents the
formation of a curvature singularity at $r$. In fact the central
singularity (i.e., $r=0$) forms at the time

\begin{equation}
t_{0}=\sqrt{\frac{2n}{(n+1)\rho_{_{0}}}}
\end{equation}

The Kretchmann scalar also diverges at this central singularity.

Now if we use the expansion (9.11) for $F(r)$ in equation (9.28)
then near $r=0$, the singularity curve can be approximately
written as (using (9.30))

\begin{equation}
t_{sf}(r)=t_{0}-\frac{F_{m}}{(n+1)F_{0}^{3/2}}r^{m}
\end{equation}

where $m\ge 2$ and $F_{m}$ is the first non-vanishing term beyond
$F_{0}$. Thus $t_{sf}(r)>t_{0}$ as $F_{m}<0$ for any $m\ge 2$.

To examine whether the singularity at $t=t_{0},r=0$ is naked or
not, we investigate whether there exist one or more outgoing null
geodesics which terminate in the past at the central singularity.
In particular, we will concentrate to radial null geodesics
only.\\

Let us start with the assumption that it is possible to have one
or more such geodesics and we choose the form of the geodesics
(near $r=0$) as
\begin{equation}
t=t_{0}+a r^{\alpha},
\end{equation}

to leading order in $t$-$r$ plane with $a>0,\alpha>0$. Now for
$t$ in the geodesic (9.32) should be less than $t_{sf}(r)$ in
(9.31) for visibility of the naked singularity so on comparison
we have
\begin{equation}
\alpha\ge m ~~~~ \text{and} ~~~~a<-\frac{F_{m}}{(n+1)F_{0}^{3/2}}.
\end{equation}

Also from the metric form (9.1), an outgoing null geodesic must
satisfy

\begin{equation}
\frac{dt}{dr}=R'
\end{equation}

But near $r=0$, the solution (9.9) for $R$ simplifies to

\begin{equation}
R=r\left[1-\frac{n+1}{2}\sqrt{F_{0}}\left(1+\frac{F_{m}}{2F_{0}}r^{m}
\right)t \right]^{\frac{2}{n+1}}
\end{equation}

Thus combining (9.32) and (9.35) in equation (9.34) we get

\begin{equation}
a \alpha r^{\alpha-1}=\frac{\left[1-\frac{n+1}{2}\sqrt{F_{0}}~
(t_{0}+a
r^{\alpha})-\frac{(2m+n+1)F_{m}}{4\sqrt{F_{0}}}r^{m}(t_{0}+a
r^{\alpha})
\right]}{\left[1-\frac{n+1}{2}\sqrt{F_{0}}\left(1+\frac{F_{m}}{2F_{0}}r^{m}
\right)(t_{0}+a r^{\alpha}) \right]^{\frac{n-1}{n+1}}}
\end{equation}

Now if there exists a self consistent solution of this equation
then it is possible to have at least one outgoing radial null
geodesic that had started at the singularity i.e., the
singularity is naked. In order to simplify the above equation we
shall use the restrictions in equation (9.33) in the following two
ways:\\

$(i)~~ \alpha>m$ :

The equation (9.36) becomes (in leading order)

\begin{equation}
a \alpha
r^{\alpha-1}=\left(1+\frac{2m}{n+1}\right)\left(-\frac{F_{m}}{2F_{0}}\right)
^{\frac{2}{n+1}}r^{\frac{2m}{n+1}}
\end{equation}

which implies
\begin{equation}
\alpha=1+\frac{2m}{n+1}~~~ \text{and}~~~
a=\left(-\frac{F_{m}}{2F_{0}}\right)^{2/(n+1)}
\end{equation}

Thus for $\alpha>m$ we have $m<\frac{n+1}{n-1}$ and $n\ge 2$.

For $n=2$, $m$ may take values 1 and 2 and we have
$\rho_{1}=F_{1}=0$ and $\rho_{2}<0,F_{2}<0$. As a result, $a$ is
real and positive from equations (9.33) and (9.38). Moreover,
these restrictions are already assumed in the power series
expansion for $\rho_{i}(r)$ so that the initial density gradient
is negative and falls off rapidly near the centre. But for $n>2$,
$m=1$ is the only possible solution for which no real positive
solution of `$a$' is permissible from equations (9.33) and (9.38).
Hence with the restriction $\alpha>m$, we have a consistent
solution of equation (9.36) only for four dimensional space-time
i.e., it is possible to have (at least) null geodesics terminate
in the past at the singularity only for four dimension and we can
have naked singularity for $n=2$.\\

$(ii)~~\alpha=m$ :

In this case equation (9.36) simplifies to

\begin{equation}
m a
r^{m-1}=\left[-\frac{F_{m}}{2F_{0}}-\frac{n+1}{2}\sqrt{F_{0}}a\right]^{\frac{1-n}{n+1}}
\left[-\frac{(2m+n+1)}{2(n+1)}\frac{F_{m}}{F_{0}}-\frac{n+1}{2}\sqrt{F_{0}}~a\right]r^{\frac{2m}{n+1}}
\end{equation}

 A comparative study of equal powers of $r$ shows that
$m=\frac{n+1}{n-1}$ and $a$ depends on $F_{m}$ and $F_{0}$. Here
for $n=2,m=3$ and this situation is already discussed by Barve et
al [1999]. For $n=3$, we have $m=2$ and from (9.39) we get
\begin{equation}
2b^{2}(4b+\xi)+(2b+\xi)^{2}=0
\end{equation}

where ~~$b=\frac{a}{\sqrt{F_{0}}}$
~~and~~$\xi=\frac{F_{2}}{F_{0}^{2}}$.

We note that for real $b$, we must have $b<-\frac{\xi}{4}$ i.e.,
$a<-\frac{F_{2}}{4F_{0}^{3/2}}$, which is essentially the
restriction in (9.33). It can be shown that the above cubic
equation has at least one positive real root if
$\xi\le-(11+5\sqrt{5})$. Thus, if
$F_{2}\le-(11+5\sqrt{5})F_{0}^{2}$ we have at least one real
positive solution for $a$ which is consistent with equation (9.36)
(or (9.39)). Further for $n>3$, we can not have any integral
(positive) solution for $m$ and hence equation (9.39) is not
consistent for $n>3$. So, it is possible for (at least) radial
null geodesics which initiate from the singularity and reach to
an external observer without get prevented by any trapped surface
for $n\le 3$. Therefore, naked singularity is possible only for
four and five dimensions and for higher dimensions ($n\ge 4$) all
singularities are covered by trapped surfaces leading to black
hole.\\

Further, to examine whether it is possible to have an entire
family of geodesics those have started at the singularity, let us
consider geodesics correct to one order beyond equation (9.32)
i.e., of the form
\begin{equation}
t=t_{0}+a r^{\alpha}+d r^{\alpha+\beta}
\end{equation}

where as before $a, d, \alpha$ and $\beta$ are positive
constants. Thus equation (9.36) is modified to
\begin{eqnarray*}
a \alpha r^{\alpha-1}+(\alpha+\beta)d
r^{\alpha+\beta-1}=~~~~~~~~~~~~~~~~~~~~~~~~~~~~~~~~~~~~~~~~~~~~~~~~~
\end{eqnarray*}
\vspace{-2mm}
\begin{equation}
\frac{\left[1-\frac{n+1}{2}\sqrt{F_{0}}~ (t_{0}+a r^{\alpha}+d
r^{\alpha+\beta})-\frac{(2m+n+1)F_{m}}{4\sqrt{F_{0}}}r^{m}(t_{0}+a
r^{\alpha}+d r^{\alpha+\beta})
\right]}{\left[1-\frac{n+1}{2}\sqrt{F_{0}}\left(1+\frac{F_{m}}{2F_{0}}r^{m}
\right)(t_{0}+a r^{\alpha}+d r^{\alpha+\beta})
\right]^{\frac{n-1}{n+1}}}
\end{equation}

So for $\alpha>m$ (retaining terms upto second order) we have

\begin{equation}
\alpha a r^{\alpha-1}+(\alpha+\beta)d
r^{\alpha+\beta-1}=\left(1+\frac{2m}{n+1}\right)\left(-\frac{F_{m}}{2F_{0}}\right)
^{\frac{2}{n+1}}r^{\frac{2m}{n+1}}+D r^{\alpha-\frac{(n-1)m}{n+1}}
\end{equation}

As before, we have the values of $a$ and $\alpha$ in equation
(9.38) and

$$\beta=1-\frac{(n-1)m}{n+1}~~~\text{and} ~~~
d=\frac{D}{2+\frac{(3-n)m}{n+1}}$$ ~~with
\begin{equation}
D=\frac{1}{2}\sqrt{F_{0}}\left(-\frac{F_{m}}{2F_{0}}\right)
^{\frac{3-n}{n+1}}\left\{(n-1)\left(1+\frac{2m}{n+1}\right)-n-1\right\}.
\end{equation}

As we have similar conclusion as before for $\alpha>m$ so we now
consider the case $\alpha=m$. But if we restrict ourselves to the
five dimensional case then $m=\alpha=2$ and we have the same cubic
equation (9.40) for $b$. Now $\beta$ can be evaluated from the
equation

\begin{equation}
2+\beta=\frac{2^{5/2}b}{(-\xi-4b)^{\frac{3}{2}}}
\end{equation}

and we must have $\beta>0$, otherwise the geodesics will not lie
in the real space-time. As there is no restriction on $d$ so it
is totally arbitrary. This implies that there exists an entire
family of outgoing null geodesics terminated in the past at the
singularity for four and five dimensions only.\\

\section{Geodesic Study in Non-Marginally Bound Case}
In the present discussion we are concerned with the gravitational
collapse, we require that $\dot{R}(t,r) < 0$. As it is possible
to make an arbitrary relabeling of spherical dust shells by
$r\rightarrow g(r)$ without loss of generality, we fix the
labeling by requiring that on the hypersurface $t=0$, $r$
coincides with the radius
\begin{equation}
R(0,r)=r
\end{equation}

So the initial density is given by
\begin{equation}
\rho(r)\equiv\rho(0,r)=\frac{n}{2}~r^{-n}F'(r)~~\Rightarrow
F(r)=\frac{2}{n}\int\rho(r) r^{n} dr
\end{equation}

Now from eq.(9.5) it can be seen that the density diverges faster
in 5D ($n=3$) as compared to 4D ($n=2$). For increasing dimensions
the density diverges rapidly. Hence there is relatively more
mass-energy collapsing in the higher dimensional space-time
compared to the 4D and 5D cases.

In this case (i.e., $f(r)\ne 0$), equation (9.17) can be rewritten
as

\begin{equation}
t=\frac{2}{(n+1)\sqrt{F}}\left[r^{\frac{n+1}{2}}~_{2}F_{1}[\frac{1}{2},a,a+1,-\frac{f
r^{n-1}}{F}]-R^{\frac{n+1}{2}}~_{2}F_{1}[\frac{1}{2},a,a+1,-\frac{f
R^{n-1} }{F}]\right]
\end{equation}

Let us assume [Dwivedi, 1992]

\begin{eqnarray}\left.\begin{array}{llll}
F(r)=r^{n-1}\lambda(r)\\\\
\alpha=\alpha(r)=\frac{r f'}{f}\\\\
\beta=\beta(r)=\frac{r F'}{F}\\\\
R(t,r)=r P(t,r)
\end{array}\right\}
\end{eqnarray}

So using equations (9.4), (9.48) and (9.49), we have the
following expressions

\begin{eqnarray*}
R'=\frac{1}{n-1}\left[P(\beta-\alpha)+\frac{1}{n}\left\{
P~_{2}F_{1}[\frac{1}{2},a,a+1,-\frac{f
P^{n-1}}{\lambda}]-P^{\frac{1-n}{2}}~_{2}F_{1}[\frac{1}{2},a,a+1,-\frac{f
}{\lambda}] \right\}\right.
\end{eqnarray*}
\vspace{-3mm}
\begin{equation}
\left.\{\alpha(n+1)-2\beta\}\sqrt{1+\frac{f P^{n-1}}{\lambda}}
+\frac{(\alpha-\beta+n-1)P^{\frac{1-n}{2}}\sqrt{1+\frac{f
P^{n-1}}{\lambda}}}{\sqrt{1+\frac{f}{\lambda}}} \right]
\end{equation}

and
\begin{eqnarray*}
\dot{R}'=\frac{1}{r}\left[\frac{1}{n+1}\left\{
~_{2}F_{1}[\frac{1}{2},a,a+1,-\frac{f
P^{n-1}}{\lambda}]-P^{-\frac{n+1}{2}}~_{2}F_{1}[\frac{1}{2},a,a+1,-\frac{f
}{\lambda}] \right\}\{\alpha(n+1)-2\beta\}\right.
\end{eqnarray*}
\vspace{-3mm}
\begin{equation}
\hspace{2.2in} \left.-\alpha\sqrt{1+\frac{f P^{n-1}}{\lambda}}
+\frac{(\alpha-\beta+n-1)P^{-\frac{n+1}{2}}}{\sqrt{1+\frac{f}{\lambda}}}
\right]
\end{equation}

When $\lambda(r)=$ constant and $f(r)=$ constant, the space-time
becomes self-similar. Now we restrict ourselves to functions
$f(r)$ and $\lambda(r)$ which are analytic at $r=0$, such that
$\lambda(0)>0$. From eq.(9.5) it can be seen that the density at
the centre $(r=0)$ is finite at any time $t$, but becomes
singular at $t=0$.

We wish to investigate whether the singularity, when the central
shell with co-moving co-ordinate ($r=0$) collapses to the centre
at the time $t=0$, is naked .The singularity is naked if there
exists a null geodesic that emanates from the singularity. Let
$K^{a}=\frac{dx^{a}}{d\mu}$ be the tangent vector to the radial
null geodesic, where $\mu$ is the affine parameter. Then we derive
the following equations:

\begin{equation}
\frac{dK^{t}}{d\mu}+\frac{\dot{R}'}{\sqrt{1+f}}K^{r}K^{t}=0
\end{equation}

\begin{equation}
\frac{dt}{dr}=\frac{K^{t}}{K^{r}}=\frac{R'}{\sqrt{1+f(r)}}
\end{equation}

Let us define $X=\frac{t}{r}$, so that $P(t,r)=P(X,r)$. So
eq.(9.48) becomes

\begin{equation}
X=\frac{2}{(n+1)\sqrt{\lambda}}\left\{_{2}F_{1}[\frac{1}{2},a,a+1,-\frac{f
}{\lambda}]-P^{\frac{n+1}{2}}~_{2}F_{1}[\frac{1}{2},a,a+1,-\frac{f
P^{n-1} }{\lambda}]\right\}
\end{equation}

The nature of the singularity can be characterized by the
existence of radial null geodesics emerging from the singularity.
The singularity is at least locally naked if there exist such
geodesics and if no such geodesics exist it is a black hole. If
the singularity is naked, then there exists a real and positive
value of $X_{0}$ as a solution to the equation

\begin{eqnarray}
\begin{array}{c}
X_{0}~=\\
{}
\end{array}
\begin{array}{c}
lim~~~~~ \frac{t}{r}\\
t\rightarrow 0~ r\rightarrow 0
\end{array}
\begin{array}{c}
=~lim~~~~~ \frac{dt}{dr}\\
~~~~~~t\rightarrow 0~ r\rightarrow 0
\end{array}
\begin{array}{c}
=~lim~~~~ \frac{R'}{\sqrt{1+f}}\\
t\rightarrow 0~ r\rightarrow 0
\end{array}
\end{eqnarray}

Define $\lambda_{0}=\lambda(0), \alpha_{0}=\alpha(0), f_{0}=f(0)$,
and $Q=Q(X)=P(X,0)$. Now from equation (9.50) it is seen that
$\beta(0)=n-1$. We would denote $Q_{0}=Q(X_{0})$, the equations
(9.54) and (9.55) reduces to

\begin{equation}
X_{0}=\frac{2}{(n+1)\sqrt{\lambda_{0}}}\left\{_{2}F_{1}[\frac{1}{2},a,a+1,-\frac{f_{0}
}{\lambda_{0}}]-Q_{0}^{\frac{n+1}{2}}~_{2}F_{1}[\frac{1}{2},a,a+1,-\frac{f_{0}
Q_{0}^{n-1} }{\lambda_{0}}]\right\}
\end{equation}
and

\begin{eqnarray*}
X_{0}=\frac{1}{(n-1)\sqrt{1+f_{0}}}\left[\frac{1}{n+1}\left\{
Q_{0}~_{2}F_{1}[\frac{1}{2},a,a+1,-\frac{f_{0}
Q_{0}^{n-1}}{\lambda_{0}}]-Q_{0}^{\frac{1-n}{2}}~_{2}F_{1}[\frac{1}{2},a,a+1,-\frac{f_{0}
}{\lambda_{0}}] \right\}\right.
\end{eqnarray*}
\vspace{-3mm}
\begin{equation}
\left.\{\alpha_{0}(n+1)-2n+2\}\sqrt{1+\frac{f_{0}
Q_{0}^{n-1}}{\lambda_{0}}}
+\frac{\alpha_{0}Q_{0}^{\frac{1-n}{2}}\sqrt{1+\frac{f_{0}
Q_{0}^{n-1}}{\lambda_{0}}}}{\sqrt{1+\frac{f_{0}}{\lambda_{0}}}}+Q_{0}(n-1-\alpha_{0})
\right]
\end{equation}
\\
TABLE: Positive values of $X_{0}$ by eliminating $Q_{0}$ from
equations (9.56) and (9.57) for different values of the parameters
$f_{0}, \lambda_{0}$ and $\alpha_{0}$ in various dimensions.
\underline{}
\begin{center}
\begin{tabular}{lcccccc}   \hline
~$f_{0}$~~~~~~$\lambda_{0}$~~~~~$\alpha_{0}$~~~~~~~~~~~~~~~~~~~~~~~~~~~~~~~~~~~~~~Positive roots ($ X_{0}$)\\
\hline\\
~~~~~~~~~~~~~~~~~~~~~~~~4D~~~~~~~~~5D~~~~~~~~~6D~~~~~~~~~7D~~~~~~~~~8D~~~~~~~~~10D~~~~~~~~~12D~~~~~~~~~16D\\
\hline\\
-.033~~.034~~~.2~~6.75507,~~4.32936,~~3.15991,~~2.44524,~~ 1.92883,~~.669541,~~~~~~~~$-$~~~~~~~~~~~~$-$\\
~~~~~~~~~~~~~~~~~~~~~1.14476~~~1.16321~~~1.1879~~~~1.22434~~~~1.29065~~~.415056\\
\\
-.1~~~~.333~~~.2~~~3.15662,~~~1.72737,~~~~$-$~~~~~~~~~~$-$~~~~~~~~~~~$-$~~~~~~~~~~~$-$~~~~~~~~~~~~~$-$~~~~~~~~~~~~~$-$\\
~~~~~~~~~~~~~~~~~~~~~1.24003~~~~1.48602\\
\\
-.301~.302~.001~~2.07773,~~~~~~~$-$~~~~~~~~$-$~~~~~~~~~~$-$~~~~~~~~~~~$-$~~~~~~~~~~~$-$~~~~~~~~~~~~~$-$~~~~~~~~~~~~~$-$\\
~~~~~~~~~~~~~~~~~~~~~1.65625\\
\\
-.5~~~~.51~~.001~~~~~~$-$~~~~~~~~~~~~$-$~~~~~~~~$-$~~~~~~~~~~$-$~~~~~~~~~~~$-$~~~~~~~~~~~$-$~~~~~~~~~~~~~$-$~~~~~~~~~~~~~$-$\\
\\
.1~~~~~.1~~~~~2~~~~~1.667,~~~~~~1.1946~~~~~~$-$~~~~~~~~~~$-$~~~~~~~~~~~$-$~~~~~~~~~~~$-$~~~~~~~~~~~~~$-$~~~~~~~~~~~~~$-$ \\
~~~~~~~~~~~~~~~~~~~~.806578~~~~.858211\\
\\
.1~~~~~.1~~~~10~~~.463679~~~~.465426~~~~.466489~~~.466499~~~.464947~~~~.45438~~~~~~.429651~~~~~.35278\\
\\
.1~~~~~~1~~~.2~~~~~~~~~$-$~~~~~~~~~~~$-$~~~~~~~~~~$-$~~~~~~~~~~~$-$~~~~~~~~~~~$-$~~~~~~~~~~~$-$~~~~~~~~~~~~$-$~~~~~~~~~~~~$-$\\
\\
.1~~~~~~5~~~~5~~~~~~~~~$-$~~~~~~~~~~~$-$~~~~~~~~~~$-$~~~~~~~~~~~$-$~~~~~~~~~~~$-$~~~~~~~~~~~$-$~~~~~~~~~~~~$-$~~~~~~~~~~~~$-$\\
\\
~1~~~.001~~~2~~~~.996033,~~~.968869,~~~.91379,~~~.848544,~~ .783787,~~~~.669541,~~~.577489,~~~~~~~$-$\\
~~~~~~~~~~~~~~~~~~~.414324~~~~.41436~~~~~.414412~~~.414489~~~ .414605~~~~.415056~~~~.416214\\
\\
10~~.001~~.2~~~~.316014,~~~.309739,~~~~~~~~$-$~~~~~~~~~~$-$~~~~~~~~~~~$-$~~~~~~~~~~~$-$~~~~~~~~~~~~$-$~~~~~~~~~~~~$-$\\
~~~~~~~~~~~~~~~~~~~.275327~~~~.275856\\
\\
10~~~~.1~~~~5~~~~.089435~~~~.089441~~~~.089435~~~.089409~~~.08935~~~~.089061~~~~~~.088347~~~~~.084352\\
\\
10~~~~~1~~~~3~~~~.128192,~~~.128966,~~~.19718,~~~~.169063,~~.142103,~~~~~~~$-$~~~~~~~~~~~~$-$~~~~~~~~~~~~$-$\\
~~~~~~~~~~~~~~~~~~~~~~~~~~~~~~~~~~~~~~~~~~~~~~.41436~~~~~.414412~~~.414489\\
\hline\\

\end{tabular}
\end{center}

From the table we conclude the following results:

(i) For same value of the parameters ($f_{0}, \lambda_{0},
\alpha_{0}$) the possibility of positive real root is more in 4D
and then it decreases with increase in dimensions.

(ii) If $\lambda_{0}$ is positive but close to zero then it is
possible to have a naked singularity.

(iii) For same values of $f_{0}$ and $\lambda_{0}$, the naked
singularity in higher dimensions will less probable as we increase
the value the parameter $\alpha_{0}$.\\

The strength of singularity [Tipler, 1987], which is the measure
of its destructive capacity, is the most important feature. A
singularity is gravitationally strong or simply strong if it
destroys by crushing or stretching any objects that fall into it.
Following Clarke and Krolak [1986],we consider the null geodesics
affinely parameterized by $\mu$ and terminating at shell focusing
singularity $r=t=\mu=0$. Then it would be a strong curvature
singularity as defined by Tipler [1980] in equations (8.23) and
(8.25).

It is widely believed that a space-time does not admit analytic
extension through a singularity if it is a strong curvature
singularity in the above sense. Now equations (8.23) and (8.25)
can be expressed as

\begin{equation}
\begin{array}{c}
lim\\
\mu\rightarrow 0\\
\end{array}
\begin{array}{c}
\mu^{2}\Psi=\\
{}
\end{array}
\begin{array}{c}
lim\\
\mu\rightarrow 0\\
\end{array}
\begin{array}{c}
\frac{nF'}{2r^{n-2} P^{n-2} R'}\left(\frac{\mu K^{t}}{R}\right)^{2}\\
{}
\end{array}
\end{equation}

Using L'Hospital's rule and using equations (9.50) and (9.51), the
equation (9.58) can be written as

\begin{equation}
\begin{array}{c}
lim\\
\mu\rightarrow 0\\
\end{array}
\begin{array}{c}
\mu^{2}\Psi=\\
{}
\end{array}
\frac{2n(n-1)\lambda_{0}Q_{0}X_{0}(\lambda_{0}+f_{0}Q_{0}^{n-1})\sqrt{1+f_{0}}}
{\left[(n-1)\lambda_{0}X_{0}\sqrt{1+f_{0}}-(n-1)\lambda_{0}Q_{0}-
\alpha_{0}f_{0}Q_{0}^{n}\right]^{2}} >0\\
\end{equation}

Thus along the radial null geodesics, the strong curvature
condition is satisfied and hence it is a strong curvature
singularity. So the formation of naked singularity will be
less probable as we increase the dimension of space-time.\\

\section{Discussion}
In this chapter, we have studied spherical dust collapse in an
arbitrary $(n+2)$ dimensional space-time. We have considered both
non-marginal and marginally bound cases separately. For
non-marginal case we have seen that naked singularity  may be
possible for all dimensions ($n\ge 2$). However, to get a definite
conclusion about naked singularity we should study the geodesic
equations for non-marginal and marginally bound cases. But we can
not proceed further due to the presence of the complicated
hypergeometric functions. Therefore, no definite conclusion is
possible for non-marginal case.

On the other hand, for marginally bound case we have definitely
concluded that naked singularity is possible only for $n\le 3$ by
studying the existence of radial null geodesic through the
singularity. Here, we should mention that this result depends
sensitively on the choice of the initial conditions. In fact, if
we do not assume the initial density to have an extremum value at
the centre (i.e., $\rho_{1}\ne 0$) the naked singularity will be
possible in all dimensions. We should mention that the naked
singularity described above is only a local feature, it is not at
all a global aspect i.e., it violates the strong form of CCC. The
numerical results which favour the formation of naked singularity
in any dimension.\\

\large \baselineskip .85cm
\chapter{Study of Higher Dimensional Szekeres' Cosmological Model}
\label{chap10}\markright{\it CHAPTER~\ref{chap10}. Study of Higher
Dimensional Szekeres' Cosmological Model}

\section{Prelude}
Usually, cosmological solutions to Einstein's field equations are
obtained by imposing symmetries [Heckmann et al, 1962] on the
space-time. One of the reasonable assumptions (in an average
sense) is the spatial homogeneity. But when we consider
cosmological phenomena over galactic scale or in smaller scale
(detailed structure of the black body radiation) then we should
drop the assumption of
homogeneity i.e., inhomogeneous solutions are useful.\\

Szekeres [1975] gave a class of inhomogeneous solutions
representing irrotational dust for the metric of the form (known
as Szekeres metric)
$$ds^{2}=dt^{2}-e^{2\alpha}dr^{2}-e^{2\beta}(dx^{2}+dy^{2})$$
Subsequently, the solutions have been extended by Szafron [1977]
and Szafron and Wainwright [1977] for perfect fluid and they
studied asymptotic behaviour for different choice of the
parameters involved. Later Barrow and Stein-Schabes [1984] gave
solutions for dust model with a cosmological constant and showed
the validity of the Cosmic `no-hair' Conjecture.\\

In this chapter, we give inhomogeneous solutions for
($n+2$)-dimensional Szekeres space-time with perfect fluid and a
cosmological constant. An asymptotic study of particular
solutions are also given.\\

\section{Basic Equations}
The metric for the ($n+2$)-dimensional Szekeres space-time is in
the form

\begin{equation}
ds^{2}=dt^{2}-e^{2\alpha}dr^{2}-e^{2\beta}\sum^{n}_{i=1}dx_{i}^{2}
\end{equation}

where $\alpha$ and $\beta$ are functions of all the ($n+2$)
space-time variables i.e., $$\alpha=\alpha(t,r,x_{1},...,x_{n}),~~
\beta=\beta(t,r,x_{1},...,x_{n}).$$ The Einstein equations for
the perfect fluid with a cosmological constant is of the form

\begin{equation}
G_{\mu\nu}=\Lambda g_{\mu\nu}+(\rho+p)u_{\mu}u_{\nu}-p g_{\mu\nu}
\end{equation}

where $\rho$ and $p$ are energy density and isotropic pressure
measured by an observer moving with with the fluid, $\Lambda$ is
the cosmological constant and $u_{\mu}$ is the fluid flow four
vector. Since $u=\partial/\partial t$, the flow lines are
geodesics and the contracted Bianchi identities imply that
pressure is a function of $t$ only i.e., $p=p(t)$. As there is no
restriction on the energy density so $\rho$ is in general a
function of all the ($n+2$) variables i.e.,
$\rho=\rho(t,r,x_{1},...,x_{n})$ and hence no equation of state
is imposed.

Now from the non-vanishing components of the field equations
(10.2) for the above metric (10.1), we have\\
\begin{eqnarray*}
n\dot{\alpha}\dot{\beta}+\frac{1}{2}n(n-1)\dot{\beta}^{2}-e^{-2\beta}\sum_{i=1}^{n}
\left\{\alpha_{x_{i}}^{2}+\frac{1}{2}(n-1)(n-2)\beta_{x_{i}}^{2}+
(n-2)\alpha_{x_{i}}\beta_{x_{i}}+\alpha_{x_{i}x_{i}}\right.
\end{eqnarray*}
\vspace{-3mm}
\begin{equation}
\left.+(n-1)\beta_{x_{i}x_{i}} \right\}+e^{-2\alpha}
\left\{n\alpha'\beta'-\frac{1}{2}n(n+1)\beta'^{2}
-n\beta''\right\}=\Lambda+\rho
\end{equation}
\vspace{-3mm}
\begin{eqnarray*}
\frac{1}{2}n(n+1)\dot{\beta}^{2}+n\ddot{\beta}-\frac{1}{2}n(n-1)e^{-2\alpha}\beta'^{2}
-e^{-2\beta}\sum_{i=1}^{n}\left\{\frac{1}{2}(n-1)(n-2)\beta_{x_{i}}^{2}+
(n-1)\beta_{x_{i}x_{i}}\right\}
\end{eqnarray*}
\vspace{-3mm}
\begin{equation}
=\Lambda-p \hspace{-3.4in}
\end{equation}
\vspace{-3mm}
\begin{eqnarray*}
\dot{\alpha}^{2}+\ddot{\alpha}+(n-1)\dot{\alpha}\dot{\beta}+\frac{1}{2}n(n-1)\dot{\beta}^{2}+
(n-1)\ddot{\beta}+e^{-2\alpha}\left\{(n-1)\alpha'\beta'-\frac{1}{2}n(n-1)
\beta'^{2}-\right.
\end{eqnarray*}
\begin{eqnarray*}
\left.(n-1)\beta''\right\}-e^{-2\beta}\sum_{i\ne
j=1}^{n}\left\{\alpha_{x_{j}}^{2}+\frac{1}{2}(n-2)(n-3)\beta_{x_{j}}^{2}+
\alpha_{x_{j}x_{j}}+(n-2)\beta_{x_{j}x_{j}}+(n-3)\alpha_{x_{j}}\beta_{x_{j}}\right\}
\end{eqnarray*}
\begin{equation}
-e^{-2\beta}\left\{(n-1)\alpha_{x_{i}}\beta_{x_{i}}+
\frac{1}{2}(n-1)(n-2)\beta_{x_{i}}^{2}\right\}=\Lambda-p
\hspace{-.6in}
\end{equation}
\vspace{-3mm}
\begin{equation}
\alpha_{x_{j}}(-\alpha_{x_{i}}+\beta_{x_{i}})+\beta_{x_{j}}(\alpha_{x_{i}}+
(n-2)\beta_{x_{i}})-\alpha_{x_{i}x_{j}}-(n-2)\beta_{x_{i}x_{j}}=0,~~~
(i\ne j)
\end{equation}
\vspace{-3mm}
\begin{equation}
\dot{\alpha}\beta'-\dot{\beta}\beta'-\dot{\beta}'=0
\end{equation}
\vspace{-3mm}
\begin{equation}
-\dot{\alpha}\alpha_{x_{i}}+\dot{\beta}\alpha_{x_{i}}-\dot{\alpha}_{x_{i}}-(n-1)\dot{\beta}_{x_{i}}=0
\end{equation}
\vspace{-3mm}
\begin{equation}
\alpha_{x_{i}}\beta'-\beta'_{x_{i}}=0
\end{equation}

where dot, dash and subscript stands for partial differentiation
with respect to $t$, $r$ and the corresponding variables
respectively (e.g., $\beta_{x_{i}}=\frac{\partial\beta}{\partial
x_{i}}$) with $i,j=1,2,...,n$.

From  equations (10.7) and (10.9) after differentiating with
respect to $x_{i}$ and $t$ respectively, we have the integrability
condition

\begin{equation}
\dot{\beta}_{x_{i}}\beta'^{2}=0,~~~~i=1,2,...,n
\end{equation}

Thus, if $\beta'\ne 0$ we must have
$\dot{\beta}_{x_{i}}=0,~~i=1,2,...,n$. In the following section we
shall consider the following possibilities $$(i)~ \beta'\ne
0,~~~~(ii)~ \beta'=0,~~\dot{\beta}_{x_{i}}=0 ~~(i=1,2,...,n) $$ to
get solutions of the field equations.\\

\section{Solutions to the Field Equations}
In this section we shall solve the field equations (10.3)-(10.9)
using the above restrictions separately.

{\bf\it Case I}:~~ $\beta'\ne 0$\\

Here due to the restrictions
$\dot{\beta}_{x_{i}}=0,~~i=1,2,...,n$ we have from the field
equations (10.7) and (10.9), the form of the metric coefficient as
\begin{equation}
e^{\beta}=R(t,r)~e^{\nu(r,x_{1},...,x_{n})}
\end{equation}
and
\begin{equation}
e^{\alpha}=R'+R~\nu'
\end{equation}

Now substituting these forms for the metric coefficient in
equation (10.4) we have the differential equations for $R$ and
$\nu$ as

\begin{equation}
R\ddot{R}+\frac{1}{2}(n-1)\dot{R}^{2}+\frac{1}{n}(p(t)-\Lambda)R^{2}=\frac{n-1}{2}f(r)
\end{equation}
and
\begin{equation}
e^{-2\nu}\sum_{i=1}^{n}\left\{(n-2)\nu_{x_{i}}^{2}+2\nu_{x_{i}x_{i}}
\right\}=n f(r)-n
\end{equation}

with $f(r)$ as arbitrary function of $r$ alone.

Equation (10.13) can be integrated once to have the first integral

\begin{equation}
\dot{R}^{2}=\frac{2\Lambda}{n(n+1)}R^{2}+f(r)+\frac{F(r)}{R^{n-1}}
-\frac{2}{n}~R^{1-n}\int p(t)R^{n}\dot{R}dt
\end{equation}

where $F(r)$ is another arbitrary function of $r$ (appears due to
integration).

Also from (10.14) the solution for $\nu$ will be

\begin{equation}
e^{-\nu}=A(r)\sum_{i=1}^{n}x_{i}^{2}+\sum_{i=1}^{n}B_{i}(r)x_{i}+C(r)
\end{equation}

with the restriction
\begin{equation}
\sum_{i=1}^{n}B_{i}^{2}-4AC=f(r)-1
\end{equation}

for the arbitrary functions $A(r),~B_{i}(r),~i=1,2,...,n$ and
~$C(r)$.

Further, to solve $R$ completely let us consider $p$ as a
polynomial in $t$ as
\begin{equation}
p(t)=p_{0}t^{-a}
\end{equation}

($p_{0}$ and $a$ are positive constants) and we have the general
solution for $R$ as

\begin{equation}R^{\frac{n+1}{2}}=\left\{\begin{array}{lll}

\sqrt{t}\left\{C_{1}J_{\xi}[\frac{2\sqrt{c}}{|a-2|}t^{-\frac{a-2}{2}}]
+C_{2}Y_{\xi}[\frac{2\sqrt{c}}{|a-2|}t^{-\frac{a-2}{2}}]\right\}\\
\\
\sqrt{t}\left\{C_{1}J_{\xi}[\frac{2\sqrt{c}}{|a-2|}t^{-\frac{a-2}{2}}]
+C_{2}J_{-\xi}[\frac{2\sqrt{c}}{|a-2|}t^{-\frac{a-2}{2}}]\right\}\\
\\
C_{1}t^{q_{1}}+C_{2}t^{1-q_{1}}
\end{array}\right.
\end{equation}

according as $\xi$ is an integer, non-integer and $a=2$. Here
$C_{1}$ and $C_{2}$ are arbitrary functions of $r$ and we have
chosen
$$\xi=\frac{1}{a-2},~~c=\frac{(n+1)p_{0}}{2n},~~q_{1}=\frac{1}{2}(1+\sqrt{1-4c}).$$

It is to be noted that to derive the above solution we have
chosen $\Lambda=0=f(r)$. However, for non-zero $\Lambda$ (but
$f(r)=0$) the solution is possible only for $a=0$ and $2$ as

\begin{equation}R^{\frac{n+1}{2}}=\left\{\begin{array}{lll}
C_{1}Cos\{t\sqrt{\frac{n+1}{2n}(p_{0}-\Lambda)}\}+
C_{2}Sin\{t\sqrt{\frac{n+1}{2n}(p_{0}-\Lambda)}\},~~ $when$~~ a=0\\
\\
\sqrt{t}\left\{C_{1}J_{\zeta}[-\frac{i
t\sqrt{\Lambda}}{\sqrt{2}}\sqrt{1+\frac{1}{n}}]+C_{2}Y_{\zeta}[-\frac{i
t\sqrt{\Lambda}}{\sqrt{2}}\sqrt{1+\frac{1}{n}}] \right\}, ~~~
$when$~~ a=2
\end{array}\right.
\end{equation}

with~ $\zeta=\frac{1}{2}\sqrt{1-\frac{2(n+1)p_{0}}{n}}$ .\\

Further, if we consider the dust model (i.e., $p(t)=0$) then the
above solution (10.19) simplifies to $R^{(n+1)/2}\propto t$. Hence
for the usual 4D (i.e., $n=2$) the scale factor $R$ grows as
$t^{2/3}$ as in the usual Friedmann model.

Now, the physical and kinematical parameters have the following
expressions

\begin{equation}
\rho=\frac{n}{2}~\frac{F'+(n+1)F\nu'}{R^{n}(R'+R\nu')}-\frac{p_{0}}{t^{a}}
\end{equation}

\begin{equation}
\theta=\frac{R\dot{R}'+(n+1)R\dot{R}\nu'+n\dot{R}R'}{R(R'+R\nu')}
\end{equation}

\begin{equation}
\sigma^{2}=\frac{n}{8(n+1)(n-1)^{2}}\left[\frac{2R^{n-1}(Rf'-2R'f)+
(n-1)(RF'-(n+1)R'F)}{\dot{R}R^{n}(R'+R\nu')}\right]^{2}
\end{equation}
\\
{\bf\it Case II}:~~ $\beta'=\dot{\beta}_{x_{i}}=0,~~ i=1,2,...,n$\\

In this case from the field equations we have the form of the
metric functions
\begin{equation}
e^{\beta}=R(t)~e^{\nu(x_{1},x_{2},...,x_{n})}
\end{equation}
and
\begin{equation}
e^{\alpha}=R(t)~\eta(r,x_{1},x_{2},...,x_{n})+\mu(t,r)
\end{equation}

Then as before from the field equation (10.4) we have similar
differential equations in $R$ and $\nu$ as
\begin{equation}
R\ddot{R}+\frac{1}{2}(n-1)\dot{R}^{2}+\frac{1}{n}(p(t)-\Lambda)R^{2}=\frac{n-1}{2}K
\end{equation}
and
\begin{equation}
e^{-2\nu}\sum_{i=1}^{n}\left\{(n-2)\nu_{x_{i}}^{2}+2\nu_{x_{i}x_{i}}
\right\}=n K
\end{equation}

with $K$, an arbitrary constant.

Here we take the solution for $\nu$ in the form

\begin{equation}
e^{-\nu}=P\sum_{i=1}^{n}x_{i}^{2}+\sum_{i=1}^{n}Q_{i}x_{i}+S
\end{equation}

where the arbitrary constants $P, Q_{i}~(i=1,2,...,n)$ and $S$ are
restricted as before

\begin{equation}
\sum_{i=1}^{n}Q_{i}^{2}-4PS=K
\end{equation}

Now to determine the function $\eta$, we have from the field
equation (10.6)

\begin{equation}
\frac{\partial^{2}(e^{-\nu}\eta)}{\partial x_{i} \partial x_{i}}=0
\end{equation}

and then from the field equation (10.5) we have the solution

\begin{equation}
e^{-\nu}\eta=u(r)\sum_{i=1}^{n}x_{i}^{2}+\sum_{i=1}^{n}v_{i}(r)x_{i}+w(r)
\end{equation}

with $u(r), v_{i}(r)~(i=1,2,...,n)$ and $w(r)$ as arbitrary
functions.

Further, to obtain the function $\mu$ we use the following
combination of the field equations namely,
\begin{equation}
\sum_{i=1}^{n}G_{x_{i}}^{x_{i}}-G_{r}^{r}=(n-1)(\Lambda-p(t))
\end{equation}

and the resulting differential equation in $\mu$ is

\begin{equation}
R\ddot{\mu}+(n-1)\dot{R}\dot{\mu}+\mu\left[\ddot{R}+\frac{2}{n}~(p(t)-\Lambda)R\right]=g(r)
\end{equation}
with
\begin{equation}
g(r)=(n-1)\left[2(uS+wP)-\sum_{i=1}^{n}v_{i}Q_{i}\right]
\end{equation}

For explicit solution if we choose $p(t)$ as in the previous case
(eq.(10.18)) then the explicit form for $R$ is same as in equation
(10.19) except here $C_{1}$ and $C_{2}$ are arbitrary constants
and we have chosen $K=\Lambda=0$. Similarly, we have the same
solutions (10.20) for non-zero $\Lambda$. But we note that the
differential equation (10.33) is not solvable for any value of
$n$. In fact only for $n=3$ (i.e., for five dimension) we have the
complete solution

\begin{equation}
\mu R=d_{1}t^{q_{2}}+d_{2}t^{1-q_{2}}
\end{equation}

for $a=2$ and
$q_{2}=\frac{1}{2}\left(1+\sqrt{1-\frac{8p_{0}}{3}}\right)$ with
$\Lambda=g(r)=0$.

Further, the physical and kinematical parameters have the
expressions as

\begin{equation}
\rho=\frac{2n\dot{\mu}\dot{R}-K}{2\mu
R}+\frac{n}{n-1}\left[\frac{\eta(R^{2}\ddot{\mu}-n\mu
R\ddot{R})-(n-1)\mu^{2}\ddot{R}}{\mu R(\mu+\eta R)}\right]-p(t)
\end{equation}

\begin{equation}
\theta=\frac{R\dot{\mu}+n\mu\dot{R}+(n+1)R\dot{R}\eta}{
R(\mu+\eta R)}
\end{equation}

\begin{equation}
\sigma^{2}=\frac{n}{2(n+1)}\left[\frac{R\dot{\mu}-\dot{R}\mu}{
R(\mu+\eta R)}\right]^{2}
\end{equation}

Finally, for simple dust case we have the solution for
$\Lambda=0$ in terms of hypergeometric function as

\begin{equation}
\sqrt{C_{1}}(t-t_{0})=R^{\frac{n+1}{2}}~~_{2}F_{1}[\frac{1}{2},\frac{n+1}{2n-2},
\frac{3n-1}{2n-2},-\frac{(n+1)^{2}}{4nC_{1}}f(r)R^{n-1}]
\end{equation}
\\
But for non-zero $\Lambda$ we can have solution only for $n=3$ as

\begin{equation}
R^{2}=C_{1}e^{t\sqrt{\frac{2\Lambda}{3}}}+C_{2}e^{-t\sqrt{\frac{2\Lambda}{3}}}-
\frac{3f(r)}{\Lambda}
\end{equation}

However, we can have an integral equation from (10.15) (with
$p=0$) as

\begin{equation}
t-t_{0}=\int\frac{dR}{\sqrt{\frac{2\Lambda}{n(n+1)}R^{2}+f(r)+\frac{F(r)}{R^{n-1}}}}
\end{equation}

and we have the following particular solutions :\\

(i)~~~~$F(r)=0$
$$
R=\sqrt{\frac{(n+1)f(r)}{2\Lambda}}~Sinh\left[(t-t_{0})\sqrt{\frac{2\Lambda}{n(n+1)}}~\right]
$$

(ii)~~~~$f(r)=0$ (i.e., $K=0$ in {\it Case II})

$$
R^{\frac{n+1}{2}}=\sqrt{\frac{n(n+1)F(r)}{2\Lambda}}~Sinh\left[(t-t_{0})\sqrt{\frac{(n+1)\Lambda}
{2n}}~\right]
$$

(iii)~~~$\Lambda=0$~~($n=3$)
$$
R^{2}=\frac{1}{3}f(r)(t-t_{0})^{2}-\frac{3F(r)}{f(r)}
$$

(iv)~~~$f(r)=\Lambda=0$
$$
R^{\frac{n+1}{2}}=\frac{n+1}{2}\sqrt{F(r)}~(t-t_{0})
$$

(v)~~~~$f(r)=F(r)=0$
$$
R=e^{(t-t_{0})\sqrt{\frac{2\Lambda}{n(n+1)}}}
$$

Also for the dust $\mu$ has the solution

\begin{equation}
\mu~R=C_{1}e^{t\sqrt{\frac{2\Lambda}{3}}}+C_{2}e^{-t\sqrt{\frac{2\Lambda}{3}}}-
\frac{3g(r)}{2\Lambda}
\end{equation}

for $n=3$.\\

\section{Asymptotic Behaviour}
We shall now discuss the asymptotic behaviour of the solutions
presented in the previous section for both perfect fluid and dust
model separately. The co-ordinates vary over the range :
$t_{0}<t<\infty;~ -\infty<r<\infty;~ -\infty<x_{i}<\infty,~
i=1,2,...,n$.\\

\subsection{Perfect Fluid Model}
As $p\ge 0,~ p\ne 0$ so we must have $\frac{1}{2}<q_{1}<1$. We
shall first consider the case when $a=2$. For large $t$ ({\it Case
I}~ i.e., $\beta'\ne 0$)
$$
R^{\frac{n+1}{2}}\sim t^{q_{1}}
$$
$$
\rho\sim\frac{n}{2}(F'+(n+1)F\nu')t^{-2q_{1}}
$$
$$
p\sim t^{-2}
$$
$$
\theta\sim t^{-1}
$$
$$
\sigma^{2}\sim t^{-2}
$$

For {\it Case II},~ we have similar behaviour for large $t$
together with $\mu\sim t^{\tilde{q}_{1}}$ where $\tilde{q}_{1}$
is the value of $q_{1}$ for $n=3$ and also we have
$\frac{1}{2}<\tilde{q}_{1}<1$. Thus as $t\rightarrow \infty$,
($p,\rho$) fall off faster compare to ($\theta,\sigma$), while
the scale factor $R$ (and $\mu$) gradually increases with time.
So the model approaches isotropy along fluid world line as
$t\rightarrow \infty$.\\

\subsection{Dust Model}
For the dust case with non-zero $\Lambda$ (and $n=3$) we have for
large $t$,
$$
\mu=R\sim e^{t\sqrt{\frac{\Lambda}{6}}}
$$
$$
\rho=\rho_{0},~~~\text{a~ constant}
$$
$$
\theta=\theta_{0},~~~\text{a~ constant}
$$
$$
\sigma=\sigma_{0},~~~\text{a~ constant}
$$

We note that for {\it Case I},~ $\rho_{0}=\sigma_{0}=0$ while for
{\it Case II}~ we have $\rho_{0}\ne 0,~\sigma_{0}\ne 0$. Thus
universe will behave locally like de-sitter model in {\it Case I}~
though the global geometry will be different. However, in {\it
Case II}~ the universe will not isotropize and for large $t$ the
measure of anisotropy and expansion scalar become finite constant.\\

\section{Discussion}
In this chapter, we have investigated cosmological solutions for
($n+2$)-dimensional Szekeres form of metric with perfect fluid
(or dust) as the matter distribution. We can classify the
solutions in two categories namely, (i) $\beta'\ne 0$ and (ii)
$\beta'=0$. The first set of solutions are known as
quasi-spherical solution while second class of solutions are
termed as quasi-cylindrical type of solutions.

However, if we assume the arbitrary functions $A(r)$, $B_{i}(r)$
and $C(r)$ to have the constant values namely
$A(r)=C(r)=\frac{1}{2}$ and $B_{i}(r)=0$ (for all $i=1,2,...,n$)
i.e., if we have chosen the arbitrary function $f(r)$ to be zero
then using the transformation
\begin{eqnarray}\begin{array}{llll}
x_{1}=Sin\theta_{n}Sin\theta_{n-1}...~~ ...
Sin\theta_{2}Cot\frac{1}{2}\theta_{1}\\\\
x_{2}=Cos\theta_{n}Sin\theta_{n-1}...~~ ...
Sin\theta_{2}Cot\frac{1}{2}\theta_{1}\\\\
x_{3}=Cos\theta_{n-1}Sin\theta_{n-2}...~~ ...Sin\theta_{2}Cot\frac{1}{2}\theta_{1}\\\\
....~~ ...~~ ...~~ ...~~ ...~~ ...~~ ...~~ ...\\\\
x_{n-1}=Cos\theta_{3}Sin\theta_{2}Cot\frac{1}{2}\theta_{1}\\\\
x_{n}=Cos\theta_{2}Cot\frac{1}{2}\theta_{1}
\end{array}\nonumber
\end{eqnarray}
the Szekeres metric (10.1) with the solutions (10.11), (10.12),
(10.16) reduces to the ($n+2$)-dimensional spherically symmetric
metric
$$
ds^{2}=dt^{2}-R'^{2}dr^{2}-R^{2}d\Omega_{n}^{2}
$$

It is to be noted that if the arbitrary functions $A, B_{i}, C$
depend on $r$, then we can not get the above spherical form of
the space-time. So these arbitrary functions play an important
roll to identify the nature of the space-time.

Finally, the study of asymptotic behaviour shows that some of the
solutions will become isotropic at late time while there are
solutions for which shear scalar remains constant throughout the
evolution. Thus Cosmic `no-hair' Conjecture is not valid for all
solutions. This violation of Cosmic `no-hair' Conjecture is not
unusual because the Szekeres metric may not have always
non-positive 3-space curvature scalar. Also in this higher
dimensional Szekeres space-time we have solutions which expands
as de Sitter in some directions but not in other directions.\\

\large \baselineskip .85cm
\chapter{Shell Focusing Singularities in Quasi-Spherical Gravitational Collapse}
\label{chap11}\markright{\it CHAPTER~\ref{chap11}. Shell Focusing
Singularities in Quasi-Spherical Gravitational Collapse}

\section{Prelude}
An extensive study [Joshi et al, 1994, 1999; Lake. 1992; Ori et
al, 1987; Harada, 1998] of gravitational collapse has been carried
out of Tolman-Bondi-Lema\^{\i}tre (TBL) spherically symmetric
space-times containing irrotational dust. Due to simplifications
introduced by the spherical symmetry several generalizations of
this model have been considered. A general conclusion from these
studies is that a central curvature singularity forms but its
local or global visibility depends on the initial data. Also the
study of higher-dimensional spherical collapse reveals the
interesting feature that visibility of singularity is impossible
in space-times with more than five dimensions with proper choice
of regular initial data [Goswami et al, 2002].\newline

By contrast, there is very little progress in studying
non-spherical collapse. The quasi-spherical dust collapse models,
given by Szekeres metric [Szekeres, 1975] were analyzed by
Szekeres himself [1975], Joshi and Krolak [1996], Deshingkar,
Joshi and Jhingan [1998] and extensively by Goncalves [2001]. The
solutions for dust and a non-zero cosmological constant were
found by Barrow and Stein-Schabes [1984]. In this work, we study
the gravitational collapse in the recently generalized
($n+2$)-dimensional Szekeres metric. As in four-dimensional
space-time, this higher dimensional model does not admit any
Killing vector and the description quasi-spherical arises because
it has invariant family of spherical
hypersurfaces.\\

In this chapter, we study the occurrence and nature of naked
singularities for a dust model with non-zero cosmological
constant in ($n+2$)-dimensional Szekeres space-times (which
possess no Killing vectors) for $n\geq 2$. We find that central
shell-focusing singularities may be locally naked in higher
dimensions but depend sensitively on the choice of initial data.
In fact, the nature of the initial density determines the
possibility of naked singularity in space-times with more than
five dimensions. The results are similar to the collapse in
spherically symmetric Tolman-Bondi-Lema\^{\i}tre space-times. We
also extend our study for global characteristic of the
singularity by studying both null and time like geodesic
originated from the singularity using the formalism of Barve et
al [1999]. The nature of the central shell focusing singularity
so formed is analyzed by studying both the radial null and
time-like geodesic originated from it.\newline

\section{Basic Equations and Regularity Conditions}
The dust solutions have been obtained for ($n+2$)-dimensional
Szekeres' space-time metric for which the line element is given
in equation (10.1). Under the assumption that $\beta ^{\prime
}\neq 0$, the explicit form of the metric coefficients are also
given in equations (10.2) and (10.3), where the form of $\nu$ is
given in equations (10.16) and (10.17) and $R$ satisfied the
differential equation
\begin{equation}
\dot{R}^{2}=f(r)+\frac{F(r)}{R^{n-1}}+\frac{2\Lambda}{n(n+1)}~R^{2}~.
\end{equation}

Here $\Lambda $ is cosmological constant, $f(r)$ and $F(r)$ are
arbitrary functions of $r$ alone.

In the subsequent discussion we shall restrict ourselves to the
quasi-spherical space-time which is characterized by the $r$ dependence of
the function $\nu $ (i.e., $\nu ^{\prime }\neq 0$). From the Einstein field
equations we have an expression for energy density for the dust, as

\begin{equation}
\rho(t,r,x_{1},...,x_{n}) =\frac{n}{2}~\frac{F^{\prime }+(n+1)F\nu ^{\prime }%
}{R^{n}(R^{\prime }+R\nu ^{\prime })}.
\end{equation}

Thus a singularity will occur when either (i) $R=0$ i.e., $\beta
=-\infty $ or (ii) $\alpha =-\infty $. Using the standard
terminology for spherical collapse, the first case corresponds to
a shell-focusing singularity, while the second case gives rise to
a shell-crossing singularity. As in a TBL space-time, the
shell-crossing singularities are gravitationally weak, and we
shall not consider them any further here. Hence, we shall restrict
ourselves to the situation with $\alpha >-\infty $.

Suppose that $t=t_{i}$ is the initial hypersurface from which the collapse
develops. For the initial data we assume that $R(t_{i},r)$ is a
monotonically increasing function of $r$. So, without any loss of
generality, we can label the dust shells by the choice $R(t_{i},r)=r$. As a
result, the expression for the initial density distribution is given by

\begin{equation}
\rho_{i}(r,x_{1},...,x_{n})=\rho(t_{i},r,x_{1},...,x_{n})=\frac{n}{2}~\frac{%
F^{\prime}+(n+1)F\nu^{\prime}}{r^{n}(1+r\nu^{\prime})}
\end{equation}

If \ we started the collapse from a regular initial hypersurface the
function $\rho _{i}$ must be non-singular (and also positive for a
physically realistic model). Furthermore, in order for the space-time to be
locally flat near $r=0,$ we must have $f(r)\rightarrow 0$ as $r\rightarrow 0$%
. Then, from equation (11.1), the boundedness of $\dot{R}^{2}$ as
$r\rightarrow
0$ demands that $F(r)\sim O(r^{m})$ where $m\geq n-1$. But, for small $r$, $%
\rho _{i}(r)\simeq \frac{nF^{\prime }}{2r^{n}}$ and consequently, for
regular $\rho _{i}(r)$ near $r=0$, we must have $F(r)\sim O(r^{n+1})$. Thus,
starting with a regular initial hypersurface, we can express $F(r)$ and $%
\rho _{i}(r)$ by power series near $r=0$ as

\begin{equation}
F(r)=\sum_{j=0}^{\infty }F_{j}~r^{n+j+1}
\end{equation}%
and
\begin{equation}
\rho _{i}(r)=\sum_{j=0}^{\infty }\rho _{j}~r^{j}.
\end{equation}

As $\nu ^{\prime }$ appears in the expression for the density as well as in
the metric coefficient, we can write

\begin{equation}
\nu ^{\prime }(r)=\sum_{j=-1}^{\infty }\nu _{j}~r^{j}
\end{equation}%
where $\nu _{_{-1}}\ge -1$.

Now, using these series expansions in equation (11.2) we have the
following relations between the coefficients,
\begin{eqnarray*}
\rho_{0}=\frac{n(n+1)}{2}F_{0},~~\rho_{1}=\frac{n}{2}\left(n+1+\frac{1}{1+
\nu_{_{-1}}}\right)F_{1},
\end{eqnarray*}
\vspace{-5mm}
\begin{eqnarray*}
\rho_{2}=\frac{n}{2}\left[\left(n+1+\frac{2}{1+
\nu_{_{-1}}}\right)F_{2}-\frac{F_{1}\nu_{_{0}}}{(1+\nu_{_{-1}})^{2}}\right],
\end{eqnarray*}
\vspace{-3mm}
\begin{equation}
\rho_{3}=\frac{n}{2}\left[\left(n+1+\frac{3}{1+
\nu_{_{-1}}}\right)F_{3}-\frac{2F_{2}\nu_{_{0}}}{(1+\nu_{_{-1}})^{2}}-
\frac{(1+\nu_{_{-1}})\nu_{_{1}}-\nu_{_{0}}^{2}}{(1+\nu_{_{-1}})^{3}}F_{1}\right],
\end{equation}

$$...~~...~~...~~...~~...~~...~~...~~...,$$
$$OR$$
\begin{eqnarray*}
\rho_{0}=\frac{n}{2}\left[\frac{F_{1}}{\nu_{0}}+(n+1)F_{0}\right],~~
\rho_{1}=\frac{n}{2}\left[\frac{2F_{2}}{\nu_{0}}+\left\{(n+1)-
\frac{\nu_{1}}{\nu_{0}^{2}}\right\}F_{1}\right],
\end{eqnarray*}
\vspace{-5mm}
\begin{eqnarray*}
\rho_{2}=\frac{n}{2}\left[\frac{3F_{3}}{\nu_{0}}+\left\{(n+1)-
\frac{2\nu_{1}}{\nu_{0}^{2}}\right\}F_{2}+\left(\frac{\nu_{1}^{2}}{\nu_{0}^{3}}-
\frac{\nu_{2}}{\nu_{0}^{2}}\right)F_{1}\right],
\end{eqnarray*}
\vspace{-5mm}
\begin{equation}
\rho_{3}=\frac{n}{2}\left[\frac{4F_{4}}{\nu_{0}}+\left\{(n+1)-
\frac{3\nu_{1}}{\nu_{0}^{2}}\right\}F_{3}+2\left(\frac{\nu_{1}^{2}}{\nu_{0}^{3}}-
\frac{\nu_{2}}{\nu_{0}^{2}}\right)F_{2}+\left(\frac{2\nu_{1}\nu_{2}}{\nu_{0}^{3}}-
\frac{\nu_{3}}{\nu_{0}^{2}}-\frac{\nu_{1}^{3}}{\nu_{0}^{4}}\right)F_{1}\right],
\end{equation}
$$...~~...~~...~~...~~...~~...~~...~~...,$$

according as~~ $\nu_{_{-1}}>-1$~ or ~$\nu_{_{-1}}=-1$.

Finally, if we assume that the density gradient is negative and
falls off rapidly to zero near the centre then we must have $\rho
_{1}=0$ and $\rho _{2}<0$. Consequently, we have the restrictions
that $F_{1}=0$ and $F_{2}<0$ for $\nu_{_{-1}}>-1$.\\

\section{Formation of Singularity and its Nature}
In order to form a singularity from the gravitational collapse of
dust we
first require that all portions of the dust cloud are collapsing i.e., $\dot{%
R}\leq 0$. Let us define $t_{sf}(r)$ and $t_{sc}(r)$ as the time
for shell-focusing and shell-crossing singularities to occur
occurring at radial coordinate $r$. This gives the relations
[Szekeres, 1975]

\begin{equation}
R(t_{sf}(r),r)=0
\end{equation}%
and
\begin{equation}
R^{\prime }(t_{sc},r)+R(t_{sc},r)\nu ^{\prime }(r,x_{1},x_{2},...,x_{n})=0.
\end{equation}%
Note that `$t_{sc}$' may also depend on $x_{1},x_{2},...,x_{n}$.

As mentioned earlier, the shell-crossing singularity is not of much physical
interest so we shall just consider the shell-focusing singularity for the
following two cases:\newline

(i) \textit{Marginally bound case} : $f(r)=0$ with $\Lambda\ne 0$

In this case equation (11.1) can easily be integrated to give

\begin{equation}
t=t_{i}+\sqrt{\frac{2n}{(n+1)\Lambda}}\left[Sinh^{-1}\left(\sqrt{\frac{%
2\Lambda r^{n+1}}{n(n+1)F(r)}}\right)-Sinh^{-1}\left(\sqrt{\frac{2\Lambda
R^{n+1}} {n(n+1)F(r)}}\right)\right]
\end{equation}%
Also, from equations (11.1) and (11.11), we have

\begin{equation}
t_{sf}(r)=t_{i}+\sqrt{\frac{2n}{(n+1)\Lambda}}~Sinh^{-1}\left(\sqrt{\frac{%
2\Lambda r^{n+1}}{n(n+1)F(r)}}\right)
\end{equation}

\vspace{7mm}

(ii) \textit{Non-marginally bound case with time symmetry} : $f(r)\neq 0,~%
\dot{R}(t_{i},r)=0, \Lambda=0$

Here, the solution of equation (11.1) gives $t$ as a function of
$r$ :

\begin{equation}
t=t_{i}+\frac{2}{(n+1)\sqrt{F}}\left[\frac{\sqrt{\pi}~\Gamma(b+1)}{%
\Gamma(b+1/2)}~ r^{\frac{n+1}{2}}-R^{\frac{n+1}{2}}~_{2}F_{1}[\frac{1}{2}%
,b,b+1,\left(\frac{R}{r}\right)^{n-1}] \right]
\end{equation}

and so the time of the shell-focusing singularity is given by

\begin{equation}
t_{sf}(r)=t_{i}+\frac{2\sqrt{\pi }}{(n+1)\sqrt{F}}\frac{\Gamma
(b+1)}{\Gamma (b+1/2)}~r^{\frac{n+1}{2}},
\end{equation}%
where $_{2}F_{1}$ is the usual hypergeometric function with $b=\frac{1}{2}+%
\frac{1}{n-1}$~. However, for the five-dimensional space-time ($n=3)$, $R$
has the particularly simple form

\begin{equation}
R^{2}=r^{2}-\frac{F(r)}{r^{2}}(t-t_{i})^{2},
\end{equation}%
and therefore
\begin{equation}
t_{sf}(r)=t_{i}+\frac{r^{2}}{\sqrt{F(r)}}.
\end{equation}

\vspace{7mm}

\section{Formation of Trapped Surfaces}
The event horizon of observers at infinity plays an important
role in the nature of the singularity. However, due to the
complexity of the calculation, we shall consider a trapped
surface which is a compact space-time 2-surface whose normals on
both sides are future-pointing converging null-geodesic families.
In particular, if ($r=$ constant, $t=$ constant) the 2-surface
$S_{r,t}$ is a trapped surface then it, and its entire future
development, lie behind the event horizon provided the density
falls off fast enough at infinity.

If $t_{ah}(r)$ is the instant of the formation of apparent horizon
then we must have [Szekeres, 1975]\\
\begin{equation}
2\Lambda
R^{n+1}(t_{ah}(r),r)-n(n+1)R^{n-1}(t_{ah}(r),r)+n(n+1)F(r)=0.
\end{equation}%
Thus, for the above two cases, the explicit expressions for $t_{ah}$ are the
following:

\begin{equation}
t_{ah}(r)=t_{i}+\sqrt{\frac{2n}{(n+1)\Lambda}}\left[Sinh^{-1}\left(\sqrt{\frac{%
2\Lambda r^{n+1}}{n(n+1)F(r)}}\right)-Sinh^{-1}\left(\sqrt{\frac{2\Lambda
R^{n+1}(t_{ah},r)} {n(n+1)F(r)}}\right)\right]
\end{equation}%
for $f(r)=0, \Lambda\ne 0$ and

\begin{equation}
t_{ah}(r)=t_{i}+\frac{2}{(n+1)\sqrt{F}}\left[ \frac{\sqrt{\pi }~\Gamma (b+1)}{%
\Gamma (b+1/2)}~r^{\frac{n+1}{2}}-F^{b}~~_{2}F_{1}[\frac{1}{2},b,b+1,\frac{F%
}{r^{n-1}}]\right]
\end{equation}%
for $f(r)\neq 0,~\dot{R}(t_{i},r)=0, \Lambda=0$ and for $n=3$ we get

\begin{equation}
t_{ah}(r)=t_{i}+r\sqrt{\frac{r^{2}}{F}-1}.
\end{equation}

From the expressions for $t_{sf}(r)$ and $t_{ah}(r)$, we note that
the shell-focusing singularity that appears at $r>0$ is in the
future of the apparent horizon in both cases. But since we are
interested in the central shell-focusing singularity (at $r=0$)
we require the time of occurrence of central shell-focusing
singularity $t_{0}(=t_{sf}(0))$. From equations (11.12), (11.14),
(11.16) and (11.17)-(11.20), taking the limit as $r\rightarrow 0$,
we have that\\
\begin{eqnarray*}
t_{ah}(r)-t_{0}=-\frac{1}{(n+1)F_{0}^{3/2}\sqrt{1+\frac{2\Lambda}{n(n+1)F_{0}%
}}}\left[F_{1}r+ \left(F_{2}-\frac{(4\Lambda+3n(n+1)F_{0})F_{1}^{2}}{%
4F_{0}(2\Lambda+n(n+1)F_{0})}\right)r^{2} +\right.
\end{eqnarray*}
\vspace{-3mm}
\begin{eqnarray*}
\left. \left(F_{3}-\frac{(4\Lambda+3n(n+1)F_{0})F_{1}F_{2}}{%
4F_{0}(2\Lambda+n(n+1)F_{0})}+ \frac{(32\Lambda^{2}+40n(n+1)\Lambda
F_{0}+15n^{2}(n+1)^{2}+F_{0}^{2})F_{1}^{3}}{24F_{0}^{2}(2%
\Lambda+n(n+1)F_{0})^{2}} \right)r^{3}\right.
\end{eqnarray*}
\vspace{-3mm}
\begin{equation}
\left.+......\frac{}{}\right]-\frac{2}{(n+1)F_{0}^{\frac{n-2}{n-1}}}\left[%
F_{0}r^{\frac{n+1}{n-1}}+\frac{1}{n-1}F_{1} r^{\frac{2n}{n-1}}+\frac{1}{n-1}%
\left(F_{2}-\frac{n-2}{2(n-1)}\frac{F_{1}}{F_{0}} \right)r^{\frac{3n-1}{n-1}%
}+...\right]
\end{equation}
for $f(r)=0, \Lambda\ne 0$ and\\
\begin{eqnarray*}
t_{ah}(r)-t_{0}=-\frac{\sqrt{\pi}}{(n+1)F_{0}^{3/2}}\frac{\Gamma(b+1)}{%
\Gamma(b+1/2)} \left[F_{1}r+\left(F_{2}-\frac{3F_{1}^{2}}{4F_{0}}%
\right)r^{2}+\left(F_{3}- \frac{9F_{1}F_{2}}{4F_{0}}\right.\right.
\end{eqnarray*}
\vspace{-3mm}
\begin{equation}
\left. \left. +\frac{5F_{1}^{3}}{8F_{0}^{2}}\right) r^{3}+...\right] -\frac{2%
}{(n+1)F_{0}^{\frac{n-2}{n-1}}}\left[ F_{0}r^{\frac{n+1}{n-1}}+\frac{1}{n-1}%
F_{1}r^{\frac{2n}{n-1}}+...\right]
\end{equation}

for $f(r)\ne 0,~ \dot{R}(t_{i},r)=0, \Lambda=0$ and \newline

for $n=3$ (with $f(r)\ne 0,~ \dot{R}(t_{i},r)=0, \Lambda=0$)
\begin{equation}
t_{ah}(r)-t_{0}=-\frac{1}{2F_{0}^{3/2}}\left[F_{1}~r+\left(F_{2}+F_{0}^{2}-
\frac{3F_{1}^{2}}{4F_{0}}\right)r^{2}+......\right]
\end{equation}

Here, $t_{0}$ is the time at which the singularity is formed at $r=0,$ and $%
t_{ah}(r)$ is the instant at which a trapped surface is formed at a distance
$r.$ Therefore, if $t_{ah}(r)>t_{0},$ a trapped surface will form later than
the epoch at which any light signal from the singularity can reach an
observer. Hence the necessary condition for a naked singularity to form is $%
t_{ah}(r)>t_{0};$ while that for a black hole to form is
$t_{ah}(r)\leq t_{0} $. It is to be noted that this criteria for
naked singularity is purely local. Hence, in the present problem
it possible to have local naked singularity or a black hole form
under the conditions shown in the Table I, for $\nu_{_{-1}}>-1$.
\[
\text {TABLE-I}
\]
\[
\begin{tabular}{|l|l|r|r|r|}
\hline\hline
& \multicolumn{1}{c|}{\emph{Naked Singularity}} & \multicolumn{1}{c|}{\emph{%
Black Hole}} \\ \hline
&  &   \\
Marginally bound case: & (i) $\rho_{1}<0,~~ \forall n$ & (i) $\rho_{1}=0,
\rho_{2}<0,n=3,$~~~~~~~~~~~~~  \\
$f(r)=0, \Lambda\ne 0$ & (ii) $\rho_{1}=0, \rho_{2}<0, n=2$ & $%
F_{2}\ge-2F_{0}^{2}\sqrt{1+\frac{\Lambda}{6F_{0}}}$~~~~~~~~~~~~~  \\
& (iii) $\rho_{1}=0, \rho_{2}<0,n=3,$ & (ii) $\rho_{1}=0, \rho_{2}<0, n\ge4$
~~~~~~~~~~~~  \\
& ~~~~~$F_{2}<-2F_{0}^{2}\sqrt{1+\frac{\Lambda}{6F_{0}}}$ & (iii) $%
\rho_{1}=0, \rho_{2}=0,\rho_{3}<0,n=2,$~~  \\
& (iv) $\rho_{1}=0, \rho_{2}=0,\rho_{3}<0,$ & $F_{3}\ge-2F_{0}^{5/2}\sqrt{1+%
\frac{\Lambda}{3F_{0}}}$~~~~~~~~~~~  \\
& ~~~~~$n=2,$ & (iv) $\rho_{1}=0, \rho_{2}=0,\rho_{3}<0, n\ge 3$
~~~\\
& ~~~$F_{3}<-2F_{0}^{5/2}\sqrt{1+\frac{\Lambda}{3F_{0}}}$ & (v) $\rho_{1}=0,
\rho_{2}=0,...,\rho_{j}=0,$~~~~~~~~  \\
&  & $\rho_{j+1}<0, j\ge 3, \forall n$~~~~~~~~~~~~~~~~\\
\hline
&  &   \\
Non-marginally bound & (i) $\rho_{1}<0, \forall n $ & (i) $\rho_{1}=0,
\rho_{2}<0, n=3,F_{2}\ge-F_{0}^{2}$\\
case with time  & (ii) $\rho_{1}=0, \rho_{2}<0, n=2$ & (ii) $%
\rho_{1}=0, \rho_{2}<0, n\ge 4$~~~~~~~~~~~~~~ \\
symmetry:~$f(r)\ne 0,$ & (iii) $\rho_{1}=0, \rho_{2}<0, n=3,$ & (iii) $%
\rho_{1}=0, \rho_{2}=0,\rho_{3}<0, n=2,$~~~   \\
$R(t_{i},r)=0, \Lambda=0$ & ~~~~~$F_{2}<-F_{0}^{2}$ & $F_{3}\ge -\frac{8}{3\pi}F_{0}^{5/2}$%
~~~~~~~~~~~~~~~~~~~~~  \\
& (iv) $\rho_{1}=0, \rho_{2}=0,\rho_{3}<0,$ & (iv) $\rho_{1}=0,
\rho_{2}=0,\rho_{3}<0, n\ge 3$~~~  \\
& ~~~~~$n=2, F_{3}<-\frac{8}{3\pi}F_{0}^{5/2}$ & (v) $\rho_{1}=0,
\rho_{2}=0,...,\rho_{j}=0,$ ~~~~~~~  \\
&  & $\rho_{j+1}<0, j\ge 3, \forall n$~~~~~~~~~~~~~~~~   \\
\hline\hline
\end{tabular}%
\]%
\\
However, if we relax the condition on the density gradient
(namely the condition that $\rho _{1}=0,~\rho _{2}<0$ i.e.,
$F_{1}=0,~F_{2}<0$ for $\nu_{_{-1}}>-1$) then it is possible to
have a naked singularity in all dimensions. These results are
identical to those obtained in the presence of spherical
symmetry. Hence the local nature of singularity is not affected
by the geometry of the space-time with respect to whether it is
spherical or quasi-spherical. However, we note that the Szekeres
space-times retain certain special geometrical properties shared
by spherical space-times in all dimensions, in particular they do
not allow gravitational radiation to be present in the
space-time.\\

\section{Radial Null Geodesics}
Now we shall consider $\Lambda=0$, $f(r)=0$ and initial time
$t_{i}=0$ so that $R(0,r)=0$. So the solution for $R$ of the
equation (11.1) becomes

\begin{equation}
R=\left[r^{\frac{n+1}{2}}-\frac{n+1}{2}\sqrt{F(r)}~t\right]^{\frac{2}{n+1}}
\end{equation}

Now with the help of equations (11.9) and (11.24), the
singularity time can be expressed as
\begin{equation}
t_{sf}(r)=\frac{2}{n+1}\frac{r^{\frac{n+1}{2}}}{\sqrt{F(r)}}
\end{equation}

Near $r=0$, above singularity curve can be approximately written
as

\begin{equation}
t_{sf}(r)=t_{0}-\frac{F_{m}}{(n+1)F_{0}^{3/2}}~r^{m}
\end{equation}

where $m~(\ge 1)$ is an integer and $F_{m}$ is the first
non-vanishing term beyond $F_{0}$ in the series form (11.4) and
$t_{0}=\frac{2}{(n+1)\sqrt{F_{0}}}$.

The equation of the outgoing radial null geodesic (ORNG) which
passes through the central singularity in the past is taken as
(near $r=0$)\\
\begin{equation}
t_{ORNG}=t_{0}+a r^{\xi},
\end{equation}

to leading order in $t$-$r$ plane with $a>0,\xi>0$. Now to
visualize the singularity the time $t$ in the geodesic (11.27)
should be less than $t_{sf}(r)$ in equation (11.26). Hence
comparing these times we have the restrictions

\begin{equation}
\xi\ge m ~~~~ \text{and} ~~~~ a<-\frac{F_{m}}{(n+1)F_{0}^{3/2}}
\end{equation}

Further, for the metric (10.1) an outgoing radial null geodesic
should satisfy

\begin{equation}
\frac{dt}{dr}=R'+R~\nu'
\end{equation}

We shall now examine the feasibility of the null geodesic
starting from the singularity with the above restrictions for the
following two cases namely, (i) $\xi>m$ and (ii) $\xi=m$.\\

When $\xi>m$  then near $r=0$ the solution for $R$ in (11.24)
simplifies to

\begin{equation}
R=\left(-\frac{F_{m}}{2F_{0}}\right)^{\frac{2}{n+1}}r^{\frac{2m}{n+1}+1}
\end{equation}

Now combining (11.27) and (11.30) in equation (11.29) we get (upto
leading order \\in $r$)
\begin{equation}
a~\xi~
r^{\xi-1}=\left(\nu_{_{-1}}+1+\frac{2m}{n+1}\right)\left(-\frac{F_{m}}{2F_{0}}\right)
^{\frac{2}{n+1}}r^{\frac{2m}{n+1}}
\end{equation}

which implies
\begin{equation}
\xi=1+\frac{2m}{n+1}~~~ \text{and}~~~
a=\frac{(\nu_{_{-1}}+1+\frac{2m}{n+1})}{(1+\frac{2m}{n+1})}\left(-\frac{F_{m}}
{2F_{0}}\right)^{2/(n+1)}
\end{equation}

We note that if $\nu_{_{-1}}=0$ then the above results are same
as TBL model in eq. (9.38), so we restrict ourselves to
$\nu_{_{-1}}\ne 0$. As $\xi>m$ so from (11.32)

$$
m<\frac{n+1}{n-1}
$$

Since $m$ is an integer so we have the following possibilities:

$ ~~~~~~~~~~~~~~~~n=2~ (4D \text{~space-time}): ~~ m=1, 2  ~~~, \xi=\frac{4}{3}, \frac{7}{3}$\\
$ ~~~~~~~~~~~~~~~~~~~~n\ge 3~ (5D \text{~or~ higher dim.
space-time}): ~~ m=1 ~~~, \xi=1+\frac{2}{n+1} $.\\

Now for $\nu_{_{-1}}>-1$, if we assume that the initial density
gradient falls off rapidly to zero at the centre then we must have
$\rho_{1}=0, \rho_{2}<0$ i.e., $F_{1}=0, F_{2}<0$. Hence the
least value of $m$ should be 2. Thus for a self consistent
outgoing radial null geodesic starting from the singularity we
must have $n=2$ only. So the singularity will be naked only in
four dimension. However, if we relax the restriction on the
initial density and simply assume that the initial density
gradually decreases as we go away from the centre (i.e.,
$\rho_{1}<0$) then $F_{1}<0$ and $m=1$ is the possible solution.
Therefore, naked singularity is possible in any dimension. However
for $\nu_{_{-1}}=-1$, the assumption about the initial density
gradient ($\rho_{1}=0, \rho_{2}<0$) does not implying $F_{1}=0$.
Hence here $m=1$ is a possible solution for all dimension with
$F_{1}<0$. Therefore naked singularity may occur in any dimension
for this choice of the parameter $\nu_{-1}$.

Now we shall study the second case namely $\xi=m$. Here also for
$\nu_{_{-1}}=0$ we have identical results as in TBL model
((9.39), (9.40)). For $\nu_{_{-1}}\ne 0$, using the same procedure
as above we get
$$
m=\frac{n+1}{n-1}
$$
and
\begin{equation}
a=-\frac{1}{m}\left(-\frac{n+1}{2}\sqrt{F_{0}}~a-\frac{F_{m}}{2F_{0}}
\right)^{\frac{1-n}{1+n}}
\left[(\nu_{_{-1}}+m)\frac{F_{m}}{2F_{0}}+\frac{1}{2}(n+1)(\nu_{_{-1}}+1)\sqrt{F_{0}}~a\right]
\end{equation}

Hence integral solution for $m$ is possible only for $n=2$ and 3.
So consistent outgoing radial null geodesic through the central
shell focusing singularity is possible only upto five dimension
i.e., we cannot have naked singularity for higher dimensional
space-time greater than five. It is to be noted that this
conclusion is not affected by the value of the parameter
$\nu_{_{-1}}$.

Now we shall examine whether the restriction (11.28) for $a$ is
consistent with the expression $a$ in equation (11.33). In fact
equation (11.33) takes the form\\
\begin{equation}
2^{\frac{2}{n+1}}bm=-\left[-(n+1)b-\zeta\right]^{-\frac{1}{m}}\left[(\nu_{_{-1}}+m)\zeta+
(n+1)(\nu_{_{-1}}+1)b\right]
\end{equation}

with the transformation
\begin{equation}
a=bF_{0}^{\frac{1}{n-1}},~~~~F_{m}=\zeta F_{0}^{\frac{m}{2}+1}.
\end{equation}

Since equation (11.34) is a real valued equation of $b$, so we
must have
$$
(n+1)b+\zeta<0~,
$$

\begin{figure}
\includegraphics{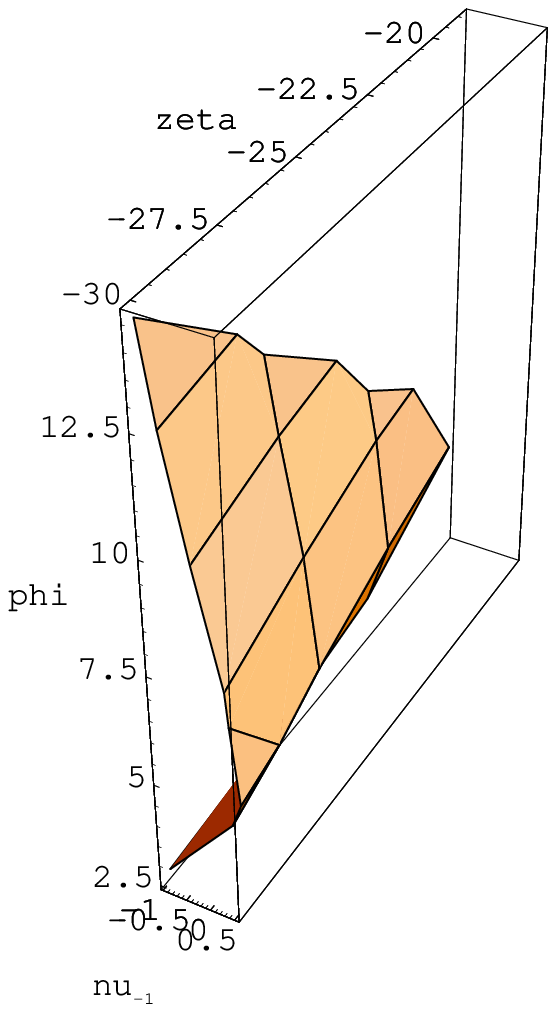}~~~~~~~~~~~
\includegraphics{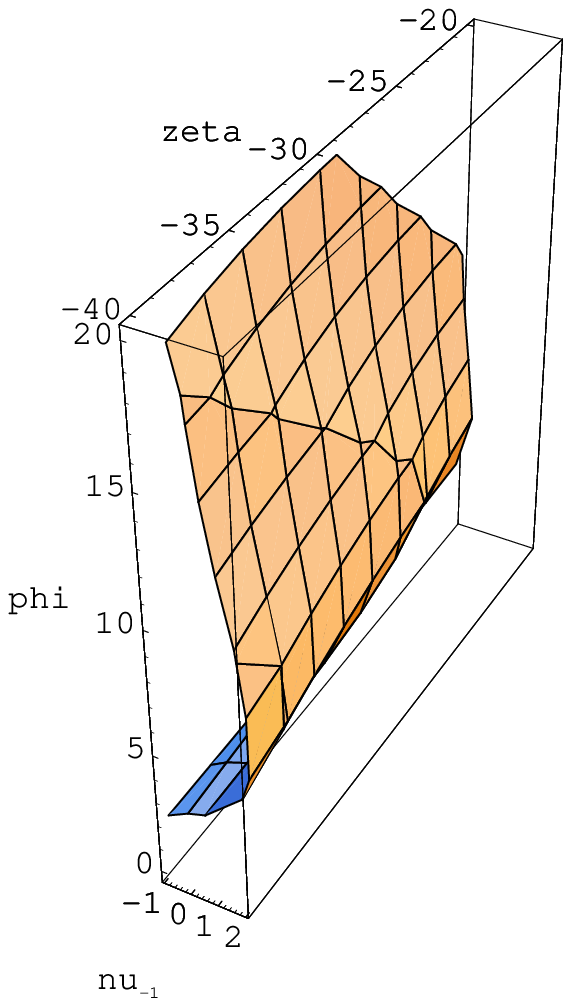}
\\\vspace{5mm}
~~~~~~~~~~~Fig 11.1~~~~~~~~~~~~~~~~~~~~~~~~~~~~~~~~~~~~~~~~~~~~~~~~~~~~~~~~~~Fig 11.2\\
\vspace{5mm}
\includegraphics{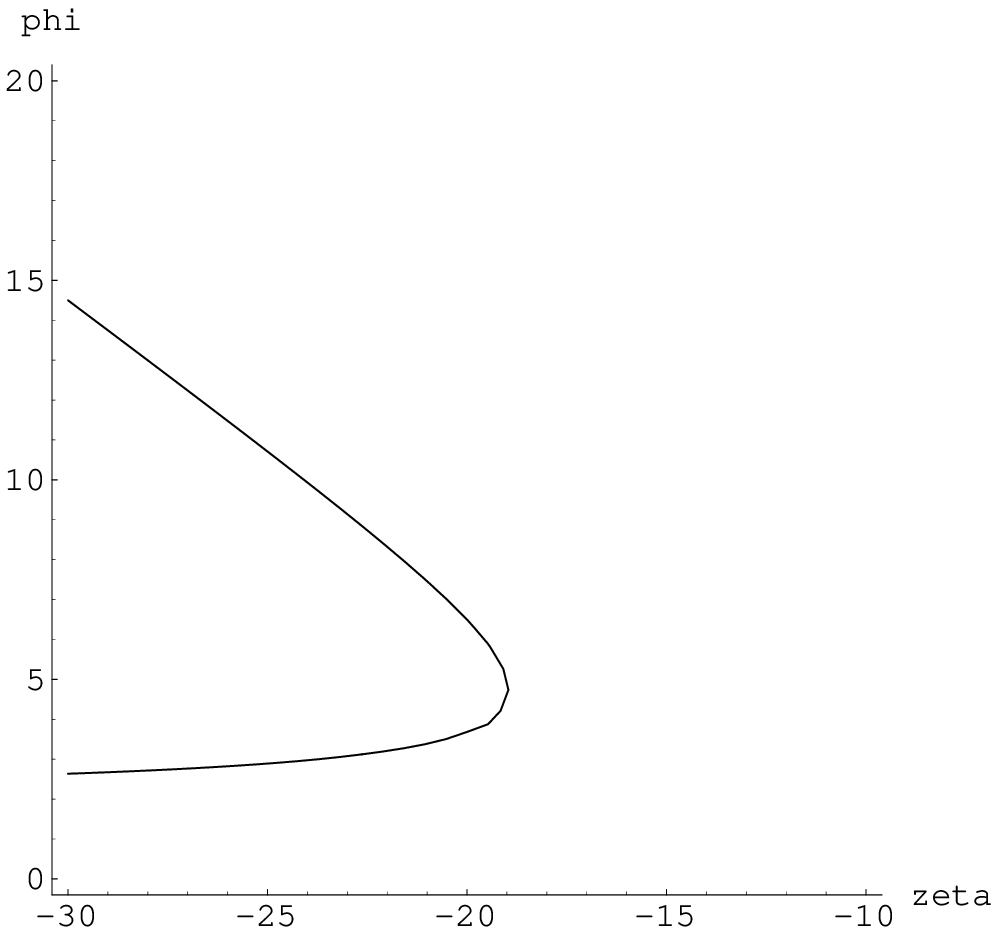}~
\includegraphics{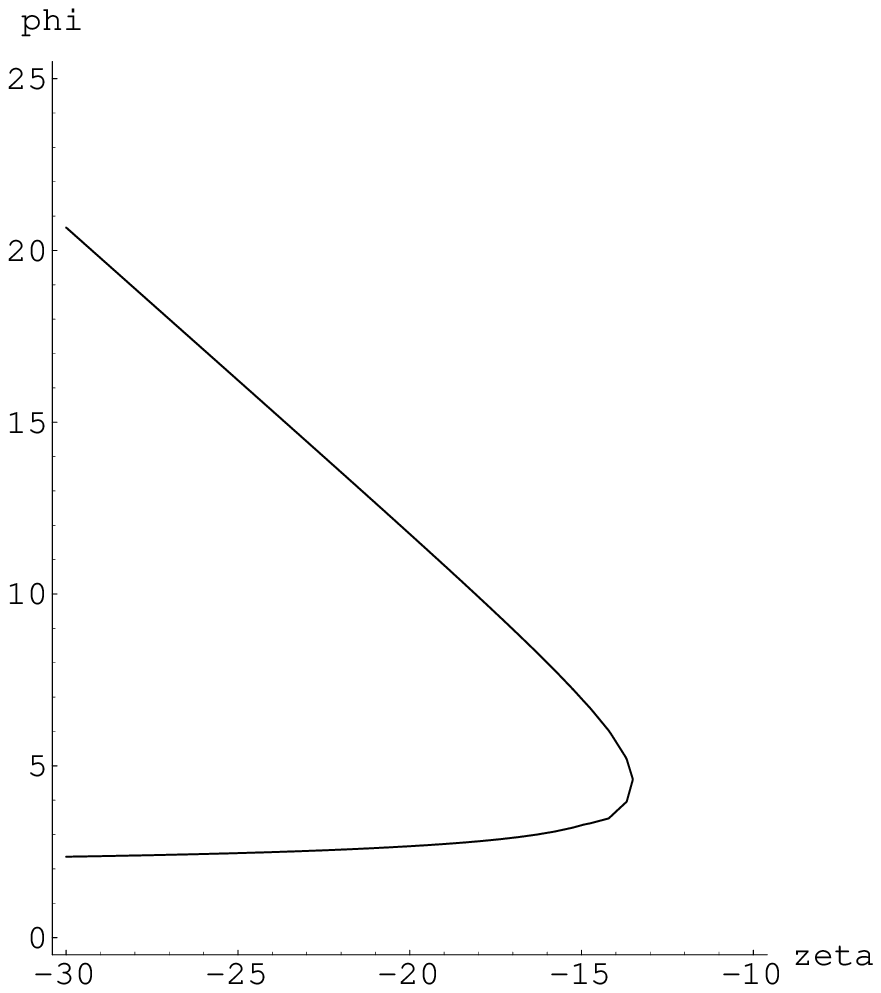}
\\\vspace{5mm}
~~~~~~~~~~~Fig 11.3~~~~~~~~~~~~~~~~~~~~~~~~~~~~~~~~~~~~~~~~~~~~~~~~~~~~~~~~~~~~~~~~~~~~~~~~~~~~~~~~~Fig 11.4\\
\vspace{4mm} {In Figs 11.1 and 11.2, we have shown the variation
of $\phi~(0<\phi<-\zeta)$ for the variations of $\nu_{_{-1}}(>-1)$
and $\zeta(<0)$ for four and five dimensions (i.e., $n=2$ and 3)
respectively. However for $\nu_{_{-1}}=-1$, the dependence of
$\phi$ over $\zeta$ has been presented in Figs 11.3 and 11.4 for
$n=2$
and 3 respectively.~~~~~~~}\\
\vspace{1mm}
\end{figure}

which using (11.35) gives us the restriction on $a$ in equation
(11.28). Hence the geodesic (11.27) will have consistent solution
for $a$ and $m$. So the above conclusion regarding the formation
of naked singularity is justified. Further, introducing the
variable $\phi$ by the relation
\begin{equation}
\phi=-(n+1)b-\zeta
\end{equation}

We have seen from equation (11.34)

\begin{equation}
4\phi^{n-1}(\zeta+\phi)^{n+1}=[2\zeta-(n-1)(\nu_{_{-1}}+1)\phi]^{n+1}
\end{equation}

with the restriction ~$0<\phi<-\zeta$.

Finally, we investigate whether there is only one null geodesic
emanating from the singularity or it is possible to have an
entire family of geodesic through the singularity. So we write
the equation for the outgoing radial null geodesics to next order
as
\begin{equation}
t=t_{0}+a r^{\xi}+h r^{\xi+\sigma}
\end{equation}

where as before $a, \xi, h, \sigma$ are all positive. Then
proceeding in the same way we have

(i) when $\xi>m$, $a$ and $\xi$ have the same expressions as in
equation (11.32) while

\begin{equation}
\sigma=\frac{m(1-n)}{1+n}+1
\end{equation}
and
\begin{equation}
h=\frac{a}{\xi+\sigma}(-\nu_{_{-1}}-\sigma)\sqrt{F_{0}}\left(-\frac{F_{m}}
{2F_{0}}\right)^{\frac{1-n}{1+n}}
\end{equation}

We note that since here $m$ can have values 1 and 2 for $n=2$ and
$m=1$ for $n\ge 3$ so $\sigma$ is always positive. But for $h$ to
be positive definite, $\nu_{_{-1}}$ is restricted within the
range $-1\le \nu_{_{-1}}\le -\sigma$ i.e., $\nu_{_{-1}}$ is
negative definite.

(ii) for $\xi=m$, $a$ and $\xi$ have the same expressions as in
equation (11.33) and $\sigma$ is obtained from the equation
\begin{equation}
m+\sigma=2^{\frac{n-1}{n+1}}~b(n+1)\left[-\zeta-(n+1)b\right]^{-\frac{2n}{n+1}}
-2^{\frac{n-1}{n+1}}~\nu_{_{-1}}~F_{0}^{-1/2}\left[-\zeta-(n+1)b\right]^{\frac{1-n}{1+n}}
\end{equation}

with expressions for $b$ and $\zeta$ from equation (11.35), while
the other constant $h$ is totally arbitrary. Thus it is possible
to have an entire family of outgoing null geodesics terminated in
the past at the singularity, provided $\sigma$ obtained from
equation (11.41) is positive definite.\\

\section{Radial Time-Like Geodesics}
We shall now investigate whether it is possible to have any
outgoing time-like geodesic originated from the singularity. For
simplicity of calculation we shall consider only outgoing radial
time-like geodesic (ORTG). Let us denote by
$K^{a}=\frac{dx^{a}}{d\tau}$, a unit tangent vector field to a
ORTG with $\tau$ an affine parameter along the geodesic. Hence
from the geodesic equation we have [Goncalves, 2001]

\begin{equation}
K^{t}\left(K^{r}\right)^{.}+2\dot{\alpha}K^{t}K^{r}+K^{r}\left(K^{r}\right)'+
\alpha'\left(K^{r}\right)^{2}=0
\end{equation}
with
\begin{equation}
K^{t}=\pm\sqrt{1+e^{2\alpha}\left(K^{r}\right)^{2}}~~.
\end{equation}

In order to satisfy the above two equations the simplest choice
for ($K^{t}, K^{r}$) is
\begin{equation}
K^{t}=\pm 1,~~~~K^{r}=0~.
\end{equation}

and so we have the solution
\begin{equation}
t-t_{0}=\pm(\tau-\tau_{0}),~~~~r=r_{0}=\text{constant}.
\end{equation}

Here $+$ (or $-$) sign corresponds to ORTG (or ingoing RTG) and
$\tau_{0}$ is the proper time at which radial time-like geodesic
passes through the central singularity.

Now similar to null geodesic let us choose the radial time-like
geodesic near the singularity to be of the form (to leading order)
\begin{equation}
t_{ORTG}(r)=t_{0}+c~r^{p}
\end{equation}

with $c$ and $p$ as positive constants. Further, consistent with
equations (11.44) and (11.45) we assume
\begin{equation}
K^{r}(t,r)=A(t-t_{0})^{\lambda}r^{\delta}
\end{equation}

where $A~(>0), \lambda$ and $\delta$ are constants. Hence we have
\begin{equation}
K^{r}(t_{ORTG},r)=A~r^{q},~~~~q=\lambda p+\delta.
\end{equation}

Also from the geodesic equation (11.46) using (11.48) we get

\begin{equation}
\frac{dt_{ORTG}}{dr}=cpr^{p-1}=\frac{K^{t}}{K^{r}}=\sqrt{A^{-2}r^{-2q}+(R'+R\nu')^{2}}
\end{equation}

Using the solution for $R$ near the singularity in equation
(11.49) and equating equal powers of $r$ we have

(i) $p=1-q$,~~$c=\frac{1}{A(1-q)}$~~ if~~
$-\frac{2m}{n+1}<q<1$.\\

(ii)
$p=1+\frac{2m}{n+1}$,~~$c=\frac{(\nu_{_{-1}}+1+\frac{2m}{n+1})}{(1+\frac{2m}{n+1})}
\left(-\frac{F_{m}}{2F_{0}}\right)^{\frac{2}{n+1}}$~~ if~~
$q<-\frac{2m}{n+1}$\\

(iii)
$p=1+\frac{2m}{n+1}$,~~$c=\frac{1}{p}\left[A^{-2}+\left(-\frac{F_{m}}{2F_{0}}
\right)^{\frac{4}{n+1}}\left(\nu_{_{-1}}+1+\frac{2m}{n+1}\right)^{2}\right]^{1/2}$~~
if~~ $q=-\frac{2m}{n+1}$\\

Therefore equation (11.46) has consistent solution for $c$ and $p$
depending on the parameters involved. So it is possible to have
outgoing radial time-like geodesic originated from the
singularity.\\

\section{Local Visibility}
We have shown the existence of outgoing both null and time-like
geodesics which were originated in the past from the singularity.
We shall now examine whether these geodesics are visible to any
non-space like observer. Now for visibility of the singularity
(local nakedness) any future directed geodesics (null or
time-like) starting from the singularity should be outside the
domain of dependence of any trapped surfaces both before and at
the time of formation of apparent horizon (AH) which is the outer
boundary of the trapped surface. For $\Lambda=0$ and $f(r)=0$ the
time of formation of apparent horizon $t_{ah}(r)$ satisfies (from
eq. (11.17))

\begin{equation}
R^{n-1}(t_{ah}(r),r)=F(r)
\end{equation}

The apparent horizon is given by the curve

\begin{equation}
t_{ah}(r)=t_{0}-\frac{1}{(n+1)F_{0}^{3/2}}\left(F_{m}r^{m}+O(r^{m+1})\right)
-\frac{2}{n+1}F_{0}^{\frac{1}{n-1}}\left(r^{\frac{n+1}{n-1}}+O(r^{\frac{n+1}{n-1}+1})\right)
\end{equation}
\\
As the apparent horizon and the singularity curve form at the
same time at $r=0$, so the visibility of the singularity is
determined by the relative slopes of the curve $t_{ah}(r)$ and
the curve for outgoing radial non-space like geodesics (denoted
by $t_{ORG}(r)$). Hence the necessary and sufficient condition
for singularity to be at least locally naked is that

\begin{equation}
\begin{array}{c}
lim\\
r\rightarrow 0\\
\end{array}
\begin{array}{c}
\left\{(\frac{dt_{ah}}{dr})/(\frac{dt_{ORG}}{dr})\right\}>1\\
{}
\end{array}
\end{equation}

Now we shall examine this condition for both ORNG and ORTG
separately.\\

\subsection{Outgoing Radial Null Geodesic}
In this case the ratio of the slopes is

\begin{eqnarray*}
\frac{(\frac{dt_{ah}(r)}{dr})}{(\frac{dt_{ORNG}(r)}{dr})}=-\frac{1}{a\xi(n+1)F_{0}^{3/2}}
\left(mF_{m}r^{m-\xi}+O(r^{m-\xi+1})\right)
\end{eqnarray*}
\vspace{-3mm}
\begin{equation}
~~~~~~~~~~~~~~~~-\frac{2}{a\xi(n-1)}F_{0}^{\frac{1}{n-1}}
\left(r^{\frac{n+1}{n-1}-\xi}+O(r^{\frac{n+1}{n-1}-\xi+1})\right)
\end{equation}

So we have the following possibilities:

(i) If $m>\frac{n+1}{n-1}$ then the above ratio of the slope is
always negative, hence the condition is violated. So we always
get black hole.

(ii) If $m<\frac{n+1}{n-1}$ then the above ratio of the slopes
approaches to $+\infty$ as $r\rightarrow 0^{+}$ for $m<\xi$ and
while the limit will be finite and greater than unity for $m=\xi$
provided $a<-\frac{m}{(n+1)\xi}\frac{F_{m}}{F_{0}^{3/2}}$. Thus
the singularity will be locally naked for any dimension if $m=1$
and $a<-\frac{1}{(n+1)\xi}\frac{F_{1}}{F_{0}^{3/2}}$ while naked
singularity appears only for four dimension if $m=2$ and
$a<-\frac{2}{3\xi}\frac{F_{2}}{F_{0}^{3/2}}$.

(iii) If $m=\frac{n+1}{n-1}$ then we have shown that naked
singularity appears only upto five dimension and the singularity
will be locally visible for
\[
F_{m}<-2F_{0}^{\frac{m}{2}+1},~~~~m<\xi
\]
\[\text{or}\]
\[
F_{m}<-2F_{0}^{\frac{m}{2}+1}-a(n+1)F_{0}^{3/2},~~~~m=\xi.
\]
\vspace{3mm}
\subsection{Outgoing Radial Time-Like Geodesic}
In this case the ratio of the slopes is

\begin{eqnarray*}
\frac{(\frac{dt_{ah}(r)}{dr})}{(\frac{dt_{ORTG}(r)}{dr})}=-\frac{1}{cp(n+1)F_{0}^{3/2}}
\left(mF_{m}r^{m-p}+O(r^{m-p+1})\right)
\end{eqnarray*}
\vspace{-3mm}
\begin{equation}
~~~~~~~~~~~~~~~~-\frac{2}{cp(n-1)}F_{0}^{\frac{1}{n-1}}
\left(r^{\frac{n+1}{n-1}-p}+O(r^{\frac{n+1}{n-1}-p+1})\right)
\end{equation}

We shall now study the following possibilities for local
visibility:

(i) If $m>\frac{n+1}{n-1}$ then as before only black hole
appears.\\

(ii) If $m<\frac{n+1}{n-1}$ we have the conclusion as in ORNG
except that instead of restricting $a$, we have here
$c<-\frac{m}{(n+1)p}\frac{F_{m}}{F_{0}^{3/2}}$.

(iii) Similar is the situation for $m=\frac{n+1}{n-1}$.

Hence the conditions for local visibility of naked singularity is
consistent for both time-like and null geodesic.\\

\section{Geodesics and the Nature of Singularity}
Here we consider $\Lambda=0$ and $f(r)=0$. So the solution of $R$
is given in equation (11.24). We now follow the geodesic analysis
of Joshi and Dwivedi [1993] for TBL model. For simplicity of
calculation we introduce the following functions:

\begin{equation}
\left.
\begin{array}{llll}
X=\frac{R}{r^{a}} &  &  &  \\
&  &  &  \\
\xi =\frac{rF^{\prime }}{F} &  &  &  \\
&  &  &  \\
\eta =\frac{rQ^{\prime }}{Q} &  &  &  \\
&  &  &  \\
\zeta =\frac{F}{r^{a(n-1)}} &  &  &  \\
&  &  &  \\
Q=e^{-\nu } &  &  &  \\
&  &  &  \\
\Theta =\frac{1-\frac{\xi }{n+1}}{r^{\frac{(a-1)(n+1)}{2}}} &  &  &  \\
\end{array}%
\right\}
\end{equation}
\\
where the constant $a$ is restricted by $a\geq 1$. In order to determine the
nature of the singularity we examine whether it is possible to have outgoing
null geodesics which are terminated in the past at the central singularity $%
r=0$. Suppose this occurs at singularity time $t=t_{0}$ at which $%
R(t_{0},0)=0$. In order to decide this we start with radial null geodesic,
given by
\begin{equation}
\frac{dt}{dr}=e^{\alpha }=\frac{R^{\prime }Q-RQ^{\prime }}{Q}.
\end{equation}%
Next, we introduce the notation $u=r^{a}$, so we write

\begin{equation}
\frac{dR}{du}=U(X,u),
\end{equation}%
where ~~~~$U(X,u)=\left( 1-\sqrt{\frac{\zeta }{X^{n-1}}}\right) \frac{H}{a}+%
\frac{\eta }{a}\sqrt{\frac{\zeta }{X^{n-3}}}$ ~~~~ with~~~~ $H=\frac{\xi }{%
n+1}~X+\frac{\Theta }{X^{\frac{n-3}{2}}}$ .\newline

We now study the limiting behaviour of the function $X$ as we approach the
singularity at $R=0,u=0$ along the radial null geodesic identified above. If
we denote the limiting value by $X_{0}$ then

\begin{eqnarray}
\begin{array}{c}
X_{0}~= \\
{}%
\end{array}
\begin{array}{c}
lim~~~~~~~~~ \frac{R}{u} \\
R\rightarrow 0~ u\rightarrow 0%
\end{array}
\begin{array}{c}
=~lim~~~~~~~~~ \frac{dR}{du} \\
~~~R\rightarrow 0~ u\rightarrow 0%
\end{array}
\begin{array}{c}
=~~lim~~~~~~~~ U(X,u) \\
R\rightarrow 0~ u\rightarrow 0%
\end{array}
\begin{array}{c}
=U(X_{0},0) \\
{}%
\end{array}%
\end{eqnarray}

Thus if this polynomial equation in $X_{0}$ has at least one
positive real root then it is possible to have a radial null
geodesic outgoing from the central singularity. More explicitly,
$X_{0}$ is the root of the following equation in $X$:\\
\begin{equation}
\left( a-\frac{\xi _{0}}{n+1}\right) X^{n}+\left( \frac{\xi _{0}}{n+1}-\eta
_{0}\right) \sqrt{\zeta _{0}}~X^{\frac{n+1}{2}}-\Theta _{0}X^{\frac{n-1}{2}}+%
\sqrt{\zeta _{0}}~\Theta _{0}=0
\end{equation}%
\\
where suffix `$o$' for the variables $\xi ,\eta ,\zeta $ and
$\Theta $ stands for the values of these quantities at $r=0$. As
exact analytic solution is not possible for $X$ so we study the
roots by numerical methods. Table II shows the dependence of the
nature of the roots on the variation of the parameters
involved.\newpage

Table II: Positive roots ($X_{0}$) of the eqn. (11.59) for
different values of the parameters namely $a,\eta _{0},\xi
_{0},\zeta _{0},\Theta _{0}$ and in different dimensions, $n$.

\begin{center}
\begin{tabular}{|l|}
\hline\hline
~$a$~~~~~$\eta_{0}$~~~~~$\xi_{0}$~~~~$\zeta_{0}$~~~~$\Theta_{0}$%
~~~~~~~~~~~~~~~~~~~~~~~~~~~Positive roots ($X_{0}$) \\ \hline
\\
~~~~~~~~~~~~~~~~~~~~~~~~~~~~~~~~~~~~~~4D~~~~~~~~5D~~~~~~~~~6D~~~~~~~~~7D~~~~~~~~8D~~~~~~~~10D~~~~~~14D
\\ \hline\hline
\\
~1~~~~-6~~~~~.05~~~.01~~~~-5~~~~.00997~~~~.09864~~~~.21131~~~~.30963~~~~.38979~~~~.50797~~~.64774
\\
\\
~2~~~~~2~~~~~~~5~~~~~3~~~~~2~~~~3.3019,~~~~1.732,~~~~1.4422,~~~1.1968,~~~~1.2457,~~~1.1699,~~~1.105,
\\
~~~~~~~~~~~~~~~~~~~~~~~~~~~~~~~~~~~3~~~~~~~~~~~1.633~~~~~1.3195~~~~1.3161~~~~~1.1345~~~~1.075~~~~~1.0334
\\
\\
~3~~~~~2~~~~~~~5~~~~~3~~~~~2~~~~~~~$-$~~~~~~~~~~$-$~~~~~~~~~~$-$~~~~~~~~~~$-$%
~~~~~~~~~~~$-$~~~~~~~~~$-$~~~~~~~~~$-$ \\
\\
~5~~~~~4~~~~~~.1~~~~~3~~~~~2~~~~1.7068,~~~1.2678,~~~~1.1613,~~~1.1148,~~~~1.0889,~~~1.0612,~~~1.0377,
\\
~~~~~~~~~~~~~~~~~~~~~~~~~~~~~~~~~~.77462~~~~.80678~~~~~.82742~~~~.84873~~~~~.86583~~~~.89102~~~.92113
\\
\\
~1~~~~~0~~~~~~~9~~~~~1~~~~~1~~~~2.035~~~~~1.6116~~~~~1.5163~~~~1.5028~~~~~1.5511~~~~~~%
$-$~~~~~~~~~$-$ \\
\\
~1~~~~~0~~~~~~~4~~~~~1~~~~~1~~~14.8239~~~~~~~$-$~~~~~~~~~~$-$~~~~~~~~~~$-$%
~~~~~~~~~~~$-$~~~~~~~~~$-$~~~~~~~~~$-$ \\
\\
~1~~~~.1~~~~~~~1~~~~~1~~~~~1~~~~~~~$-$~~~~~~~~~~~$-$~~~~~~~~~~$-$~~~~~~~~~~$%
- $~~~~~~~~~~~$-$~~~~~~~~~$-$~~~~~~~~~$-$ \\
\\
~1~~~~~1~~~~~~~1~~~~~1~~~~~1~~~~1.3104,~~~1.1547,~~~~1.0934,~~~1.0627,~~~~1.045,~~~~~1.0265,~~1.0124,
\\
~~~~~~~~~~~~~~~~~~~~~~~~~~~~~~~~~~~1~~~~~~~~~~~1~~~~~~~~~~~~1~~~~~~~~~~1~~~~~~~~~~~~1~~~~~~~~~~~1~~~~~~~~~1
\\
\\
~4~~~~~0~~~~~~~5~~~~.1~~~~~4~~~~~~~$-$~~~~~~~~~~~$-$~~~~~~~~~~~$-$~~~~~~~~~$%
- $~~~~~~~~~~~$-$~~~~~~~~~$-$~~~~~~~~~$-$ \\
\\
~1~~~~~0~~~~~~~1~~~~~1~~~~~9~~~~2.25,~~~~~~2.5336,~~~~2.2120,~~~1.9675,~~~~1.7967,~~~1.5829,~~1.3756,
\\
~~~~~~~~~~~~~~~~~~~~~~~~~~~~~~~~~~1.5963~~~~1.1725~~~~~1.1045~~~~1.0752~~~~~1.0588~~~~1.041~~~~1.0256
\\
\\
~1~~~~~0~~~~~~~1~~~~~5~~~~~1~~~~~~~$-$~~~~~~~~~~~$-$~~~~~~~~~~$-$~~~~~~~~~~$%
- $~~~~~~~~~~~$-$~~~~~~~~~$-$~~~~~~~~~$-$ \\
\\
~1~~~~~0~~~~~~~1~~~~.1~~~~~1~~~~.90297,~~~.87007,~~~~.86041,~~~.85671,~~~~.85235,~~~~~~%
$-$~~~~~~~~~$-$ \\
~~~~~~~~~~~~~~~~~~~~~~~~~~~~~~~~~~.10733~~~~.36226~~~~~.54185~~~~.66031~~~~.74455
\\
\\ \hline\hline
\end{tabular}
\end{center}

\vspace{5mm}

From the definition of the functions in equation
(11.55) it is clear that the parameters $a,\xi _{0},\zeta _{0}$
are always positive while $\eta _{0},\Theta _{0}$ can have
positive as well as negative values. From the Table we see that
as we decrease the value of $a$, keeping the other parameters
fixed, then the formation of a naked singularity is more probable
in higher dimensions. The situation for $\zeta _{0}$ is similar. But for $%
\xi _{0}$ and $\eta _{0}$ the condition is reversed: black hole
formation becomes more probable as we increase the dimension of
the space-time if we decrease the value of $\xi _{0}$ (or $\eta
_{0}$) and keep the other constants fixed. However, no such
conclusions can be drawn about the variation of $\Theta _{0}$.\\

\section{Strength of the Naked Singularity}
A singularity is called \textit{gravitationally strong,} or simply \textit{%
strong,} if it destroys by crushing or tidally stretching to zero
volume all objects that fall into it; it is called \textit{weak}
if no object that falls into the singularity is destroyed in this
way. A precise characterization of Tipler strong [Tipler, 1987]
singularities has been given by Clarke and Krolak [1986], who
proposed the strong focusing condition. A sufficient condition
for a strong curvature singularity is that, for at least one non
space-like geodesic with affine parameter $\lambda \rightarrow 0$
on approach to the singularity, we must have

\begin{equation}
\begin{array}{c}
lim \\
\lambda \rightarrow 0 \\
\end{array}%
\begin{array}{c}
\lambda ^{2}R_{ij}K^{i}K^{j}>0 \\
\\
\end{array}%
\end{equation}%
where $K^{i}=\frac{dx^{i}}{d\lambda }$ is the tangent vector to
the radial null geodesic.

Our purpose here to investigate the above condition along
future-directed radial null geodesics that emanate from the naked
singularity. Now equation (11.60) can be expressed as (using
L'H\^{o}pital's rule)

\begin{equation}
\begin{array}{c}
lim \\
\lambda \rightarrow 0 \\
\end{array}%
\begin{array}{c}
\lambda ^{2}R_{ij}K^{i}K^{j} \\
\\
\end{array}%
\begin{array}{c}
=\frac{n\zeta _{0}(H_{0}-\eta _{0}X_{0})(\xi _{0}-(n+1)\eta _{0})}{%
2X_{0}^{^{n}}\left( N_{0}+\eta _{0}\sqrt{\frac{\zeta _{0}}{X_{0}^{n-1}}}%
\right) ^{2}} \\
\end{array}%
\end{equation}%
where $H_{0}=H(X_{0},0),N_{0}=N(X_{0},0)$.

The singularity is gravitationally strong in the sense of Tipler if

\[
\xi _{0}-(n+1)\eta _{0}>max\left\{ 0,-\frac{(n+1)\Theta _{0}}{X_{0}^{\frac{%
n+1}{2}}}\right\}
\]%
or
\[
\xi _{0}-(n+1)\eta _{0}<min\left\{ 0,-\frac{(n+1)\Theta _{0}}{X_{0}^{\frac{%
n+1}{2}}}\right\}
\]

If the above condition is not satisfied for the values of the
parameters then\\ $
\begin{array}{cl}
lim  \\
\lambda \rightarrow 0
\end{array}
\begin{array}{c}
\lambda ^{2}R_{ij}K^{i}K^{j}\leq 0{}
\end{array}
$ and the singularity may or may not be Tipler strong.\\

\section{Discussion}
In this chapter, we have studied gravitational collapse in
$(n+2)$-dimensional space-time using a higher-dimensional
generalization of the quasi-spherical Szekeres metrics with
non-zero cosmological constant. We have examined the local nature
of the central shell-focusing singularity by a comparative study
of the time of the formation of trapped surface and the time of
formation of the central shell-focusing singularity. If we assume
the initial density gradient falls off rapidly and vanishes at
$r=0,$ then naked singularity formation is possible only up to
space-time dimension five for $\nu_{_{-1}}>-1$. However, if we
drop the above restriction on the initial density distribution
then the Cosmic Censorship Conjecture may be violated in any
dimension ($n\geq 2$). Following the approach of Barve et al
[1999] for null geodesic we have shown that if we choose the
parameter $\nu_{_{-1}}$ to be greater than $-1$, then it is
possible to have a class of outgoing radial null geodesic for
four and five dimensions only with the restriction that initial
density falls off rapidly to the centre. However for
$\nu_{_{-1}}=-1$, naked singularity is possible in all dimensions
irrespective of the assumption on the initial density. Thus we
deduce that the nature of the central singularity depends
sensitively in these metrics on the choice of the initial data
(particularly on the choice of initial density profile). This is
also confirmed by our geodesic study following the approach of
Joshi and Dwivedi [1993] where we have shown numerically that the
nature of singularity depends on the value of the defining
parameters at $r=0$. Finally, we have examined the strength of
the naked singularity using the criterion introduced by Tipler
[1987]. We found that the naked singularity will be a strong
curvature singularity depending on the appropriate choice of the
value of the parameters at $r=0$. On the other hand, for
time-like geodesic the results are very similar to the study of
null geodesic. Regarding local visibility we have consistent
results for both ORNG and ORTG with some restrictions on the
parameters involved in the equation of the geodesics. Therefore
we conclude that it is possible to have at least local naked
singularity for the given higher dimensional non-spherical
space-time. These investigations provide some insights into the
phenomenon of gravitational collapse in a situation without any
imposed Killing symmetries. However, the collapses are special in
other senses which permit exact solutions to be found. In
particular, there is an absence of gravitational radiation in
these space-times [Bonnor, 1976]. An investigation of its role is
a challenge for future analytic and computational investigations.\\

\large \baselineskip .85cm
\chapter{Shell Crossing Singularities in Quasi-Spherical Szekeres Models
} \label{chap12}\markright{\it CHAPTER~\ref{chap12}. Shell
Crossing Singularities in Quasi-Spherical Szekeres Models }

\section{Prelude}

In the study of gravitational collapse, we always encounter with
two types of singularities $-$ shell focusing singularity and
shell crossing singularity. In Tolman-Bondi-Lema\^{\i}tre (TBL)
dust model, these two kinds of singularities will corresponds to
$R=0$ and $R'=0$ respectively. A shell focusing singularity (i.e.,
$R=0$) on a shell of dust occurs when it collapses at or starts
expanding from the centre of matter distribution. The instant at
which a shell at the radial co-ordinate $r$ will reach the centre
of the matter distribution should be a function of $r$. So
different shells of dust arrive at the centre at different times
and there is always a possibility that any two shells of dust
cross each other at a finite radius in course of their collapse.
In this situation the comoving system breaks down, both the
matter density and kretchman scalar diverge [Seifert, 1979;
Hellaby et al, 1984, 1985] and one encounters the shell crossing
singularity. If one treats it as the boundary surface then the
region beyond it is unacceptable since it has negative density.
It is therefore, of interest to find conditions which guarantee
that no shell crossing will occur.\\

Goncalves [2001] studied the occurrence of shell crossing in
spherical weakly charged dust collapse in the presence of a
non-vanishing cosmological constant. The positive cosmological
constant model conceively prevent the occurrence of shell
crossing thereby allowing at least in principle for a singularity
free `bounce' model. Nolan [1999, 2003] derive global weak
solutions of Einstein's equations for spherically symmetric
dust-filled space-times which admit shell crossing singularities.
Recently, Hellaby et al [2002] investigate the anisotropic
generalization of the wormhole topology in the Szekeres model. We
have studied the shell crossing singularity in Szekeres model of
the space-time both from physical and geometrical point of view.
We study the physical conditions and geometrical features of
shell crossing singularities.\\

\section{Physical Conditions for Shell Crossing Singularity}
Here we consider ($n+2$) dimensional Szekeres space-time model
with metric ansatz is given in equation (10.1). Under the
assumption that $\beta ^{\prime }\neq 0$, the explicit form of
the metric coefficients are also given in equations (10.2) and
(10.3), where the form of $\nu$ is given in equations (10.16) and
(10.17) and $R$ satisfied the differential equation (11.1) for
non-zero cosmological constant $\Lambda$.

We shall now make a comparative study of shell focusing and shell
crossing singularity time and find conditions in favour (or
against) of formation of shell crossing singularity for the
following different choices:\newline

(i)~~ $f(r)=0$, $\Lambda= 0$:

In this case equation (11.1) can be integrated to give
\begin{equation}
R^{\frac{n+1}{2}}=r^{\frac{n+1}{2}}-\frac{n+1}{2}\sqrt{F(r)}~(t-t_{i})
\end{equation}

So from (11.9), $R(t_{sf}(r),r)=0$ gives the result
\begin{equation}
t_{sf}(r)=t_{i}+\frac{2}{(n+1)\sqrt{F(r)}}~r^{\frac{n+1}{2}}
\end{equation}

Now to avoid the shell crossing singularity either all shells
will collapse at the same time (i.e., $t_{sf}(r)$ is independent
of $r$) or larger shell will collapse at late time (i.e.,
$t_{sf}(r)$ is a monotone increasing function of $r$). These two
conditions can be combined as
$$
t'_{sf}(r)\ge 0
$$

or equivalently from equation (12.2)
\begin{equation}
\frac{F'(r)}{F(r)}\le \frac{n+1}{r}
\end{equation}

Now combining equations (11.10), (12.1) and (12.2) we have

\begin{equation}
t_{sc}(r)-t_{sf}(r)=\frac{2r^{\frac{n+1}{2}}\left\{\frac{n+1}{r}-\frac{F'(r)}{F(r)}
\right\}} {(n+1)\sqrt{F(r)}\left\{\frac{F'(r)}{F(r)}+(n+1)\nu'
\right\}}
\end{equation}

But if it is so happen that $R'+R\nu'=0$ is a regular extremum for
$\beta$, then we must have finite $\rho$. This implies from
equation (11.2) that $F'+(n+1)F\nu'=0$. Hence, if there is no
shell crossing singularity corresponding to equation (11.10) we
must have two possibilities:
\begin{equation}
\text{either~~~~(a)}~~\frac{F'(r)}{F(r)}+(n+1)\nu'=0~~~~\text{and}~~~
\frac{n+1}{r}-\frac{F'(r)}{F(r)}=0
\end{equation}

\begin{equation}
\text{or~~~~~~~~(b)}~~\frac{F'(r)}{F(r)}+(n+1)\nu'=0~~~~\text{and}~~~
\frac{n+1}{r}-\frac{F'(r)}{F(r)}>0
\end{equation}

For the first choice $t_{sf}(r)$ is constant, so all shells
collapse simultaneously while for the second choice $t_{sf}(r)$ is
a monotonic increasing function of $r$ and there is an infinite
time difference between the occurrence of both type of
singularities.

The value of $R$ at $t=t_{sc}(r)$ is

\begin{equation}
\left\{R(t_{sc},r)\right\}^{\frac{n+1}{2}}=\frac{r^{\frac{n+1}{2}}
\left\{\frac{F'(r)}{F(r)}-\frac{n+1}{r}\right\}}
{\left\{\frac{F'(r)}{F(r)}+(n+1)\nu' \right\}}
\end{equation}

Therefore as an complementary event, the conditions for occurrence
of shell crossing singularity are
$R'+R\nu'=0,~\rho=\infty,~\dot{R}<0,~R>0,~t'_{sf}(r)<0$.

As $t'_{sf}(r)<0$ implies
\begin{equation}
\frac{F'(r)}{F(r)}>\frac{n+1}{r}
\end{equation}

so $R>0$ demands
\begin{equation}
\frac{F'(r)}{F(r)}+(n+1)\nu'>0
\end{equation}

Hence we have
\begin{equation}
F(r)\sim r^{l} ~~\text{and}~~ e^{\nu}\sim r^{p}
\end{equation}

for shell crossing singularity with $l>(n+1)$ and $(n+1)p>-l$.\\

(ii)~~ $f(r)=0$, $\Lambda\ne 0$:

This choice will give the solution to equation (11.1) as

\begin{equation}
t=t_{i}+\sqrt{\frac{2n}{(n+1)\Lambda}}\left[Sinh^{-1}\left(\sqrt{\frac{%
2\Lambda r^{n+1}}{n(n+1)F(r)}}\right)-Sinh^{-1}\left(\sqrt{\frac{2\Lambda
R^{n+1}} {n(n+1)F(r)}}\right)\right]
\end{equation}%
At the shell focusing time $t_{sf}(r)$,~$R=0$, hence we have

\begin{equation}
t_{sf}(r)=t_{i}+\sqrt{\frac{2n}{(n+1)\Lambda}}~Sinh^{-1}\left(\sqrt{\frac{%
2\Lambda r^{n+1}}{n(n+1)F(r)}}\right)
\end{equation}
\\
Since to avoid the shell crossing singularity, $t_{sf}(r)$ should
be an increasing (or a constant) function of time i.e.,
$t'_{sf}(r)\ge 0$, which implies the same condition (12.3) as in
case (i) and is independent of $\Lambda$ (whether zero or not).
Further, substitution of equation (12.11) in equation (11.10) will
give
\begin{equation}
t_{sc}(r)-t_{sf}(r)=tanh^{-1}\left[\frac{\sqrt{\frac{2(n+1)\Lambda}{n}}
~t'_{sf}(r)}{\left\{\frac{F'(r)}{F(r)}+(n+1)\nu'\right\}}\right]
\end{equation}

But if there is no shell crossing singularity corresponding to
$R'+R\nu'=0$ (then it will correspond to an extremum of $\beta$)
then $\rho$ must be finite. This will be possible only when
$F'+(n+1)F\nu'=0$. But from eq.(12.13) it is permissible only when
$t'_{sf}(r)=0$ i.e., $t_{sf}(r)$ is independent of $r$.\\

(iii)~~ $f(r)\neq 0, \Lambda=0,~ \dot{R}(t_{i},r)=0$ (\textit{time
symmetry}):

In this case explicit solution is possible only for five
dimension (i.e., for $n=3$) and the result as
\begin{equation}
R^{2}=r^{2}-\frac{F(r)}{r^{2}}(t-t_{i})^{2},
\end{equation}%
But the shell focusing condition $R(t_{sf}(r),r)=0$ gives
$$
t_{sf}(r)=t_{i}+\frac{r^{2}}{\sqrt{F(r)}}.
$$

So $t'_{sf}(r)\ge 0$ will give
\begin{equation}
\frac{F'}{F}\le \frac{4}{r}
\end{equation}

Here the time difference between the two types of singularities is
\begin{equation}
t_{sc}(r)-t_{sf}(r)=\frac{r^{2}}{\sqrt{F}}\left[\sqrt{\frac{\nu'+\frac{1}{r}}
{\frac{F'}{2F}+\nu'-\frac{1}{r}}}~~-1 \right]
\end{equation}

The r.h.s. of equation (12.16) always positive by the inequality
(12.15).\\

\section{Geometrical Features of Shell Crossing Singularity}
Now we shall discuss the shell crossing singularity from
geometrical point of view. In fact geometrically, a shell crossing
singularity (if it exists) is the locus of zeros of the function
$R'+R\nu'$. Now writing explicitly the function $R'+R\nu'$ using
the solution (10.16) for $\nu$ we have\\
\begin{equation}
R'+R\nu'=e^{\nu}\left[(R'A-RA')\sum_{i=1}^{n}x_{i}^{2}+\sum_{i=1}^{n}(R'B_{i}-RB'_{i})~x_{i}+
(CR'-C'R) \right]
\end{equation}

The discriminants of this quadratic function in $x_{i}$'s
($i=1,2,...,n$) with respect to these variables are respectively
\begin{eqnarray*}
\Delta
x_{i}=-4\left[(R'A-RA')^{2}\sum_{k=i+1}^{n}x_{k}^{2}+(R'A-RA')
\sum_{k=i+1}^{n}(R'B_{k}-RB'_{k})~x_{k}\right.
\end{eqnarray*}
\vspace{-3mm}
\begin{equation}
\left.+(R'A-RA')(CR'-C'R)-\frac{1}{4}\sum_{l=1}^{i}(R'B_{l}-RB'_{l})^{2}
\right],~~(i=1,2,...,n-1)
\end{equation}

and
\begin{equation}
\Delta
x_{n}=(R'A-RA')^{2}\left[R^{2}\left(\sum_{i=1}^{n}B_{i}'^{2}-4A'C'\right)-R'^{2}\right]
\end{equation}

Thus $R'+R\nu'$ will have the same sign for all $x_{i}$'s when
$\Delta x_{n}<0$ (because then also $\Delta x_{1},... ,\Delta
x_{i}<0, ~i=1,2,...,n-1~~\forall~ x_{i+1},...,x_{n}$). Hence
$R'+R\nu'$ has the same sign for all $x_{i}$'s (i.e., no shell
crossing) if and only if

\begin{equation}
\frac{R'^{2}}{R^{2}}>\sum_{i=1}^{n}B'^{2}_{i}-4A'C'=\psi(r)~\text{(say)}
\end{equation}

But if $\frac{R'^{2}}{R^{2}}=\psi(r)$ then $\Delta x_{n}=0$ and
so $\Delta x_{i}=0$ ($i=1,2,...,n-1$) at just one value of
$x_{n}$. At this value of $x_{n}$, $R'+R\nu'=0$ at one value of
$x_{i}$~ ($i=1,2,...,n-1$). So here shell crossing is a single
point in the constant ($t, r$)-hypersurface ($n$ dimensional). In
other worlds, it is a curve in the $t$-constant ($n+1$)-D
hypersurface and a 2 surface in ($n+2$)-D space-time.

When $\frac{R'^{2}}{R^{2}}<\psi(r)$ then $\Delta x_{n}>0$ and
from equation (12.17), $R'+R\nu'=0$ will represent the equation of
a $n$-hypersphere in the $n$ dimensional $x_{i}$'s plane. Further,
for $\Delta x_{n}>0$ the two limiting values $x_{n}$ at which
$\Delta x_{i}$ ($i=1,2,...,n-1$) changes sign are

\begin{equation}
x_{n_{1,2}}=\frac{-(R'B_{n}-RB'_{n})\pm
\sqrt{R^{2}\left(\sum_{i=1}^{n}B_{i}'^{2}-4A'C'\right)-R'^{2}}}{2(R'A-RA')}
\end{equation}

Then for every $x_{n}$ such that $x_{n_{1}}<x_{n}<x_{n_{2}}$,
there are two values of $x_{n-1}$ (only one if $x_{n}=x_{n_{1}}$
or $x_{n_{2}}$) such that $R'+R\nu'=0$ and so on. These values of
$x_{i}$'s ($i=1,2,...,n-1$) are
\begin{eqnarray*}
x_{i_{1,2}}=\frac{1}{2(R'A-RA')}\left[-(R'B_{i}-RB'_{i})\pm\left\{
R^{2}\left(\sum_{k=1}^{n}B_{k}'^{2}-4A'C'\right)-R'^{2}\right.\right.
\end{eqnarray*}
\vspace{.5mm}
\begin{equation}
\left.\left.-4\left((R'A-RA')
\sum_{l=i+1}^{n}x_{l}+\frac{1}{2}\sum_{l=i+1}^{n}(R'B_{l}-RB'_{l})
\right)^{2}\right\}^{1/2}\right]
\end{equation}

These values of $x_{i}$'s will lie on the hypersphere with centre

$$
(x_{1_{c}},...,x_{n_{c}})=\left(-\frac{(R'B_{1}-RB'_{1})}{2(R'A-RA')},......,
-\frac{(R'B_{n}-RB'_{n})}{2(R'A-RA')} \right)
$$
and radius
\begin{equation}
r_{c}=\frac{\sqrt{R^{2}\left(\sum_{i=1}^{n}B_{i}'^{2}-4A'C'\right)-R'^{2}}}{2(R'A-RA')}
\end{equation}

This hypersphere is different from the hypersphere with $\nu'=0$
i.e.,

\begin{equation}
A'(r)\sum_{i=1}^{n}x_{i}^{2}+\sum_{i=1}^{n}B'_{i}(r)x_{i}+C'(r)=0
\end{equation}

So the shell crossing set intersects with the surface of constant
$r$ and $t$ along the line (curve) $\frac{R'}{R}=-\nu'$=constant.

Now for positive density we note that $F'+(n+1)F\nu'$ and
$R'+R\nu'$ must have the same sign. We now consider the case
where both are positive (when both are negative, we just reverse
the inequalities). When both are zero then it can happen for a
particular value of $x_{i}$'s ($i=1,2,...,n$) if
$\frac{F'}{(n+1)F}=\frac{R'}{R}=-\nu'$, which can not hold for
all time. This is possible for all $x_{i}$ if $F'=R'=\nu'=0$.
This implies that at some $r$, $F'=f'=A'=C'=B'_{i}=0$
($i=1,2,...,n$). Hence we choose
\begin{equation}
\frac{F'}{(n+1)F}>-\nu' ~~~~\text{and}~~~~ \frac{R'}{R}>-\nu'
\end{equation}

Also from the solution (10.16) we have

$$
-\nu'=\frac{A'(r)\sum_{i=1}^{n}x_{i}^{2}+\sum_{i=1}^{n}B'_{i}(r)x_{i}+C'(r)}
{A(r)\sum_{i=1}^{n}x_{i}^{2}+\sum_{i=1}^{n}B_{i}(r)x_{i}+C(r)}
$$
Now writing in a quadratic equation in $x_{1}$ we have for real
$x_{1}$,

$$
\nu'^{2}+\left\{2(A'+A\nu')\sum_{k=2}^{n}x_{k}+\sum_{k=2}^{n}(B'_{k}+B_{k}\nu')
\right\}^{2}\le \sum_{i=1}^{n}B_{i}'^{2}-4A'C'
$$
So
$$
\nu'^{2}|_{max}=\sum_{i=1}^{n}B_{i}'^{2}-4A'C'
$$
Hence from (12.25) we have
\begin{equation}
\frac{F'}{(n+1)F}\ge
\sqrt{\sum_{i=1}^{n}B_{i}'^{2}-4A'C'}~,~~~\forall ~r
\end{equation}

which implies $F'\ge 0$, $\forall~r$. Now for $R'+R\nu'>0$, we
shall study the three possible choices separately.\\

(i)~ $f(r)=0$, $\Lambda=0$:

Here the solution for $R$ can be written as
$$
R^{\frac{n+1}{2}}=\frac{(n+1)}{2}\sqrt{F(r)}~(t-a(r))
$$

So as $t\rightarrow a$,~~
$R^{\frac{n-1}{2}}R'+R^{\frac{n+1}2{}}\nu'\rightarrow
-\sqrt{F(r)}~a'(r)$ and as $t\rightarrow\infty$,
~~$\frac{R'}{R}+\nu'\rightarrow \frac{F'}{(n+1)F}+\nu'$. Hence
for $R'+R\nu'>0$ we must have $a'<0$ and
$\frac{F'}{(n+1)F}>\sqrt{\sum_{i=1}^{n}B_{i}'^{2}-4A'C'}$~.\\

(ii)~ $f(r)=0$, $\Lambda\ne 0$:

The solution for $R$ can be written as
$$
R^{\frac{n+1}{2}}=\sqrt{\frac{n(n+1)F(r)}{2\Lambda}}~Sinh\left[\sqrt{\frac{(n+1)\Lambda}
{2n}}~(t-a(r))\right]
$$

In this case as $t\rightarrow
a$,~~$R^{\frac{n-1}{2}}R'+R^{\frac{n+1}2{}}\nu'\rightarrow
-\sqrt{F(r)}~a'(r)$ and as $t\rightarrow\infty$,
~~$\frac{R'}{R}+\nu'\rightarrow
\frac{F'}{(n+1)F}-\sqrt{\frac{2\Lambda F}{n(n+1)}}~a'(r) +\nu'$.
Thus for $R'+R\nu'>0$ we must have $a'(r)<0$ and
$\frac{F'}{(n+1)F}-\sqrt{\frac{2\Lambda
F}{n(n+1)}}~a'(r)>\sqrt{\sum_{i=1}^{n}B_{i}'^{2}-4A'C'}~.$\\

(iii)~~$f(r)\neq 0, \Lambda=0,~ \dot{R}(t_{i},r)=0,~n=3$:

Here the solution for $R$ is
$$
R^{2}=r^{2}-\frac{F(r)}{r^{2}}(t-t_{i})^{2}~.
$$

The limiting value of $\frac{R'}{R}+\nu'$ ~as~$t\rightarrow
\infty$ will be $\frac{F'}{2F}+\nu'-\frac{1}{r}$. Hence for
$R'+R\nu'>0$ we should have $\frac{F'}{2F}+\frac{1}{r}>
\sqrt{\sum_{i=1}^{n}B_{i}'^{2}-4A'C'}$~.\\

\section{Discussion}
In this chapter,  a details study of shell crossing singularity
has been done for dust model with or without cosmological
constant for Szekeres model of ($n+2$)-D space-time. The physical
conditions however do not depend on $\Lambda$ (whether zero or
not) and the form of the conditions are identical for the three
cases presented there. For geometrical conditions the locus of
shell crossing depends on the discriminant of the co-ordinate
variables $x_{i}$'s ($i=1,2,...,n$). If both $\frac{R'}{R}$ and
$\frac{F'}{(n+1)F}$ are greater than $\sqrt{\sum
B'^{2}_{i}-4A'C'}$ then there will be no shell crossing
singularity even if $R'+R\nu'=0$. Here $\rho$ is finite and
$R'+R\nu'=0$ will correspond to a real extrema for $\beta$. On
the other hand if $\frac{R'}{R}=\sqrt{\sum B'^{2}_{i}-4A'C'}$
then shell crossing singularity is a 2-surface in
$(n+2)$-dimensional space-time. For $\frac{R'}{R}<\sqrt{\sum
B'^{2}_{i}-4A'C'}$, the shell crossing set lie on a
$n$-hypersphere and it intersects with constant $(t,~r)$ along
the curve $\frac{R'}{R}=-\nu'=$constant.\\

\addcontentsline{toc}{part}{Short Discussions and Concluding
Remarks}
 \baselineskip
.81cm

\markright{}

\begin{center}
 { \huge {\bf Short Discussions and Concluding
Remarks} }
\end{center}
\vspace{.7in}

The thesis consists of two parts - Part A and part B. Part A
consists of seven chapters dealing with cosmological solutions
both in Einstein gravity and in Brans-Dicke theory of gravity with
Varying speed of light and also quintessence problem has been
discussed for these solutions. In the last chapter (chapter 7) of
part A, validity of Cosmic No-Hair Conjecture has been studied in
Brane world scenario. Part B consists of five chapters dealing
with spherical gravitational collapse in higher dimensional
Tolman-Bondi-Lema\^{\i}tre models and Quasi-spherical
gravitational collapse in higher dimensional
Szekeres models.\\

In chapter 1, the cosmological implications of the theory of
varying speed of light and a brief idea of Quintessence models
have been discussed.\\

In chapter 2, Brans-Dicke cosmology in an anisotropic
Kantowski-Sachs space-time model have been studied in varying
speed of light theory. The flatness problem has been solved, when
Brans-Dicke scalar field $\phi$ is constant or variable,
specially for radiation dominated era both perturbatively and
non-perturbatively. The other problems like lambda, quasi-lambda
and quasi-flatness problems have been solved for varying speed of
light and their asymptotic behaviour has been
examined.\\

Chapter 3 deals with Friedmann-Robertson-Walker space-time models
of a perfect or causal viscous fluid distribution with a scalar
field $\phi$ having potential $V(\phi)$ for isotropic space-time
model, considering variation of the velocity of light. For
perfect fluid distribution and radiation dominated era, it has
been shown that $V$ is monotonically decreasing function of
$\phi$ and $\phi$ increases with $c$. For viscous fluid
distribution, $V$ decreases exponentially with $\phi$ when
co-efficient of bulk viscosity $\eta\sim\rho^{\frac{1}{2}-m}$ and
$V$ is monotonically decreasing function of $\phi$ when
$\eta\sim\rho$. Therefore, for perfect and causal viscous fluid
distribution, the potential function $V$ is monotonically
decreasing with Brans-Dicke scalar field $\phi$.\\

Quintessence problem in anisotropic cosmological models for
Brans-Dicke theory has been discussed in chapter 4. It has been
shown that it is possible to have an accelerated expanding
universe without any matter due to Brans-Dicke field. In order to
solve the field equations, it has been assumed power-law form of
scale factors. It has also been shown that the anisotropy
character is responsible for getting an accelerated model of the
universe. The flatness problem has been discussed in this context
using conformal transformation in Jordan frame.\\

Chapter 5 deals with anisotropic cosmological models for
Brans-Dicke theory but with a scalar field which is
self-interacting. Assuming the power-law form for the scale
factors and brans-Dicke scalar field, it has been studied the
accelerating or decelerating nature of the universe for various
values of the parameters. For the present universe (i.e., for dust
models) it has been shown in principle that perfect fluid with
barotropic equation of state can be considered to study the
quintessence problem. In addition to the non-decelerating
solution, flatness problem has been solved potentially.\\

An extensive analysis of quintessence problem for Brans-Dicke
theories with varying speed of light have been studied in chapter
6. Here also without any quintessence matter, an accelerated form
of universe is possible. For the power-law  forms  and
exponential  forms  of  the cosmological scale  factors  and
scalar  field, the  velocity  of  light is in the  same form  and
identified  the  cases  where  the quintessence problem  can be
solved  for  some  restrictions on  parameters involved. Along
with providing a non-decelerating solution, it can solve flatness
problem also.\\

In chapter 7, Cosmic No-Hair Conjecture in Brane Scenarios have
been studied. It has been examined that for perfect fluid model
and for general form of the energy momentum tensor for any matter
fields (scalar field, perfect fluids, kinematic gases,
dissipative fluids, etc.) whether Cosmic No-Hair Conjecture in
brane scenario is satisfied if it is valid in general relativity.\\

In chapter 8, a brief review of spherical and non-spherical
gravitational collapse for dust and perfect fluid models have
been discussed.\\

Chapter 9 deals with Tolman-Bondi-Lema\^{\i}tre dust collapse
model in $(n+2)$ dimensional space-time. The local nature of the
central shell-focusing singularity by a comparative study of the
time of the formation of trapped surface and the time of
formation of the central shell-focusing singularity has been
examined. When the initial density gradient falls off rapidly and
vanishes at $r=0$, then naked singularity formation is possible
only up to space-time dimension five for marginally bound case
and any dimension in non-marginally bound case. If the above
restriction on the initial density distribution has been relaxed
then naked singularity is possible in any dimension both for
marginally and non-marginally bound cases. The geodesic study
also supports the above results for marginally bound case. Also
the numerical results for geodesic study in non-marginally bound
case favour the formation of naked singularity in any dimension.\\

Cosmological solutions for ($n+2$)-dimensional Szekeres form of
metric with perfect fluid (or dust) as the matter distribution
and cosmological constant have been studied in chapter 10. The
solutions have been found in two cases namely, quasi-spherical
and quasi-cylindrical. For specific values of the arbitrary
functions (namely, $A(r)=C(r)=\frac{1}{2}$ and $B_{i}(r)=0$,
$\forall~ i$) the Szekeres metric reduces to spherically symmetric
metric for a suitable transformation of co-ordinates. The study of
asymptotic behaviour shows that some of the solutions will become
isotropic at late time while there are solutions for which shear
scalar remains constant throughout the evolution. Thus Cosmic
`no-hair' Conjecture is not valid for all solutions.\\

In chapter 11, gravitational collapse in $(n+2)$-dimensional
space-time using a higher-dimensional generalization of the
quasi-spherical Szekeres metrics with non-zero cosmological
constant have been studied. The local nature of the central
shell-focusing singularity by a comparative study of the time of
formation of trapped surface and the time of formation of the
central shell-focusing singularity has been examined. When initial
density gradient falls off rapidly and vanishes at $r=0$, then
naked singularity formation is possible only up to space-time
dimension five for $\nu_{_{-1}}>-1$ and any dimensional
space-time for $\nu_{_{-1}}=-1$. If the above restriction on the
initial density distribution has been relaxed then the Cosmic
Censorship Conjecture may be violated in any dimension. Also both
by geodesic study and numerical study, it has been deduced that
the nature of the central singularity depends sensitively on the
choice of the initial data. The radial null and radial time-like
geodesics and their local visibility have been studied. Therefore,
it is possible to have at least local naked singularity for the
given higher dimensional quasi-spherical space-time.\\

Finally, in chapter 12, a details study to shell-crossing
singularity has been done for dust model with or without
cosmological constant for Szekeres model in $(n+2)$ dimensional
space-time. The physical conditions do not depend on cosmological
constant and for the geometrical conditions, the locus of
shell-crossing depends on the discriminant of the co-ordinate
variables. Also the conditions for no shell-crossing
singularities have been studied.\\

\addcontentsline{toc}{part}{\bf References of the Papers Presented
in the Thesis}
 \baselineskip
.81cm

\markright{}

\begin{center}
 { \huge {\bf References of the Papers Presented
in the Thesis} }
\end{center}
\vspace{.8in}

\begin{enumerate}

\item  {\bf Brans-Dicke Cosmology in an Anisotropic Model when
       Velocity of Light Varies:} - {\bf\it Subenoy Chakraborty , Narayan Chandra
       Chakraborty and Ujjal Debnath}\\
       {\it International Journal of Modern Physics D}, {\bf 11} 921-932
       (2002).\\
       Pre-print: {\it gr-qc}/0305064.\\

\item  {\bf The Cosmology in a Perfect or Causal Viscous Fluid with Varying Speed of Light:} - {\bf\it
       Subenoy Chakraborty, Narayan Chandra Chakraborty and Ujjal Debnath}\\
       {\it Physica Scripta}, {\bf 68} 399-404 (2003).\\

\item  {\bf Quintessence Problem and Brans-Dicke Theory:} - {\bf\it Subenoy Chakraborty, Narayan Chandra
       Chakraborty and Ujjal Debnath}\\
       {\it Modern Physics Letters A}, {\bf 18} 1549-1555
       (2003).\\
       Pre-print: {\it gr-qc}/0306040.\\

\item  {\bf A Quintessence Problem in Self-Interacting Brans-Dicke Theory:} - {\bf\it \\Subenoy Chakraborty,
       Narayan Chandra Chakraborty and Ujjal Debnath}\\
       {\it International Journal of Modern Physics A}, {\bf 18} 3315-3323
       (2003).\\
       Pre-print: {\it gr-qc}/0305103.\\

\item  {\bf A Quintessence Problem in Brans-Dicke Theory with Varying Speed of Light:} - {\bf\it Subenoy
       Chakraborty , Narayan Chandra Chakraborty and ~~Ujjal Debnath}\\
       {\it International Journal of Modern Physics D}, {\bf 12} 325-335
       (2003).\\
       Pre-print: {\it gr-qc}/0305065.\\

\item  {\bf Does Cosmic No-Hair Conjecture in Brane Scenarios follow from General Relativity?~:} - {\bf\it
       Subenoy Chakraborty and Ujjal Debnath}\\
       {\it Classical and Quantum Gravity}, {\bf 20} 2693-2696
       (2003).\\
       Pre-print: {\it gr-qc}/0212045.\\

\item  {\bf Naked Singularities in Higher Dimensional Gravitational Collapse:} - {\bf\it Asit Banerjee,
       Ujjal Debnath and Subenoy Chakraborty}\\
       {\it International Journal of Modern Physics D}, {\bf 12}
       1255-1264 (2003).\\
       Pre-print: {\it gr-qc}/0211099.\\

\item  {\bf Gravitational Collapse in Higher Dimensional Space-Time:} - {\bf\it Ujjal Debnath and
       Subenoy Chakraborty}\\
       {\it General Relativity and Gravitation}, {\bf 36}
       1243-1253 (2004).\\
       Pre-print: {\it gr-qc}/0211102.\\

\item  {\bf The Study of Gravitational Collapse Model in Higher Dimensional Space-Time:} - {\bf\it
       Ujjal Debnath and Subenoy Chakraborty}\\
       {\it Modern Physics Letters A}, {\bf 18} 1265-1271
       (2003).\\
       Pre-print: {\it gr-qc}/0212062.\\

\item  {\bf A Study of Higher Dimensional Inhomogeneous Cosmological Model:} - {\bf\it Subenoy
       Chakraborty and Ujjal Debnath}\\
       {\it International Journal of Modern Physics D}, {\bf 13}
       1085-1093 (2004).\\
       Pre-print: {\it gr-qc}/0304072.\\

\item  {\bf Quasi-Spherical Gravitational Collapse in Any Dimension:} - {\bf\it Ujjal Debnath,
       Subenoy Chakraborty and John D. Barrow}\\
       {\it General Relativity and Gravitation}, {\bf 36} 231-243 (2004).\\
       Pre-print: {\it gr-qc}/0305075.\\

\item  {\bf Naked Singularities in Higher Dimensional Szekeres Space-Time:} - {\bf\it Ujjal Debnath
       and Subenoy Chakraborty}\\
       {\it Journal of Cosmology and Astroparticle Physics}, {\bf 05} 001
       (2004).\\
       Pre-print: {\it math-ph}/0307024.\\

\item  {\bf Shell Crossing Singularities in Quasi-Spherical Szekeres Models:} - {\bf\it \\Subenoy
       Chakraborty and Ujjal Debnath}\\
       (Communicated)\\

\end{enumerate}

\addcontentsline{toc}{part}{\bf  Bibliography}
 \baselineskip
.81cm \markright{ Bibliography}

\end{document}